\input harvmac
\input epsf
\noblackbox

\def\us#1{\underline{#1}}
\def\nup#1({Nucl.\ Phys.\ $\us {B#1}$\ (}
\def\plt#1({Phys.\ Lett.\ $\us  {B#1}$\ (}
\def\cmp#1({Comm.\ Math.\ Phys.\ $\us  {#1}$\ (}
\def\prp#1({Phys.\ Rep.\ $\us  {#1}$\ (}
\def\prl#1({Phys.\ Rev.\ Lett.\ $\us  {#1}$\ (}
\def\prv#1({Phys.\ Rev.\ $\us  {#1}$\ (}
\def\mpl#1({Mod.\ Phys.\ Let.\ $\us  {A#1}$\ (}
\def\ijmp#1({Int.\ J.\ Mod.\ Phys.\ $\us{A#1}$\ (}
\def\tit#1|{{\it #1},\ }
\def\subsubsec#1{\ \br \noindent {\it #1} \br}
\def\IP{{\bf P}}\def\IC{{\bf C}}
\def\br{\hfill\break}
\def\cx#1{ {\cal #1}}
\def\bx#1{ {\bar{#1}}}

\def\mx#1#2{ \la #1 , #2 \ra}
\def\al{\alpha}\def\be{\beta}
\def\ga{\gamma}\def\de{\delta}\def\ep{\epsilon}
\def\p{\partial}
\def\d{\nabla}
\def\Om{\Omega}
\def\la{\langle} \def\ra{\rangle}

\def\ns{\nu^\star}
\def\x{{\bf X}}\def\xs{{\bf X}^\star}
\def\y{{\bf Y}}\def\ys{{\bf Y}^\star}
\def\z{{\bf Z}}
\def\IR{ {\bf R}}
\def\cs{\IC^\star}
\def\IZ{ {\bf Z}}
\def\IF{{\bf F}}
\def\del{\Delta}\def\dels{\Delta^\star}
\def\pt{{\bf P}(\del)}\def\pts{{\bf P}(\dels)}
\def\hoo{h^{1,1}}\def\hdd{h^{d-1,1}}
\def\hot{h^{1,2}}\def\htt{h^{2,2}}
\def\rs{{\cal R}}
\def\co#1{{\cal O}^{(#1)}}
\def\cc#1{{\cal C}^{(#1)}}\def\ccp#1{{\cal C}^{(#1)\prime }}
\def\ca#1{{\cal Q}^{(#1)}}
\def\gc#1{\gamma^{(#1)}}\def\gcs#1{{{^\star\gamma}}^{(#1)}}
\def\per#1{\Pi ^{(#1)}}\def\pert#1{\tilde{\Pi} ^{(#1)}}
\def\nv{{\bf n}}
\def\nrho{{\bf r}}
\def\Ds{\Delta_{\xs}}
\def\IR{ {\bf R}}
\def\cs{\IC^\star}
\def\IZ{ {\bf Z}}
\def\del{\Delta}\def\dels{\Delta^\star}
\def\pt{{\bf P}(\del)}\def\pts{{\bf P}(\dels)}
\def\hoo{h^{1,1}}\def\hdd{h^{d-1,1}}
\def\hot{h^{1,2}}\def\htt{h^{2,2}}
\def\rs{{\cal R}}
\def\co#1{{\cal O}^{(#1)}}
\def\ca#1{{\cal Q}^{(#1)}}
\def\gc#1{\gamma^{(#1)}}\def\gcs#1{{{^\star\gamma}}^{(#1)}}
\def\per#1{\Pi ^{(#1)}}\def\pert#1{\tilde{\Pi} ^{(#1)}}

\def\pit{\tilde{\pi}}
\def\Dh{\hat{D}}
\def\IT{{\bf T}}
\def\sI{{\bf S}^1}

\def\mcs{ {\cx M}_{CS}}\def\mkm{ {\cx M}_{KM}}
\def\ycl{Y^{\rm cl.}}\def\kcl{K^{\rm cl.}}
\def\bx#1{\bar{#1}}
\def\fff{F^{(4)}}\def\ttt{T^{(3)}}
\lref\agm{P. S. Aspinwall, B. R. Greene, and D. R. Morrison, {\it The
    monomial-divisor mirror map}, Internat. Math. Res. Notices (1993),
    319-337, alg-geom/9309007}
\lref\morn{D. R. Morrison, {\it
Compactifications of moduli spaces inspired by mirror
      symmetry}, Journees de Geometrie Algebrique d'Orsay (Juillet 1992),
      Asterisque, vol. 218, Societe Mathematique de France, 1993,
      pp. 243-271, alg-geom/9304007;
{\it 
The geometry underlying mirror symmetry}, alg-geom/9608006;
{\it 
Mathematical aspects of mirror symmetry}, Complex
      Algebraic Geometry--Park City (J. Kollar, ed.), IAS/Park City
      Math. Series, vol. 3, 1996, pp. 265-340, alg-geom/9609021}
\lref\schubert{S. Katz and S.A. Str\o mme, {\it Schubert: a Maple
package for intersection theory}, ftp.math.okstate.edu/pub/schubert}
\lref\puntos{J. DeLoera, {\it PUNTOS}, ftp.geom.umn.edu/pub/software}
\lref\hkty{S. Hosono, A. Klemm, S. Theisen and S.T. Yau,
\cmp167 (1995) 301,\nup433 (1995) 501}
\lref\lsw{W. Lerche, D.-J. Smit and N. Warner, 
\nup372 (1992) 87}
\lref\gmp{B. R. Greene, D. R. Morrison and M. R. Plesser,
\cmp173 (1995) 559}
\lref\bat{V. Batyrev, Journal Alg. Geom. 3 (1994) 493, Duke Math. Journ., 69 (1993) 349}
\lref\batm{V. V. Batyrev and D. I. Dais, {\it
Strong McKay correspondence, string theoretic Hodge numbers
                  and mirror symmetry}, 
Topology 35    (1996), 901--929,alg-geom/9410001}
\lref\batmII{V. V. Batyrev and L. A. Borisov,
{\it Mirror duality and string-theoretic Hodge numbers},
alg-geom/9509009}
\lref\batII{V. Batyrev, {\it
Quantum Cohomology Rings of Toric Manifolds}, alg-geom/9310004}
\lref\swsix{N. Seiberg and E. Witten, \nup471 (1996) 121 }
\lref\dj{ M. Jinzenji and M. Nagura, \ijmp11 (1996) 171;\br
M. Nagura, \mpl10 (1995) 1677;\br
K. Sugiyama,  hep-th/9410184;hep-th/9504114;hep-th/9504115}
\lref\ste{J. Louis, J. Sonnenschein, S. Theisen and S.
                  Yankielowicz, {\it
Non-perturbative properties of heterotic string vacua
                  compactified on  $K3 \times T^2$}, hep-th/9606049}
\lref\morI{P. S. Aspinwall, B. R. Greene and D. R. Morrison, \nup416 (1994) 414}
\lref\bongI{S. Hosono and B. H. Lian, {\it GKZ Hypergeometric Systems and 
Applications to Mirror Symmetry}, hep-th/9602147}
\lref\wittop{E. Witten, \cmp 117 (1998) 353; {\it 
Mirror manifolds and topological field theory}. In: Essays on 
Mirror Manifolds (ed. S.-T. Yau), Hong Kong: Int. Press, 1992, pp. 120-158}
\lref\flatmet{R. Dijkgraaf, H. Verlinde and E. Verlinde, \nup352 (1991) 59}
\lref\candI{P. Candelas, X. C. De La Ossa, P. S. Green
                  and L. Parkes,\nup359 (1991) 21}
\lref\morII{P. S. Aspinwall and D. R. Morrison, \cmp151 (1993) 245}
\lref\morIII{P. S. Aspinwall, B. Greene and D. R. Morrison, \nup420 (1994) 184}
\lref\sheI{S. Katz, {\it Rational curves on Calabi-Yau manifolds: Verifying
                  predictions of mirror symmetry}, alg-geom/9301006}
\lref\sw{N.\ Seiberg and E.\ Witten, \nup426(1994) 19;
\nup431(1994) 484}
\lref\kmI{A. Klemm and P. Mayr, \nup469 (1996) 37}
\lref\stroII{A. Strominger, \nup451 (1995) 96}
\lref\kmv{A. Klemm, P. Mayr and C. Vafa, {\it BPS States of Exceptional Non-Critical Strings}, hep-th/9607139}
\lref\bkk{P. Berglund, A. Klemm and S. Katz, \nup456 (1995) 153}
\lref\klm{A. Klemm, W. Lerche and P. Mayr, \plt357 (1995) 313}
\lref\kklmv{S. Kachru, A. Klemm, W. Lerche, P. Mayr and C. Vafa,
\nup459 (1996) 537}
\lref\klmvw{A. Klemm, W. Lerche, P. Mayr, N. Warner and C. Vafa,
{\it Self-Dual Strings and $N=2$ Supersymmetric Field Theory}, hep-th/9604034}
\lref\mv{D. R. Morrison and C. Vafa, {\it
Compactifications of F-Theory on Calabi--Yau Threefolds --I,II},
hep-th/9602114; hep-th/9603161}
\lref\eestring{O. J. Ganor and A. Hanany,{\it 
 Small $E_8$ Instantons and Tensionless Non-critical
                  Strings}, hep-th/9602120}
\lref\ald{G. Aldazabal, A. Font, L. E. Ib\'a\~nez and A. M. Uranga,
{\it New Branches of String Compactifications and their F-Theory
                  Duals}, hep-th/9607121}
\lref\dkv{M. R. Douglas, S. Katz and C. Vafa, {\it Small Instantons, del Pezzo Surfaces and 
Type I' theory}, hep-th/0609071;\br
D. R. Morrison and N. Seiberg, {\it Extremal Transitions and Five-Dimensional
Supersymmetric Field Theories}, hep-th/9609070}
\lref\ganorI{O. J. Ganor, {\it 
A Test Of The Chiral E8 Current Algebra On A 6D Non-
                  Critical String}, hep-th/9607020}
\lref\ganorII{O. Ganor, {\it 
Compactification of Tensionless String Theories}, hep-th/9607092}
\lref\witm{E. Witten, \nup471 (1996) 195}
\lref\stromI{A. Strominger, \cmp133 (1990) 163}
\lref\candsg{P. Candelas, \nup298 (1988) 458;\br
 P. Candelas and C. C. de la Ossa,\nup355 (1991) 455}
\lref\witspp{E. Witten, {\it Non-Perturbative Superpotentials 
In String Theory}, hep-th/9604030}
\lref\donI{R. Donagi, A. Grassi and E. Witten, 
{\it A Non-Perturbative Superpotential With $E_8$-Symmetry},
hep-th/9607091}
\lref\antI{B. de Wit, V. Kaplunovsky, J. Louis and
                  D. L\"ust, \nup451 (1995) 53;\br
I. Antoniadis, S. Ferrara, E. Gava, K. S. Narain
                  and T. R. Taylor, \nup447 (1995) 35}
\lref\aft{I. Antoniadis, S. Ferrara and T. R. Taylor, \nup460 (1996) 489}
\lref\wittalk{E. Witten, {\it 
Some comments on string dynamics}, hep-th/9507121}
\lref\kmp{S. Katz,  D. R. Morrison and M. R. Plesser, {\it
Enhanced Gauge Symmetry in Type II String Theory}, hep-th/9601108}
\lref\vafaF{C. Vafa, \nup469 (1996) 403}
\lref\vwI{C. Vafa and E. Witten, {\it 
Dual string pairs with N=1 and N=2 supersymmetry in four-
                  dimensions}, hep-th/9507050}
\lref\kv{S. Kachru and C. Vafa, \nup450 (1995) 69}
\lref\sethi{S. Sethi, C. Vafa and E. Witten, {\it 
Constraints on low-dimensional string compactifications}, hep-th/9606122}
\lref\ff{
I. Brunner and R. Schimmrigk, {\it
F-Theory on Calabi-Yau Fourfolds}, hep-th/9606148;\br
M. Bianchi, S. Ferrara, G. Pradisi, A. Sagnotti
                  and Ya. S. Stanev,{\it
Twelve-Dimensional Aspects of Four-Dimensional N=1 Type I    Vacua},
hep-th/9607105}
\lref\fhsv{S. Ferrara, J. A. Harvey, A. Strominger
                  and C. Vafa, \plt361 (1995) 59}
\lref\berI{M. Bershadsky, K. Intriligator, S. Kachru, 
D. R. Morrison, V. Sadov and C. Vafa,
{\it  Geometric Singularities and Enhanced Gauge Symmetries},  hep-th/9605200}
\lref\bsv{M. Bershadsky, V. Sadov and C. Vafa, \nup463 (1996) 398}
\lref\katzvafa{S. Katz and C. Vafa, {\it Matter from geometry},  
hep-th/9606086}
\lref\kw{W. Lerche and N. Warner, {\it 
Exceptional SW geometry from ALE fibrations}, hep-th/9608183}
\lref\dmw{M. J. Duff, R. Minasian and E. Witten,
\nup465 (1996) 413}
\lref\ksil{S. Kachru and E. Silverstein, {\it
 Singularities, Gauge Dynamics, and Nonperturbative Superpotentials in String
     Theory}, hep-th/9608194}
\lref\std{G. Aldazabal, A. Font, L.  Ib\'a\~nez and F.  Quevedo, \plt380 (1996)
33;\br 
G. Lopes Cardoso, G. Curio, D. L\"ust and  T. Mohaupt,
{\it  Instanton Numbers and Exchange Symmetries in $N=2$ Dual String Pairs},
 hep-th/9603108} 
\lref\witbag{E. Witten and J. Bagger, \plt115 (1982) 202}
\lref\witsi{E. Witten, \nup460 (1996) 541}
\lref\reid{M. Reid in Journ\'ees
de G\'eometrique Alg\'ebraic d' Angers, Juillet 1979,
Sijthoff \& Noordhoff (1980) 273}
\lref\witnew{E. Witten, {\it Physical Interpretation Of Certain Strong Coupling Singularities},
hep-th/9609159 }
\lref\bkkm{P. Berglund, S. Katz, A. Klemm and P. Mayr,
{\it New Higgs Transitions between Dual N=2 String Models},
hep-th/9605154}
\lref\ov{H. Ooguri and C. Vafa, {\it Summing up D-Instantons}, hep-th/9608079}
\lref\bebe{K. Becker and M. Becker, {\it M-Theory on Eight-Manifolds}, hep-th/9605053}
\lref\bebeII{K. Becker, M. Becker, D.R. Morrison, H. Ooguri, Yaron Oz and Zheng Yin,
{\it Supersymmetric Cycles in Exceptional Holonomy Manifolds and Calabi--Yau 4-Folds},
hep-th/9608116}
\lref\bebes{K. Becker, M. Becker and A. Strominger, \nup456 (1995) 130}
\lref\seibIII{N. Seiberg, {\it IR Dynamics on Branes and Space-Time Geometry}, hep-th/9606017}
\lref\swIII{N. Seiberg and E. Witten, {\it 
Gauge Dynamics And Compactification To Three Dimensions}, hep-th/9607163}
\lref\sIII{K. Intriligator and N. Seiberg, {\it 
Mirror Symmetry in Three Dimensional Gauge Theories}, hep-th/9607207;\br
N. Seiberg and S. Shenker, {\it Hypermultiplet Moduli Space and 
String Compactification to Three Dimensions}, hep-th/9608086}
\lref\cec{S. Cecotti, S. Ferrara and L. Girardello, \ijmp4 (1989) 2475}
\lref\cre{E. Cremmer, B. Julia and J. Scherk, \plt76 (1978) 409}
\lref\dlm{M. J. Duff, J. T. Liu and R. Minasian, \nup452 (1995) 261}
\lref\nis{H. Nishino, \mpl7 (1992) 1805}
\lref\stroIII{A. Strominger and E. Witten, \cmp101 (1985) 341}
\lref\ferII{S. Ferrara, R. Khuri and R. Minasian, \plt375 (1996) 81}
\lref\hk{A. Hanany and I. R. Klebanov, {\it On tensionless strings in 3+1 dimensions},
hep-th/9606136}
\lref\town{P. Townsend, \plt373 (1996) 68}
\lref\hs{J. A. Harvey and A. Strominger, \nup449 (1995) 535; \nup458 (1996) 456}
\lref\witflux{E. Witten, {\it On Flux Quantization In M-Theory And The Effective Action},
hep-th//9609122}

\Title{\vbox{
\hbox{CERN-TH/96-269}
\hbox{\tt hep-th/9610162}
}}{Mirror Symmetry, $N=1$ Superpotentials and } 
\vskip-1cm
\centerline{{\titlefont Tensionless Strings on Calabi--Yau Four-Folds}}

\bigskip
\centerline{P. Mayr}
\bigskip
\bigskip
\centerline{\it
Theory Division, CERN, 1211 Geneva 23, Switzerland}

\vskip .3in
We study aspects of Calabi--Yau four-folds as compactification
manifolds of F-theory, using  mirror symmetry of toric hypersurfaces. 
Correlation functions of the topological field theory are determined
directly in terms of a natural ring structure of divisors 
and the period integrals, and subsequently used to extract invariants 
of moduli spaces of rational curves subject
to certain conditions. We then turn to the discussion of 
physical properties of the space-time theories, for a number of examples
which are dual to $E_8\times E_8$ heterotic $N=1$ theories. Non-critical
strings of various kinds, with low tension for special values of the moduli,
lead to interesting physical effects. We give a complete classification
of those divisors in toric manifolds that contribute to the 
non-perturbative four-dimensional superpotential; the 
physical singularities associated to it are related to the
apppearance of tensionless strings. In some
cases non-perturbative effects generate an everywhere non-zero 
quantum tension leading to a combination of a conventional field
theory with light strings hiding at a low energy scale related
to supersymmetry breaking.

\goodbreak

\Date{\vbox{\hbox{\sl {October 1996}}\hbox{CERN-TH/96-269}}}

\parskip=4pt plus 15pt minus 1pt
\baselineskip=15pt plus 2pt minus 1pt
\newsec{Introduction}
Dualities between perturbatively different string theories 
in various dimensions
have led to a considerable improvement of the understanding 
of their non-perturbative aspects. In particular the duality \kv,\fhsv\
between type II strings on Calabi--Yau three-folds and heterotic
string on K3$\times \IT^2$, leading to $N=2$ supersymmetric
theories in four dimensions, makes possible the exact determination of
string theory space-time instanton effects, reducing to the 
exact field theory result of \sw\ after taking appropriate limits \kklmv.
The underlying duality (at present) is however 
now understood as the duality between F-theory in 8 dimensions 
on K3 and heterotic string on $\IT^2$ \vafaF\mv; four-dimensional
type II/heterotic duality than follows from
further fibration over $\IP^1\times \IT^2$,
using variants of the adiabatic argument introduced in \vwI. 

Alternatively one can get theories with minimal $N=1$ supersymmetry
in four dimensions by fibering the eight dimensional duality 
such as to obtain a Calabi--Yau four-fold $\x$ on the F-theory side and
a Calabi-Yau three-fold $\z$ on the heterotic side \vafaF\witspp;
Calabi--Yau four-fold compactifications of F-theory 
have been discussed recently in \sethi,\ff\ksil. While the
geometrical data of the compactification manifolds are largely
fixed by the adiabatic arguments, the choice of the appropriate 
vector bundle on the heterotic side - the generalization of the 
choice of the instanton numbers in K3 $\times \IT^2$ 
compactifications - is not yet known in general, given \x\
on the $F$-theory side. Independently of this question one might
ask to what extent more refined geometrical data of the four-fold -
such as period integrals and the correlation functions calculated
by the topological field theory \wittop\ - will descend to
relevant physical quantities of the $N=1$ compactification.
To address this question it is then natural to attempt to take
advantage of the previously detailed studies of $N=2$ dual pairs
by choosing four-folds obtained as fibrations of three-folds over a further
$\IP^1$.

It is useful to think about the various dual descriptions as obtained
from limits of two--dimensional compactifications. Specifically, after
compactification on $\sI$, F-theory on $\x\times \sI$ is dual to
M-theory on $\x$ and after further compactification on $\sI$ we 
have a duality between F-theory on $\x\times \IT^2$ and type IIA
on $\x$, which is the valid view for the discussion of 
periods and mirror symmetry in a geometrical string theory compactification 
on the four-fold. There are two particularly  interesting limits to consider
starting from this theory: first we can undo the $\sI$ or $\IT^2$ 
compactification by taking special limits in the Calabi--Yau moduli
space. In this case we go back to the four-dimensional $N=1$ theories,
e.g. heterotic string on $\z$. The second is to take the large base space
limit of $\x$. In this case one flows to a theory which looks locally like
$N=2$ in four dimensions, e.g. heterotic sting on K3 times the extra
torus; in fact we will see that one obtains precisely the $N=2$
periods in this limit. It is suggestive to think about the 
world sheet instantons associated to the base $\IP^1$ 
departing from the large base space limit as $N=2$ breaking corrections.
In sect. 2 we discuss 
the behavior of the four-fold periods in the large base limit.

The derivation of the period integrals and the correlation functions
of the topological field theory rely on methods of mirror symmetry 
between Calabi--Yau four-folds. A concept of mirror symmetry for 
Calabi--Yau $d$-folds for $d>3$ has been defined in \gmp\ for one moduli 
cases\foot{For a discussion of mathematical aspects of mirror
symmetry see \agm\morn.}.
For the more complicated four-folds which are relevant in the present context,
we develop the appropriate framework in terms of toric geometry in sect. 3,
defining the fundamental correlation functions of the topological field 
theory directly in terms of the period integrals and a natural ring
structure present in the toric variety.
Other then in $d=3$ and in the one moduli cases considered 
in \gmp, the 3-pt\ functions calculate a whole set of  invariants $N_\al$, 
counting the Euler number of the moduli space of rational curves subjected to 
constraints on the location of the curves in the manifold, which arise
from operators associated to codimension 2 submanifolds in $\x$.

In the second part we apply these methods to elliptically 
fibred four-folds which are fibrations of 
Calabi--Yau three-folds which have itself well-known
$E_8 \times E_8$
heterotic $N=2$ duals in four dimensions. In sect. 4 we determine the
correlation functions and the invariants associated to them and 
describe the geometrical meaning of the K\"ahler moduli which 
relates them to the moduli of the heterotic dual. In sect. 5 we
make some verifications on the numbers calculated in the four-fold
by imposing appropriate constraints and comparing the result with the
known numbers of rational curves in the fibre. We discuss some
interesting properties of the couplings and their role in the
space-time effective theory.

We then turn to the question of whether a superpotential is generated
in the four-dimensional $N=1$ supersymmetric F-theory compactification.
In sect. 6 we analyze possible wrappings of five-branes in an M-theory 
compactification, following \witspp. 
A complete classification of appropriate divisors (six-cycles)
is given using intersection theory on the toric hypersurface;
 it turns out that a superpotential is indeed generated generically.
We then ask about the physical effects related to these superpotential 
terms. As might be expected from the conjectured duality to $E_8\times E_8$
heterotic string on a threefold, tensionless strings and compactifications
of them play an important role. In sect. 7 we investigate 
singularities in the complex structure moduli space related to 
fibration singularities and possible gauge symmetry enhancement and
describe the geometrical properties of the relevant divisors which
provide the link to physical properties.
In some cases the instanton generated superpotential can be interpreted 
as world sheet instantons of the magnetic non-critical string in six 
dimensions. Special singularities which appear in the moduli space when
the five brane intersects or coincide with tensionless strings from
three branes wrappings are discussed in sect. 8. 
A new kind of theories arises if non-perturbative
effects generate a everywhere non-zero tension for the string with 
``classically'' zero tension. 
In this case one obtains in the appropriate scaling limit
a conventional field theory, however with a hidden string
at a non-perturbatively generated low energy scale related
to the scale of supersymmetry breaking.

\newsec{Periods on the four-fold}
One of the first questions about mirror symmetry 
of four-folds and its use to determine non-perturbative effects in 
F-theory compactifications is, which kind of non-perturbative effects
are expected to be treated by the topological sigma model and
which kind are not. In three-fold compactifications  mirror symmetry allows
to determine the exact K\"ahler moduli 
space $\mkm$ of the type IIA theory
on $\y$ from the map to the complex structure moduli space $\mcs$
of the type IIB theory on the mirror $\ys$. 
From the brane point of view this is 
possible since the complex structure moduli are associated to 
3-cycles on $\ys$, however there is no two-brane available in type IIB
which can be wrapped on these 3-cycles to generate an instanton effect. 
Therefore the classical computation
in the type IIB theory is exact and using the mirror map one obtains
information about the world sheet instanton corrected K\"ahler moduli
space of the type IIA theory on $\y$. The same can not be said
about the other moduli space - $\mkm$ of the type IIB
theory on $\y$ or $\mcs$ of the type IIA theory on $\ys$ - 
since the
latter theory has Dirichlet two branes which do generate complex structure 
moduli dependent instanton effects. Moreover the string coupling constant
is a hypermultiplet and there are perturbative corrections in the 
type IIA string theory.

We will be primarily interested in the K\"ahler moduli space of 
type IIA compactified on the four-fold $\x$, including the
corrections to the correlation functions calculated by the 
isomorphisms of the two topological theories, called the 
$A$ and the $B$ model. It would be
interesting to know possible factorization properties of the full
non-perturbative moduli space, a problem which is of course closely
related to a similar question about (0,2) moduli spaces. Generally
we expect that different than in the three dimensional case
there are corrections that are not taken into account by
conventional mirror symmetry based on the isomorphism of
two-dimensional topological theories. However the information provided by
the exact mirror map should be enough to pin down the individual origin
of an instanton effect (thus counting D-branes states ) 
from the scaling behavior whereas the exact contribution will 
contain an additional sum of corrections as e.g. in the case of D2 
brane instantons in type IIA theory \ov. Moreover 
it is an interesting question, what is the freedom that is not fixed by the 
holomorphic bundle structure starting from the apparently rather complete
information on the type IIA side. We will see later that the answer
is related to the the cohomology $H^{2,2}$ of the four-fold, 
which is special in many respects.

In the following paragraph 
we consider F-theory compactification on four-folds $M$ obtained from 
fibering elliptic Calabi--Yau three-folds $\y$ over a two-sphere, $D$, 
of volume $t_D$. If $\y$ 
has a K3 fibration in addition to the elliptic fibration, this theory
is expected to have a heterotic $N=1$ dual by fibre-wise application
of the 8 dimensional duality between F-theory on K3 and heterotic string 
on $T^2$ \vafaF. 

\subsec{Periods in the large base space limit}
It is instructive to consider the large base space limit
of F-theory on $\x \times T^2$; in this case one expects to recover $N=2$
supersymmetric IIA on $\y$ in four dimensions 
or the dual representation, heterotic string on 
$K3\times T^2$. Since we want to use mirror symmetry to extract
physical couplings from the integrals over the holomorphic $(d,0)$
form on a Calabi-Yau manifold $\xs$ it is useful to make precise
this limit on the period integrals.

The observables of the $B$ model on the mirror manifold $\xs$
are in correspondence with elements of the middle cohomology
of $\xs$, $\oplus_k H^{d-k,k}(\xs)$, or rather a subspace of it
in the case of four-folds as will be discussed in more detail in
the next section. Mirror symmetry relates the correlation function
of the $B$ model on $\xs$ to those of the $A$ model on $\x$. 
In the $A$ model observables are associated to cohomology elements 
in $H^{k,k}(\x)$ or equivalently dimension $2k$
homology cycles $\in H_{k,k}(\x)$. The relation between 
the periods of the four-fold and the periods of the three-fold fibre $\y$
can be best understood in this last representation.

In the $A$ model on the three-fold $\y$, the $2+h^{\prime 1,1}$ periods\foot{A prime
refers to the fibre data in the following.}
are related to the 0 cycle $\ccp 0 $, 
$h^{\prime 1,1}$ 2-cycles $\ccp 2 $,
$h^{\prime 1,1}$ 4-cycles $\ccp 4 $ and one 6-cycle $\ccp 6  = \y$
in $\oplus _k H_{k,k}(\y)$.
If we fibre $\y$ over a $\IP^1$ to get a four-fold $\x$, 
we can think of the elements $\cc k \in H_{k,k}(\x)$  
as obtained from joining elements $\ccp l  $ 
with 0 and 2 cycles in the base
$\IP^1$. In this way we get 
$(1,h^{1,1}=h^{\prime 1,1}+1,h_V^{2,2}=
2 h^{\prime 1,1} , h^{3,3}=h^{\prime 1,1}+1,1)$ homology cycles $\cc k $ 
of  the four-fold, generating the so-called vertical primary subspace of
$H^{k,k}(\x)$.

In the $B$ model on $\ys$, expanded in special coordinates $t_\al$ around
a large complex structure point, the $2+2h^{\prime 1,3}
(\ys)=2+2h^{\prime 1,1}(\y)$ 
periods in special coordinates are of the form 
\eqn\periii{\eqalign{
H_{3,0}(\ys) \to 1&, \quad
H_{2,1}(\ys) \to t_\al, \quad\cr
H_{1,2}(\ys) \to \cx F_\al&, \quad
H_{0,3}(\ys) \to \cx F_0 = 2\cx F -\sum_k^{h^{\prime 1,1}} t_\al \cx F_\al\ .
}}
where $F_\al = {\partial\over \partial t_\al} \cx F$ with $\cx F$ the $N=2$ 
prepotential. Note that in the $A$ model the leading classical
terms\foot{ That is powers of the $t_\al$ rather than instanton
corrections $\sim q_\al = e^{2\pi i t_\al}$}
of the periods can be interpreted as the 
volume of the homology cycles. In the large base space limit 
world sheet instanton corrections from the base $\IP^1$ are suppressed
and integrating over the homology cycles $\cc k $ of the four-fold reduces
to an integration over the three-fold cycles $\ccp k $, possibly multiplied by
the classical volume of the base $\IP^1$, if $\cc k $ is obtained
from $\ccp {k-1}$ by joining the whole base. The periods of the four-folds
in the large base space limit are then simply given by combining these
factors with the three-fold result \periii:
\eqn\periiii{
(1,t_D) \times (1,\ t_\al,\ \cx F_\al,\ \cx F_0)
}
From the definition of the homology cycles on $\x$ it is clear
that non-vanishing intersections involve only pairs of 
elements which intersect on $\y$; more precisely there
is a set of $a$ cycles with $\int _{\gamma_a}  \Omega \sim
(1,\ t_\al,\ \cx F_\al,\ \cx F_0)$
and a set of $b$ cycles with $\int _{\gamma_b} \Omega \sim 
t_D \times (1,\ t_\al,\ \cx F_\al,\ \cx F_0) $
with non-vanishing intersection only between $a$ and $b$ cycles and
the intersection form given by that of the Calabi--Yau fibre.
This implies in particular that the K\"ahler potential of the
four-fold $\cx K$ reduces that of the three-fold compactification plus
a constant 
$$
\lim _{t_D \to \infty }e^{- \cx K}  = 
\lim _{t_D \to \infty }
\int_{\xs} \Omega \wedge \bar{\Omega} = 
(t_D-\bar{t}_D) \big( 2 (\cx F - \bar{\cx F}) - \sum_\al(t_\al-\bar{t}_\al)
(\cx F_\al+\bar{\cx F}_\al)\big )
$$
Starting from the precise relation in the large basis limit
of the four-fold periods of the $N=1$ compactification
and the $N=2$ structure of a compactification on the  three-fold fibre,
it is suggestive to treat the instanton corrections associated 
with the base space modulus as $N=2$ breaking deformations 
of a $N=2$ theory. 

Note also that the periods of the four-fold \periiii\ are algebraically
dependent; this is not only true in the large base space limit
but simply a consequence of 
\eqn\trx{
\int \Omega \wedge \Omega = 0
}
which provides a non-trivial {\it algebraic} relation between the entries 
of the period vector. This is different than 
in the odd-dimensional case, where  the first non-trivial equation derived 
from \trx\ involves a derivative acting on one $\Omega$ and leads to a 
differential equation relating the periods.

If instead of fibering the three-fold $\y$ over a $\IP^1$ we consider a four-fold
of the type $\y \times \IT^2$, eq. \periiii\ becomes exact. Let 
$\y_{\IF_1}$ be the elliptically fibred three-fold with base $\IF_1$; there is a point
in the moduli space with the appearance of $E_8$ tensionless strings \mv.
Then we have precisely the same situation as in \ganorII, where a torus 
compactification of this string is considered, leading to $N=2$ in four
dimensions. In this $N=2$ four-fold compactification the gauge coupling is
determined from the Calabi--Yau 
periods in the usual way (taking into account the 
F-theory limit); moreover it is easy to see from the results in \kmv\
that the relevant periods at the tensionless string point are precisely 
those over the shrinking del Pezzo inside $\y$, implying its appearance
in the final result of ref. \ganorII.

\newsec{Mirror map and Yukawa couplings}

The description of moduli spaces of $d$-dimensional Calabi--Yau 
manifolds in terms of a holomorphic section $\Omega$ of the Hodge
bundle and period integrals over this holomorphic $(d,0)$ form has
been given in \stromI,\candsg.
The concept of a mirror map relating n-point functions of 
$A$ and $B$ type topological field theories
associated to a $d$-dimensional Calabi--Yau manifold $\x$ and its mirror 
$\xs$ has been defined in \gmp, see also \dj. In this section we provide 
the general framework for the description of four-folds with an arbitrary 
number of  moduli in terms of toric geometry.

\subsec{Toric description of $\x$ and $\xs$}
\def\IR{ {\bf R}}
\def\cs{\IC^\star}
\def\IZ{ {\bf Z}}
\def\del{\Delta}\def\dels{\Delta^\star}
\def\pt{{\bf P}(\del)}\def\pts{{\bf P}(\dels)}
\def\hoo{h^{1,1}}\def\hdd{h^{d-1,1}}
\def\hot{h^{1,2}}\def\htt{h^{2,2}}
\def\rs{{\cal R}}
\def\co#1{{\cal O}^{(#1)}}
\def\ca#1{{\cal Q}^{(#1)}}
\def\gc#1{\gamma^{(#1)}}\def\gcs#1{{{^\star\gamma}}^{(#1)}}
\def\per#1{\Pi ^{(#1)}}\def\pert#1{\tilde{\Pi} ^{(#1)}}

Batyrev \bat\ has given a construction of mirror pairs $\x,\ \xs$ of 
d-dimensional Calabi--Yau manifolds as hypersurfaces in (d+1)-dimensional 
toric varieties $\pt,\ \pts$,
where $\del$ and $\dels$ denote the reflexive polyhedra $\subset \IR^{d+2}$
defining
the combinatorial data of $\pt$ and $\pts$. We will use this description 
of Calabi--Yau four-folds in the following.

Let $\ns_i$ denote the integral vertices of $\dels$. The toric variety
$\pts$ contains a canonical torus $(\IC^\star)^{d+1}$ with coordinates 
$X_i$. Then $\xs$ is defined as the zero set of the Laurent polynomial 
\eqn\polyI{
f_{\dels}(X,a) = \sum_{\ns_i} a_iX^{\ns_i} \ ,
\qquad X^{\ns_i} \equiv \prod_k X_k^{\nu_i^{\star\ (k)}}
}
where the coefficients $a_i$ are parameters characterizing the complex 
structure of $\xs$. In \bat\batm\foot{See also \batmII.} Batyrev shows 
that the Hodge numbers $h^{p,1}$ are determined by the polyhedron 
data as
{\ninepoint{
\eqn\hon{\eqalign{
\hoo(\x)&=\hdd(\xs)=
l(\dels)-(d+2)-\sum_{{\rm codim} S^\star=1}l^\prime(S^\star)+
\sum_{{\rm codim} S^\star=2}l^\prime(S^\star)\cdot l^\prime(S)\ , \cr
h^{p,1}(\x)&=\sum_{{\rm codim} S^\star=p+1} 
l^\prime(S)\cdot l^\prime(S^\star)\ ,\ \ {1<p<d-1}\ ,\cr
\hdd(\x)&=\hoo(\xs)=
l(\del)-(d+2)-\sum_{{\rm codim} S=1}l^\prime(S)+
\sum_{{\rm codim} S=2}l^\prime(S)\cdot l^\prime(S^\star) \ ,}}}}
where $S$ denotes faces of $\del$ and $S^\star$ the dual face of 
$S$. $l$ and $l^\prime$ are the numbers of integral points on a 
face and in the interior of a face, respectively. 

If the manifold has $SU(4)$ holonomy 
rather than a subgroup, then $h^{2,0}=h^{1,0}=0$ and 
the remaining non-trivial  hodge number $h^{2,2}$ is determined by 
$h^{2,2}=12+{2\over 3} \chi+2 h^{1,2}$
\sethi. The Euler number $\chi$ is $\chi=2(2+\hoo+h^{1,3}-2\hot)+ \htt$.

The target space toric variety $\pts$ can be described as 
$$
\pts = (\IC^m-F)/(\cs)^{m-d-2} \ ,
$$
a generalization of projective space with $m-d-2$ scaling symmetries
$x^i \to \lambda^{\al_i} x^i$ acting on the $m$ coordinates $x_i$ of 
$\IC^m$ and a disallowed set $F$ which consists of the unions
of intersections of coordinate hyperplanes $D_i \equiv \{x_i=0\}$
as determined by the so called primitive collections (see e.g. \batII,\bongI);
e.g. for ordinary projective space one has $\lambda^{\al_i}=\lambda$ and
$F=\{ x_i=0,\  \forall i\}$.

Apart from the combinatorial data $\dels$, a specific phase of
the Calabi--Yau manifold $\xs$ depends on the choice of  a
regular triangularization of $\dels$. This 
defines in turn a choice of set of generators for the $h^{1,1}$ relations 
between the integral vertices\foot{Here we restrict ourselves to the set of
vertices $\ns_i$ which lie on edges or faces of $\dels$; in the
general case it can be necessary to consider also vertices corresponding
to those interior points on faces of $\dels$ 
which represent automorphisms of $\pts$ \bongI.} $\ns_i$, 
$l^{(i)}_j \ns_j = 0,\ i=1 \dots h^{1,1}$, called the
the Mori vectors $l^{(i)}$.
The K\"ahler cone of the mirror $\x$ is then the dual of
the cone generated by the Mori generators. Starting from these
data one obtains a system of differential equations, the Picard-Fuchs system,
for the periods over the holomorphic $(d,0)$ form on $\xs$ \hkty.
The period integrals on $\xs$ are then given as linear combinations of the 
solutions to the Picard-Fuchs system.

There is a natural ring structure on $\pts$ from taking unions and 
intersections of toric divisors $D_i$. The intersection ring $\rs$
is defined as the quotient ring $\rs=\IZ[D_i]/\cx I$, where
$\cx I$ is the ideal generated by linear relations $\sum_i^m\langle m, \ns_i 
\rangle
D_i=0$ and a set of non-linear relations $\prod_i^m D_i^{\xi_i}=0$;
the latter is called the Stanley-Reisner ideal and determines the
disallowed set $F$.

In the next section we will relate the elements of $\rs$ at degree $k$
(where $k$ is here the complex codimension of a homology element)
to observables $\co k \in H_V \subset \oplus_k H^{k,k}(\x)$ of the
$A$ type model on $\x$; here $H_V$ is the so called primary vertical subspace
of $\oplus_k H^{k,k}(\x)$ \gmp\ which is the subspace of $\oplus_k H^{k,k}(\x)$
generated by wedge products of elements in $H^{1,1}(\x)$.
The ideal $\cx I$ determines the dimension of the ring at degree $k$;
in fact, for Calabi--Yau 
fibered four-folds one has $\dim_k(\rs)\equiv d_k(\rs)=
(1,\hoo,2\hoo-2,\hoo,1)$ for $k=0 \dots 4$.

Another distinguished set of generators of $\rs$ is determined by the
divisors as defined by the K\"ahler cone of $\x$. Let $J_\al$ be the
$(1,1)$ forms dual to the special flat coordinates $t_\al$
on the K\"ahler moduli space, centered at a large radius structure limit
of maximal unipotent monodromy. Let 
$J=\sum_\al t^\al J_\al$ be the K\"ahler form, and $K_\al$ the divisors
dual to the $J_\al$. We can use equivalently $K_\al$ as generators of
the intersection ring $\rs$. In particular, if $R_0$ is the
intersection form of $\x$
\eqn\iform{
R_0 = \sum_{\al \geq \be \geq \ga \geq \de} k_{\al \be \ga \de}
K_\al K_\be K_\ga K_\de \ ,
}
where the convention is that $k_{\al\be\ga\de}$ is the value of the
integral $\int_\x J_\al \wedge J_\be \wedge J_\ga \wedge J_\de$, 
then the top element of dimension 4 of $\rs$ is simply $R_0$ while
the volume of $\x$ is obtained by replacing the divisors $K_\al$ 
by the coordinates $t_\al$ in \iform\ and relaxing the condition on the
summation indices in \iform.

Other topological invariants of $\x$ are defined by integrating
elements of $\rs$ wedged with the Chern classes $c_i$ of $\x$,
$i=2 \dots 4$:
\eqn\toppart{\eqalign{
R _{22} &= \int_M c_2^2 \cr
R _2 &= \int J \wedge J  \wedge c_2, \cr
R _3 &= \int J \wedge  c_3\ , \cr
R _4 &= \int_M c_4 = \chi}}
with  the obvious index structures. For $SU(4)$ holonomy, 
$R_{22}=480+{\chi \over 3}$ \sethi.

\subsec{The $A$ model}
Mirror symmetry implies that the correlation functions 
of two topological field theories defined on a Calabi--Yau manifold $\xs$
and its mirror $\x$ are isomorphic. 
The correlation functions of the first theory, 
the $B$ model defined on $\xs$, depend on the 
complex structure (CS) moduli of $\xs$ in a purely geometrical (classical)
way and
can be calculated straightforwardly. On the other hand 
the correlation functions of second theory, the $A$ model defined on $\x$,
depend on the K\"ahler moduli (KM) of $\x$ in a complicated 
way due to the presence of world-sheet instanton corrections.
Mirror symmetry allows to determine these A model correlation functions 
by construction of the explicit mirror map from the complex structure moduli
space of the B model to the K\"ahler moduli space of the A model.
\br

\subsubsec{Choice of a basis for the A model}
To match the moduli space of the CS moduli space of the $B$ model 
to the K\"ahler moduli space of the $A$ model we first chose a basis in the
$A$ model in the following way. The 
basis for the primary vertical subspace $H_V \subset \oplus_k H^{k,k}$ 
with the
most natural geometrical interpretation is given by forms Poincare dual
to submanifolds of complex codimension $k$ \wittop.
Specifically we will chose 
a basis generated by the (1,1) forms $J_k$ dual to the special coordinates
on the KM space, $t_\al$, and wedge products of them:
\eqn\basis{\eqalign{
&\co0 = 1,\ \ \co1_\al = J_\al ,\ \ 
\co2_\al=E_\al^{(2) \ \be\ga}J_\be J_\ga,\cr
&\co3_\al=E_\al^{(3) \ \be\ga\de}J_\be J_\ga J_\de,\ \ 
\co4_\al=E_\al^{(4) \ \be\ga\de\epsilon} J_\be J_\ga J_\de J_\epsilon\ .}}
As mentioned above the dimension of the basis $\co i,\ i>1$ is 
reduced by the intersection properties of the dual homology elements $K_\al$
determining the range of the lower index of the coefficients $E_\al$.
E.g., in the case of Calabi--Yau fibrations, 
where $\dim(H_V^{2,2})=\dim(H_V^{3,3})=2 h^{1,1}-2$, the intersection 
of the divisor $D$ dual to the base $\IP^1$ with itself is empty, $D\cdot D = 0$,
implying that there can appear at most one power 
of $J_D$ in the definition of the $\co i_\al$ (this is the same kind of  argument that 
ensures the linear coupling of 
the K\"ahler coordinate identified as the dilaton in K3 fibrations).
In general the $E^{(i)}$ are chosen such that the elements $\co i$ generate
the degree $i$ subspace of $\rs$.\br

\subsubsec{Topological metric, operator product expansions and correlation functions}
The 2-pt functions define the flat
metric on $H_V$ in terms of integrals of the basis elements $\co i$ over $\x$:
\eqn\metricI{
\eta^{(i)}_{\al\be} = \mx {\co i_\al}{\co {d-i}_ \be} =  \int \co i_\al 
 \co {d-i} _\be}
In fact the metric is constant in the flat variables $t_\al$ \flatmet\ and non-zero
only for pairs $\co i,\  \co j $ of operators with $i+j=d$.

The r.h.s. is then 
determined by the coefficients $E^{(i)}$ of a given basis \basis\ together
with the intersection numbers $k_{\al\be\ga\de}$.

The factorization properties of the topological field theory ensure that
all correlation functions can be expressed in terms of the fundamental
2-pt and 3-pt functions. Similarly as in the case of three-folds there is only
one independent type of 3-pt functions, namely  $\langle \co1 \co1 \co2 \rangle$, 
which contain the full information about the moduli dependence
$$
Y_{\al\be\de}= \langle  \co1_\al \co1_\be \co2_\ga \rangle = 
\int \co1 _\al \co1 _\be \co2 _\ga + {\rm inst. corr.}
$$
The $Y$ are determined in terms of the operator product coefficients
$C^{(1)}$, $C^{(2)}$:
\eqn\opI{
\co1_\al \cdot \co i _\be = C^{(i)\ \ga}_{\al\be} \co {i+1}_ \ga
}
to be
$$
Y_{\al\be\ga} = C^{(1) \ \mu}_{\al\be} \eta^{(2)}_{\mu\ga}=
 C^{(2) \ \mu}_{\al\ga} \eta^{(1)}_{\be\mu}
$$

While the 2-pt and 3-pt functions are the fundamental objects of the
underlying topological theory, the simplest object on the four-fold
which can be defined entirely in terms of the marginal operators $\co1_\al$
are the 4-pt functions $K$
$$
K_{\al\be\ga\de}=\la \co1 _\al \co1 _\be \co1 _\ga \co 1 _\de \ra = 
\int_{X_4} \co 1_\al \co1 _\be \co1 _\ga \co 1_\de + {\rm inst. corr.}
$$
whose classical piece is given by the intersection numbers 
$k_{\al\be\ga\de}$ of 
$\cx R_0$. Factorization in terms of 2-pt and 3-pt functions yields
\eqn\facI{
K_{\al\be\ga\de}=
(C^{(1)}_\al \cdot \eta^{(2)} \cdot C^{(1)T}_\ga)_{\be\de} = 
(C^{(1)}_\al \cdot C^{(2)}_\ga \cdot \eta^{(1)\ T})_{\be\de} \ ,
}
where we use a matrix notation $(C^{(i)}_\al)_\be^{\ \mu}$.
Non-trivial conditions on the ring coefficients $C^{(1)},\ C^{(2)}$ follow from 
associativity of the operator products:
\eqn\asso{\eqalign{
(C^{(1)}_\al \cdot \eta^{(2)} \cdot C^{(1) \ T}_\ga)&=
(C^{(1)}_\al \cdot \eta^{(2)} \cdot C^{(1) \ T}_\ga)^T\ ,\cr
C^{(1)\ \rho}_{\al\be} C^{(2)\ \mu}_{\ga\rho}&=
C^{(1)\ \rho}_{\al\ga} C^{(2)\ \mu}_{\be\rho}  \ ,}
}
where the second identity follows from the first using 
$(C^{(2)}_\al)^T = (\eta^{(1)})^{-1 \, T} \cdot C^{(1)}_\al\cdot \eta^{(2)}$.
This identity provides highly non-trivial relations between
the instanton corrected correlation functions.

\subsec{Basis for the $B$ model and the mirror map}
The next step to find the mirror map is the construction of a basis of
the observables of the $B$ model which matches the properties of the 
above chosen basis for the $A$ model:
\eqn\propI{
\co1 _\al \cdot \co i _\be = C^{(i) \ \ga} _{\al\be} \co {i+1} _\ga,\qquad 
\mx {\co i _\al}{\co j _\be}=\delta_{i+j,d}\eta^{(i)}_{\al\be} \qquad
\co1_\al \cdot \co4 = 0
}
The appropriate basis for the $B$ model can be defined \gmp\ using the
Gauss--Manin connection $\d$, the flat metric-compatible connection 
on the Hodge bundle $\cx H$ over the CS moduli space $\cx M _{CS}$.
The following construction is a generalization of 
the procedure in \gmp; we can therefore focus on the complications introduced 
by the higher dimensional moduli space as compared to the one moduli
case considered in \gmp. In particular we will define the 3-pt functions
directly in terms of the period integrals over the holomorphic $(d,0)$
form and the intersection ring $\rs$.

The fundamental step in the construction of the $B$-model 
basis in \gmp\ is the replacement of the operator product
involving a charge one operator $\co1$ with the action of
the unprojected Gauss-Manin connection $\d$
$$
\co1 _\al \cdot \co i _\be  \to \d_\al \ca j _\be
$$
where now $\ca i$ denote the basis elements of the $B$ model and the 
directional derivative is defined in terms of the parametrization 
of the deformations corresponding to  marginal operators $\ca 1 _\al$ 
by the special flat coordinates $t_\al$.
This definition implies a holomorphic dependence of the basis $\ca i$ as opposed
to the other natural choice, a basis of elements of pure type $(d-k,k)$.
We will now define a basis matching the property \propI\ of the
$A$ model basis using the intersection ring $\rs$ and a map $m:\ D_i \to \theta_i$ \bongI,
where $\theta_i = z_i{\partial\over \partial z_i}$ are the logarithmic derivatives with respect to the complex structure moduli of the $B$ model \hkty
\eqn\algcoo{
z_i = (-)^{l_0^{(i)}}\prod_i a_i^{l^{(k)}_i}
}
Let $\gc i _\al$ be a basis of topological homology cycles spanning the 
primary horizontal subspace $H_H(\x) \subset \oplus_kH^{d-k,k}(\x)$ 
and $\gcs i _\al$ be the dual cohomology elements fulfilling
\eqn\topbas{
\int_{\gc i _\al} \gcs j _\be = \delta^{ij}\delta_{\al\be},\qquad
\int \gcs i _\al \gcs j _\be = \cases 
{0, \ i+j > d;\cr M^{(i,j)}_{\al\be},\ i+j=d;\cr}
}
Furthermore let $\ca i _\al $ be a set of elements of $F^{d-i}$, where 
$F^p=\oplus_{k \geq p} H^{k,d-k}(\xs)$ are the holomorphic bundles of
forms with anti-holomorphic degree at most $d-k$, fulfilling
\eqn\diagI{
\int_{\gc i _\al} \ca i _\be = \cases{\delta_{\al\be},\ i=j;\cr 0,\ i<j \cr}
}
The following relations are elementary:
\eqn\kkk{\eqalign{
\ca i _\al &= \gcs i _\al + \sum_{k>i} \tilde{a}^{(i,k)} _{\al\be} \gcs k _\be,\ \ 
\ \ \ \tilde{a}^{(i,k)} _{\al\be}\equiv \int_{\gc k _\be} \ca i _\al\ ,\cr
\d_\al \ca i _\be &= (\d_\al \tilde{a}^{(i,i+1)}_{\be\ga}) \ca {i+1} _\ga\ ,\cr
\mx {\ca i _\al} {\ca j _\be} &= M_{\al\be}^{(i,j)}\delta_{i+j,d}\ ,\cr
Y_{\al\be\ga}&=\langle \ca1 _\al \ca1 _\be \ca 2 _\ga \rangle = 
(\d _\al \tilde{a}^{(1,2)}_{\be\de}) M_{\de\ga}^{(2,2)} \ .
}}
Comparing \kkk\ with \metricI,\ \opI\ it follows that $\ca i$ is the basis matching the 
properties \propI\ provided that 
\eqn\condI{
\d_\al \tilde{a}^{(i,i+1)}_{\be\ga}=C^{(i)\ \ga}_{\al\be},\qquad
M_{\al\be}^{(i,j)}=\eta^{(i)}_{\al\be} \ .
}
We now construct the basis $\{ \ca i _\al,\ \gc i _\al\}$ in two steps:\br
a) First chose a basis for the $\ca i _\al$. Of course we have 
$\ca 0 = \Omega \in H^{d,0}(\x)$.
The $\ca i _\al$ are then obtained by choosing $d_k(\rs)$ independent
generators\foot{We use here $\cx O_\al^{(k)}$ to denote the element 
in $H^{k,k}(\x)$ that corresponds 1-1 to the operator $\cx O_\al^{(k)}$.}
$\cx 
O_\al^{(k)}=E_\al^{(k)\al_1\dots\al_k} J_{\al_1}
\wedge\dots\wedge J_{\al_k} \in H^{k,k}(\x)$ 
of the ring $\rs$ at degree $k$ and defining
$$
\ca k _\al = L^{(k)}_\al \Omega\ , \qquad 
L^{(k)}_\al = \cx O^{(k)}_\al(J_\ga \to \theta_{z_\ga}) \
$$
where $L^{(k)}_\al$ are differential operators of degree $k$ obtained from the
map $\tilde{m}: K_\al \to \theta_{z_\al}$ which follows from 
$m:D_i \to \theta_{a_i}$ by transformation to  
the $K_\al$ basis.
The topological metric
in this basis is then given by $\eta^{(i)}_{\al\be}=\int \cx O_\al^{(i)} 
\wedge \cx O_\be^{(j)}$.\br
b) We have to find the basis of cycles $\{\gc i _\al\}$ that satisfies \diagI\
with the given basis $\{\ca i_\al \}$. This can be done by fixing the leading logarithmic
behavior of the period integrals 
$$
\per i _\al = \int_{\gc i _\al} \Omega = \int_{ \gc i _\al } \ca 0  \ ;
$$
the exact periods $\per i _\al$ are then determined by the solution
of the Picard-Fuchs system with the appropriate leading behavior.

At the large complex structure point of the CS moduli space $\cx M _{CS}$
$z_i = 0\ \forall i$, the solutions to the Picard--Fuchs system have the leading 
behavior $\sim (\ln z)^k,\ k=0\dots 4$, where $z$ stands for any of the
CS moduli $z_i$. In fact there are precisely $d_k(\rs)$ solutions $S_\al^{(k)}$ with
leading behavior $(\ln z)^k$, as a consequence of the relation between the 
intersection ring $\rs$ and the ring of differential operators $\theta_i$ 
\hkty\bongI.
The basis $\{\gc i _\al\}$ with the property $\diagI$ is then fixed by
the condition 
\eqn\normI{
\per k _\al = S^{(k)}_\al + \dots, 
\qquad L^{(k)}_\al S_\be^{(k)}=\delta_{\al\be} +\dots \ .
}
where the ellipsis denote terms involving powers $(\ln z)^l$ with $l< k$
and polynomial corrections.
For convenience 
we state explicitly the expression for
the leading piece of $\Pi_\al^{(k)}$, as obtained from \normI\
by trivial matrix multiplication:
$$
S_\al^{(k)}=\sum {1\over k!} \tilde{E}^{\be_1\dots\be_k}_\al
\prod_{n=1}^{k} \ln(z_{\be_n}),\qquad \tilde{E}^{\be_1\dots\be_k}_\al=
E^{\star\ga_1\dots\ga_{d-k}}_\al k_{\be_1\dots\be_k\ga_1\dots\ga_{d-k}}
$$
where $k_{\be_1\dots\be_d}$ is the intersection form given in \iform,
and $\cx O^{\star(d-k)}_\al\equiv E^{\star\ga_1\dots\ga_{d-k}}J_{\ga_1}
\wedge\dots\wedge 
J_{\ga_{d-k}}$ is the Poincare dual of $\cx 
O^{(k)}_\al$ (as is obvious from the relation 
$L^{(k)}_\al=\cx O^{(k)}_\al(J_\ga \to \theta_{z_\ga})$). The
exact expressions are then determined as the linear combination
of the solution to the Picard-Fuchs system with the appropriate
leading behavior. For more details we refer to app. C.

\subsubsec{3-pt functions}
All the fundamental 3-pt correlators are then determined explicitly 
in terms of the period integrals on
the middle dimensional cohomology of the Calabi--Yau
4-fold $\xs$. Namely, from 
\eqn\relII{
\d_\al \ca 0 = \ca 1,\quad \d_\al \ca 1 _\be = C^{(1)\ \ga }_{\al\be} \ca 2 _\ga,\quad
\d_\al \ca 2 _\be = C^{(2)\ \ga }_{\al\be} \ca 3 _\ga,
}
and \diagI\ we obtain the final formula for the Yukawa coupling $Y_{\al\be\ga}$:
\eqn\yukI{
Y_{\al\be\ga}=\d_\al \d_\be \pert 2 _\ga 
}
where $\pert 2 _\al = \per 2 _\be \eta^{(2)}_{\be\al}$ with leading behavior
$\pert 2_\al = {1\over2}
 C^{(1)\ga}_{\be\de}\eta^{(2)}_{\ga_\al}\ln(z_\be)\ln(z_\de)+
\cx O(z)$.

Integrating the relations \relII\ over the cycles $\gc i _\al$ we obtain the
Picard-Fuchs equation satisfied by the periods:
$$
\d_{\lambda}(C^{(3)})^{-1}_{\nu\al} \d_\al (C_{\hat{\epsilon}}^{(2)})^{-1}_{\be\de}
\d _{\hat{\epsilon}}(C_{\hat{\rho}}^{(1)})^{-1}_{\de\mu}\d_{\hat{\rho}}\d_\mu \per i _\al = 0\ ,
$$
where hatted indices are not summed over. This is the holomorphic form
of the differential equation reflecting the restricted K\"ahler structure of 
the CS moduli space $\cx M_{CS}$ of the four-fold. The corresponding
linear system is the system obtained by integrating \relII\ over the manifold.

Finally note that the full intersection matrix of the period vector 
$\per i _\al$ is obtained
from the topological metrics $\eta^{(i)}$ as
\eqn\sympmet{
\pmatrix{0&0&0&0&1\cr 0&0&0&\eta^{(1)}&0\cr 0&0&\eta^{(2)}&0&0\cr
0&(\eta^{(1)})^T&0&0&0\cr 1&0&0&0&0}}

\subsec{Counting of rational curves}
One of the most striking aspects of the calculation of the world-sheet
instanton corrected 3-pt couplings of the $A$ model on a Calabi--Yau three-fold 
$\x$ via mirror symmetry is the interpretation of the integral coefficients 
of the $q=e^{2\pi i t}$-expansion in terms of the number $N(\nv)$ of rational curves of
multi-degree $\nv = (n_1,\dots,n_{\hoo})$ on $\x$ \candI:
\eqn\inst{\eqalign{
Y_{\al\be\ga}&=Y_{\al\be\ga}^{(0)}+\sum_\nv N^\prime(\nv) \prod_{\de=1}^{\hoo} 
q_\de^{n_\de},\qquad \qquad \qquad q_\al=e^{2\pi i t_\al}\cr
&= 
Y_{\al\be\ga}^{(0)}+\sum_\nv N(\nv) M_{\al\be\ga}(\nv)
{\prod_{\de=1}^{\hoo} q_\de^{n_\de}\over (1-\prod_{\de=1}^{\hoo} q_\de^{n_\de})}
,\qquad M_{\al\be\ga}(\nv)=n_\al n_\be n_\ga
}
}
The factor $(1-\prod_{\de=1}^{\hoo} q_\de^{n_\de})$ in the denominator 
of \inst\ takes into account the contribution of multiple coverings.
This interpretation has been justified in the framework of the topological 
sigma model in \morII. The generalization of these argument to the case
of $d>3$ dimensional Calabi--Yau manifolds has been given in \gmp; in 
particular it was shown that the multiple covers contribute in the
analogous way as in three dimensions.

On the other hand, the additional factor $M_{\al\be\ga}(\nv)=n_\al n_\be n_\ga$
in \inst\ gets modified. Let us recall the relevant fact of the
definition of the correlation functions in the topological field theory.
By definition the local operators $\co i _\al(P),\ P \in \Sigma$
have delta function support on maps $\Phi: \Sigma \to \x$
with the property $\Phi(P) \in H^{(i)}_\al$; here $\Sigma$ is
the 2d world sheet and $H^{(i)}_\al$ a codimension $i$
homology cycle of the Calabi--Yau target space $\x$.
For the case of a $\co 1 $ operator, $H^{(i)}$ is a divisor,
in fact in our choice of basis one of the divisors $K_\al$
as defined by the K\"ahler cone of $\x$. The $n_\al$ factors arise
from the multiple intersections of a curve $C$ with that divisor 
and count the degree of the curve with respect to it
In the case of the 3-pt functions on the four-folds 
there are two charge one operators $\co 1 _\al,\ \co 1 _\be$ involved, so one
gets to factors of the relevant degrees,  $n_\al$ and $n_\be$, of the curve 
\sheI.

The third operator has charge two and is associated to a codimension
two homology cycle - in our basis a linear combination of
intersections of two K\"ahler divisors $K_\al$. Differently from 
the previous case the condition $\Phi(P)\in H^{(i)}_\al$ is a 
real constraint on the curve $C$; we have to adjust the position of 
the curve in the manifold to satisfy this condition. As a consequence
the numbers $N(\nv)$  appearing in the 3-pt functions do count the
appropriate Euler number of the moduli space of rational curves 
subject to a constraint. The constraint varies with the choice of 
$H^{(2)}_\al$ and thus with the choice of $\co 2 _\al$. Therefore
we do not expect to get the same numbers $N(\nv)$ from 3-pt functions
involving different operators $\co 2 _\al$.

This is actually a nice circumstance for the present case of Calabi--Yau
fibrations. In general, rational curves of the fiber get moduli in the
four-fold from ``moving them over the base''. Therefore the
Gromov--Witten invariant of a curve in the three-fold is generically not
the same as the invariant of the same curve in the four-fold. However
we will show that one can always 
fix the curves by choosing the appropriate operators $\co 2 _\al$;
in this case the numbers $N(\nv)$ of the four-fold indeed 
coincide with those of the Calabi--Yau fiber.
The appropriate choice
for the $\co 2$ is clear: one of the intersecting divisors
will be the Calabi--Yau fiber $D$ itself - tautologically this imposes no 
constraint on the curves of the fiber. The second divisor $K_\al$ which we
intersect with $D$ to obtain a codimension 2 cycle plays the role
of the third charge one operator in the three-fold calculation and
contributes naturally another factor of $n_\al$, counting the
number of intersections of $C$ and $K_\al$ in the fiber.
\vfill
\newsec{Three-fold fibered four-folds: toric construction and calculation of 
invariants}

In the following we use the above construction to analyze
some examples of four-folds with four K\"ahler moduli which have at the
same time phases which allow elliptic, K3 and Calabi--Yau three-fold fibrations.
This structure will allow various interpretations
in terms of compactifications of $F$-theory, type IIA/IIB and heterotic/type I
theories.
In particular much is known about the theories on elliptic three-folds
$\y_{\IF_n}$
with base ${\bf F}_n,\ n=0,1,2$ which can be fibered 
to four-folds with four K\"ahler moduli, one from the base and three 
from the Calabi--Yau fiber $\y_{\IF_n}$. In addition to the
type of three-fold fiber there is the freedom to chose the bundle structure
involving the base $\IP^1$; this will determine in particular the
Calabi--Yau three-fold $\z_{\IF_n}$ of the dual heterotic theory. 
In table 1 we collect examples of four-folds with four moduli
arising in this way; the precise definition in terms of reflexive 
polyhdedra is given in the text and in appendix D.

Schematically the fibration structure is of the form indicated 
in the first column of table 1. The three complex dimensional 
base of the elliptic fibration can be thought of a collection
of three $\IP^1$ factors with non-trivial bundle structure.
The Chern classes $c_1$ of these bundles are described by the 
vector (a,b;c). Here the first two entries refer to the bundle
structure of the ``top'' $\IP^1$ (the base of the elliptically fibered
K3) over the other two $\IP^1$ factors, and the third one to the
structure of the remaining $\IP^1$ over the base\foot{For a
consistent notation, a choice
has been made for those cases, where the base of the three-fold 
fibration can be chosen in different ways.}.
$$
\vbox{\offinterlineskip
\halign{\hfil$#$\hfil\cr \circ \cr  {a \atop \ }
 \swarrow \searrow {b \atop \ } \cr \circ \ {\rightarrow\atop c}\  \circ\cr}}
$$
In the following columns we give the hodge numbers and the Euler number.
$h^{1,1}_{\rm np.}$ and $h^{1,2}_{\rm np.}$ denote the number of
deformations which can not be realized as polynomial deformations
in the toric description; they will play a role later on.
In general there is more then one Calabi--Yau phase; in this case
the stated fibration structure is present at least in one of 
those phases.
\vskip 12pt
\vbox{$$
\vbox{\offinterlineskip\tabskip=0pt
\halign{\strut
\vrule#
&\hfil~$#$~\hfil&\vrule#
&\hfil~$#$~\hfil
&\vrule#
&\hfil ~$#$~\hfil
&\hfil ~$#$~\hfil&\hfil ~$#$~\hfil
&\hfil ~$#$~\hfil&\hfil ~$#$~\hfil
&\hfil ~$#$~\hfil&\hfil ~$#$~\hfil
&\vrule#&\hfil~$#$~\hfil&\vrule#\cr
\noalign{\hrule}
& &&{\rm type}  &&h^{1,1}&h^{1,1}_{\rm np}
&h^{1,3}&h^{1,3}_{\rm np.}&h^{1,2}&h^{2,2}&\chi&&&\cr
\noalign{\hrule}
&1&&(0,0;0)
& & 4&0& 2916&0& 0& 11724& 17568&&&\cr
&2&&(1,0;0)
& & 4&0& 2916&0& 0& 11724& 17568&&IV&\cr
&3&&(2,0;0)
& & 4&0& 2916&1& 0& 11724& 17568&&&\cr
&4&&(1,1;0)
& & 4&0& 3156&0& 0& 12648& 19008&&II&\cr
&5&& (1,2;0)
& & 4&0& 3396&0& 0& 13644& 20448&&&\cr
&6&& (2,2;0)
& & 4&0& 3877&0& 1& 15566& 23328&&&\cr
&7&& (1,0;1)
& & 4&0& 2796&0& 0& 11244& 16848&&&\cr
&8&& (1,2;2)
& & 4&0& 3156&1& 0& 12648& 19008&&III&\cr
&9&& (2,4;2)
& & 4&0& 3877&1& 1& 15566& 23328&&I&\cr
&10&&(2,2;1)
& & 4&0& 3396&0& 0& 13644& 20448&& V&\cr
\noalign{\hrule}}
\hrule}
$$
\vbox{\ninepoint{
\vskip 0pt
\noindent
{\bf Table 1}: 
Examples of 4 moduli four-folds 
from fibrations of elliptically fibered three-folds
with base ${\bf F}_n,\ n=0,1,2$.
}
\vskip7pt}}
In the following we consider in some detail the four-folds denoted by $I-V$ in
the last column. The first one turns out to have a particularly simple
singularity structure which we will investigate in sect. 8. 
Singularities in the three-fold compactification have been
analyzed in \kklmv,\kmI,\kmp,\mv,\witm,\klmvw\ste\bkkm\witnew. 
The other examples have as the
fiber the elliptically fibered three-fold with base $\IF_1$; the 
three-fold compactification has an interesting point in the moduli space
with the appearance of tensionless strings with $E_8$ current algebra
\eestring \mv, which has recently 
attracted considerable interest \ganorI\kmv\dkv.

For the invariants we will restrict to the first three
cases; the superpotentials and physical singularities will be discussed 
in sects. 7 and 8 together with 
the geometrical properties which are relevant for the physical 
interpretation\foot{The relevant toric data are given in appendix D.}. 
In the following we sketch
the calculation for the first example. The character of the rest of 
this section is necessarily technical; the discussion of the 
results is therefore given in the next section. 

\subsec{The four-fold $X_I$}
The first example we consider is a four-fold with a elliptic fibered
three-fold with base $\IF_2$ fibered over a $\IP^1$ such that the
last fibration has itself the structure of $\IF_2$:
\eqn\cstarXI{
\vbox{\offinterlineskip
\halign{$#$ \cr 
\hskip 22pt\circ \cr 
\ \scriptstyle{2}  \swarrow \vert \cr
\hskip 2pt \ominus \hskip 13pt \vert \ \scriptstyle{4} \cr
\ \scriptstyle{2}  \searrow \hskip 1pt \downarrow\cr
\hskip 22pt\circ\cr}\vskip 15pt}
\qquad \qquad
\vbox{\offinterlineskip\tabskip=0pt
\halign{\strut\vrule#
&\hfil~$#$
&\vrule#&~
\hfil ~$#$~
&\hfil ~$#$~
&\hfil $#$~
&\hfil $#$~
&\hfil $#$~
&\hfil $#$~
&\hfil $#$~
&\hfil $#$~
&\hfil $#$~
&\vrule#\cr
\noalign{\hrule}
&  &&x_1&x_2&x_3&x_4&x_5&x_6&x_7&x_8&x_9&\cr
\noalign{\hrule}
&\lambda_1  && 0& 0& 0& 2& 3& 1& 0& 0& 0&\cr
&\lambda_2  && 0& 0& 1& 4& 6& 0& 1& 0& 0&\cr
&\lambda_3  && 0& 1& 2& 8& 12& 0& 0& 1& 0&\cr
&\lambda_4  && 1& 2& 4& 16& 24& 0& 0& 0& 1&\cr
\noalign{\hrule}}
\hrule}
}
The line in the middle $\IP^1$ denotes the codimension of the divisor
$x_8=0$ which will play a role in the discussion of $N=1$ superpotentials.
The table on the right side specifies the $C^\star$ actions which is part of
the definition of 
the toric variety; one can easily see the fibration structure
$$
{\bf T}^2(x_4,x_5,x_6) \to \IP^1_A(x_3,x_7) \to \IP_B^1(x_2,x_8) \to \IP_D(x_1,x_9)
$$
where the last two bundles $\IP^1_A \to \IP^1_B$ and $\IP^1_B \to \IP^1_D$
have the structure of rational surfaces of type $\IF_2$.
The dual heterotic theory has therefore $\IF_2$ as the base.

\subsubsec{Basic data and properties\foot{The calculation of the 
triangulations and the intersection calculus has been performed 
using the maple codes puntos \puntos\ and schubert \schubert.}}
The dual polyhedron for the four-fold $X_I$ is the convex hull of the
negative unit vectors in $\IR^5$,  $\ns_i, \ i=1\dots 5$ and
$$
\ns_6 = (0,0,0,2,3),\ 
\ns_7= (0,0,1,4,6),\ 
\ns_8=(0,1,2,8,12),\ 
\ns_9= (1,2,4,16,24)
$$
There is a single Calabi--Yau phase which is an elliptic and K3 fibration.
The generators $l^{(\al)}$ of its Mori cone are
\eqn\moriXI{\eqalign{
l^{(1)}=(0,0, 1, 0, 0, 0, 0, -2, 1, 0),\ 
\ &l^{(2)}=(-6,0, 0, 0, 2, 3, 1, 0, 0, 0)\cr
\ l^{(3)}=(0,1, 0, 0, 0, 0, 0, 0, -2, 1),
\ &l^{(4)}=(0,0, 0, 1, 0, 0, -2, 1, 0, 0)\ .}}
The topological intersection numbers are:
\eqn\topXI{\eqalign{
R _0 =& (64 K_2^4+2 K_4 K_2 K_1^2+4 K_2^2 K_1^2+16 K_2^3 K_1+
8 K_4 K_2^2 K_1+32 K_4 K_2^3+\cr&16 K_2^2 K_4^2+8 K_4^3 K_2+4 K_4^2 
K_2 K_1)+(2 K_2^2 K_1+2 K_4^2 K_2+4 K_4 K_2^2+\cr&K_4 K_2 K_1+8 K_2^3)
 K_3\cr
R _2 =&(184 K_1 K_2+364 K_2 K_4+96 K_1 K_4+192 K_4^2+\cr
&48 K_1^
2+728 K_2^2)+(48 K_4+24 K_1+92 K_2) K_3\cr
R _3 =& (-3856 K_2-1920 K_4-960 K_1)-480 K_3\ .}}

The coefficients of $K_3$ in the above expression 
are precisely the intersection invariants of the Calabi--Yau three-fold fiber.

\subsubsec{Cohomology classes of genus zero curves}
To relate physical properties to the geometry of the Calabi--Yau 
manifold one has to know the location of the relevant
objects, e.g. divisors and curves, in the manifold. Let us first
determine the classes of the curves $C_i$ associated to the Mori generators $l^{(i)}$
in  \moriXI. This will be important for two reasons: firstly the volume of 
$C_i$ is proportional to the special K\"ahler coordinate $t_i$. Because of
the multiple fibration structure it is straightforward to establish a rough
correspondence with some of the heterotic moduli: the base of the elliptically
fibered K3 is related to the dilaton by the eight-dimensional duality, while
the two-dimensional base of the K3 fibration on the F-theory side maps to the 
base of the heterotic side. Secondly this information will be necessary in
order to understand the singularities which arise when this 2-cycle shrinks
to zero size. Physical phenomena such as spectra and the behavior of 
non-perturbative corrections are closely related to 
the position and the embedding of the vanishing cycle in the Calabi--Yau
four-fold.

Let $C_\al$ denote the curve dual to the Mori generator $l^{(\al)}$ and
$D_i$ denote the toric divisor $\{x_i=0\}$. As explained in \bkk\ the
curve $C_\al$ can be  determined from i) the intersection of $C_\al$ with the $D_i$'s,
ii) the Stanley-Reisner ideal (SR) and iii) the explicit classes as defined by the 
polynomial constraint in the Batyrev-Cox variables. The Stanley Reisner ideal 
is given by 
$$
SR:\  \{x_1 x_9, x_2 x_8, x_3 x_7, x_4 x_5 x_6\}
$$
and defines the disallowed set $F= \{x_1=x_9=0\}\cup \{x_2=x_8=0\}, \{x_3=x_7=0\}, 
\{x_4=x_5=x_6=0\}$ in the definition of the toric variety. 
The typical Batyrev-Cox polynomial is given by
\eqn\BCXI{\eqalign{
x_7^{12} x_8^{24} x_6^6 x_9^{48}+x_7^{12} x_8^{24} x_6^6 x_1^{48}+x_7^{12} x_2^{24} x_6^6+
x_3^{12} x_6^6+x_4^3+x_5^2}}
respecting the $C^\star$ actions \cstarXI.
From the above data it is now straightforward to determine the 
volumes which are measured by the K\"ahler coordinates $t_\al$: 
$t_1$ is the area of the fiber $\IP^1_B$, 
$t_2$ is the area of a rational curve in the elliptic fiber, 
$C_3$ is the the area of the base of the Calabi--Yau fibration $\IP^1_D$  and
$C_4$ is the the area of the fiber $\IP_A^1$. 

\subsubsec{Counting of rational curves on $X_I$}
From the Mori generators \moriXI\ one obtains the Picard-Fuchs system as in \hkty:
\eqn\pfI{\eqalign{
\cx L_1 &=   z_1 (2 \theta_1+1-\theta_4) (-\theta_4+2 \theta_1)-
\theta_1 (\theta_1-2 \theta_3)\ , \cr
\cx L_2 &=   12 z_2 (5+6 \theta_2) (1+6 \theta_2)-\theta_2 (\theta_2-2 \theta_4)\ ,\cr
\cx L_3  &=  z_3 (\theta_1-2 \theta_3) (\theta_1-2 \theta_3-1)-\theta_3^2 \ ,\cr
\cx L_4 &= z_4 (\theta_2-2 \theta_4) (\theta_2-2 \theta_4-1)+\theta_4 (-\theta_4+2 \theta_1)\ ,}}
where $\theta_i = z_i\partial_{z_i}$ and $z_i$ are the 
algebraic coordinates on the 
complex structure moduli space \algcoo.
The period vector is generated by the 16 independent solution to this system; 
specifically, at $z_i=0$  there is one series solution $w_0$, four linear 
logarithmic solutions $w_1^i \sim \ln z_i$,
six double logarithmic solutions $w_2^i$, four triple logarithmic solutions $w_3^i$
and one quartic logarithmic solutions $w_4$ (see appendix C for more details).
As described in sect. 3 we can determine the 3-pt functions from the 
intersection ring $\rs$, whose basis we take to be
\eqn\ringXI{
\vbox{
\halign{
\hfil ~$#$ ~\hfil
&\hfil ~$#$~\hfil\cr
{\rm deg} & {\rm generators} \cr 
1& K_1, K_2,K_3,K_4\cr 
2&K_1 K_2,\ 2 K_1^2+K_1 K_3,\ K_2 K_3,\cr 
 &K_4 K_1+2 K_4^2,\ 2 K_2^2+K_2 K_4,\ K_3 K_4\cr
3&2 K_1 K_2^2+8 K_2^3+K_1 K_2 K_4+4 K_2^2 K_4+2 K_2 K_4^2,\cr
 &2 K_4 K_1^2+K_3 K_4 K_1+4 K_1 K_4^2+2 K_3 K_4^2+8 K_4^3,\cr
 &2 K_3 K_2^2+K_3 K_2 K_4,\ 2 K_2 K_1^2+K_2 K_1 K_3\cr
}}}
The top element at degree 4 is obtained from the intersection form $R_0$ as 
in eq. \iform. 
The leading logarithmic terms of the double periodic solutions $\Pi^{(2)}_\al$
are then determined by \normI, while 
the 2-pt functions  $\eta^{(1)},\ \eta^{(2)}$ in the basis \ringXI\
follow from \topXI:
\eqn\topmetI{
\eta^{(1)} = \pmatrix{
178& 0& 5& 0 \cr  712& 89& 20& 10 \cr  89& 0& 0& 0 \cr  356& 0& 10& 5}
\qquad
\eta^{(2)}=\pmatrix{
4& 0& 2& 10& 40& 1\cr 
0& 0& 0& 0& 25& 0\cr 
2& 0& 0& 5& 20& 0\cr 
10& 0& 5& 0& 100& 0\cr
40& 25& 20& 100& 400& 10\cr
1& 0& 0& 0& 10& 0}
}
Using eq. \yukI\ we obtain the 3-pt functions $Y_{\al\be\ga}$ as power
series in the $q_\al$ where $q_\al=e^{2i\pi t_\al}$. The associativity relations
\asso\ provide a set of quite non-trivial relations amongst these 3-pt
functions, which can be seen to be satisfied. The results are displayed in the
tables in app. F in 
terms of invariants $N_\al(\nv)$ explained in the next section.

\newsec{Results and verification}

We have collected the results for the 3-pt functions $Y_{\al\be\ga}$ for 
the four-fold examples under consideration in app. F in 
terms of invariants $N_\al(\nv)$ defined in the following way.
Taking into account the multiple covering
formula and the two factors from the intersections
of a curve $C$ with the divisors $H^{(1)}_\al,\ H^{(1)}_\be$
we write similar as in \inst:
\eqn\instII{\eqalign{
Y_{\al\be\ga}&=Y_{\al\be\ga}^{(0)}+\sum_\nv N^\prime(\nv) \prod_{\de=1}^{\hoo} 
q_\ga^{n_\ga},\cr
&= 
Y_{\al\be\ga}^{(0)}+\sum_\nv N_\ga(\nv) M_{\al\be}(\nv)
{\prod_{\de=1}^{\hoo} q_\de^{n_\de}\over (1-\prod_{\de=1}^{\hoo} q_\de^{n_\de})}
,\qquad M_{\al\be}(\nv)=n_\al n_\be 
}
}
The result for the numbers $N_\ga(\nv)$ obtained from the various $Y_{\al\be\ga}$
in this way depends 
then only on the index $\ga$ of the charge two operator $\co 2 _\ga$,
reflecting the fact that the numbers $N_\ga(\nv)$ contain 
a factor counting the intersection of the rational curve $C$ with the codimension
2 cycle $H^{(2)}_\ga$ associated to $\co 2 _\ga$; $N_\ga(\nv)$
divided by this factor is expected to agree with $(-1)^{{\rm dim}\cx M}$ 
times the Euler number of the moduli space $\cx M$ of rational 
curves on $\x$, subject to the constraint to intersect $H^{(2)}_\ga$. 
To be more precise, there is a single normalization factor $c_\gamma$ for each
$\gamma$ which we have not yet fixed by the constraints, 
which corresponds to multiplication
of $\co 2 _\gamma$ by a constant $\co 2 _\gamma \to c_\ga \co 2 _\ga$. 
We can fix these constant by explicit
knowledge of one non-zero number $N_\gamma(\nv)$ for each $\gamma$, 
as we will do below by comparison with the fiber data.

Let us verify the above interpretation for the example $X_I$ by comparing the
invariants $N_\ga(\nv)$ with the Gromov--Witten invariants of 
the three-fold fiber, calculated in \hkty. As explained at the end of sect. 3, 
we can use the constraint that a curve $C$ has to intersect $H^{(2)}_\ga$ to 
fix possible moduli for the same curves in the three-fold fibered four-fold.
Remind that $H^{(2)}_\ga$ is obtained as the intersection of two divisors.
The first divisor we chose is of course the fiber itself, which can
be seen to be $D_1=D_9=K_3$. The second divisor will be one of the other
$K_i, \ i=1,2,4$. For the curves in the fiber the constraint to intersect
$H^{(2)}_\ga$ obtained in this way reduces to the constraint to
intersect $K_i$, moreover the number of intersections is the degree 
$n_i$. From \ringXI\ we see that we have to consider 3-pt functions 
involving  $\co 2 _\gamma$ for $\gamma=2,3,6$, respectively.
The result agrees with the three-fold invariants in \hkty\ for the 
normalization of $\co 2$ operators $c_\ga= ({1\over 5},1,1)$.

To be more precise, note that the operator $\co 2 _2$ in \ringXI\ 
is not precisely $K_1K_3$ but $K_1 K_3 + 2 K_1^2$. 
Although it turns out to be sufficient in  this geometrical simple example 
to chose the proper leading piece in the large base space limit, 
this is not 
the case in general. However we can determine the correct choice of 
$H_\al^{(2)}$ in the following way.

Consider the large base limit of the 4-pt functions $K_{\al\be\ga\de}$
on $\x$.
From their definition we expect that in the limit $t_D \to \infty$, where the 
instanton effects from the base are switched off,  the
4-pt functions with first index equal to $D$ become the 3-pt functions on the
fiber $\y$. For the same reason, 
the operator product coefficients $C^{(1)\ \ga}_{\al \be}$ with first index $D$ 
become constant. We have:
\eqn\fourptlim{
 Y^{\rm 3f}_{\al\be\ga} = \lim_{q_D=0} K_{D\al\be\ga} = 
\lim_{q_D=0} Y_{\be\ga\mu} N_\al^{\ \mu},\qquad 
N_\al^{\ \mu} = \lim_{q_D=0} C_{D\al}^{(1) \mu }
}
stating that the coefficients of the $q_D$ independent terms in the 
new basis obtained by multiplication  of $Y_{\al\be\mu}$ with
the constant matrix $N_{\al}^{\ \mu}$ agree with those of the three-fold fiber.
The corresponding operators $O^{(2)}_\al$ follow then from transforming 
them into the same basis. The matrices  $N_{\al}^{\ \mu}$ determining
the appropriate choice of $\co 2 _\al$ in the cases $X_{II}$ and $X_{III}$ 
are given in the appendix.

\subsec{Space-time effective action}
The limiting behavior \fourptlim\ can also be understood from the 
relation of the correlation functions $K_{\al\be\ga\de}$ and
$Y_{\al\be\ga}$ to couplings in the space time effective action. 
In the following we will perform a preliminary analysis of the 
properties of these couplings and discuss some of their possible implications;
we defer a more complete analysis and the precise relation to the
four-dimensional effective theory of the heterotic dual to a future 
publication.

Consider the eleven dimensional supergravity limit of \cre,
with field content the eleven-dimensional vielbein $e_M^a$, the gravitino
$\Psi_M$ and the antisymmetric gauge tensor $A^{(3)}_{MNR}$,
augmented by a term taking care of the five-brane sigma model anomaly \dlm:
 \eqn\newterm{
\delta S = {1\over 2} \int A^{(3)} \wedge X_8 (R)
}
where $X_8(R)$ is the eight-form anomaly polynomial, quartic in the Riemann tensor $R$.
The equation of motion for $\fff$, including a source term for the membrane and five brane
reads
\eqn\feom{
d\star \fff = -{1\over 2} \fff \wedge \fff + X_8 (R)  + \sum_i Q^i_{2B} \delta^{(8)}_{i}  + 
\sum_i Q^i_{5B} \delta^{(5)}_{i}\wedge T^{(3)} 
}
where $\ttt$ is the self-dual three-form field strength on the five brane, 
$\delta^{(n)}$ is a $n$-form integrating to one on the transverse directions to a brane
and $Q_{2B},Q_{5B}$ the charges of the branes.

Implications of the term \newterm\ for M-theory compactifications to three dimensions
have been discussed in \bebe\sethi\witflux. Integrating \feom\ over the four-fold
imposes the constraint that the integral on the rhs has to be zero. This
integral is also the coefficient of the $A^{(3)}$ tadpole in the three dimensional space-time
which has again to be zero for a consistent vacuum \sethi. The integral of the anomaly polynomial 
$X_8(R)$ over $\x$\ is proportional to the Euler number, $\int_\x X_8(R)=-{1\over 24}\chi$
\bebe, and gives generically a non-zero contribution. 
It can be cancelled by the contributions
from membranes filling space-time and $\fff$ flux on the manifold. In fact it has been shown in 
\witflux\ that there has to be non-zero $\fff$ flux in the case that the first Pontryagin class $p_1$
of $\x$ is not divisible by four.

Different supersymmetric vacua on $\x$ are then characterized
by the values of the non-trivial flux of the internal components
$F_{m\bx m n \bx n}$ of $\fff$ which can be expanded as 
\eqn\flux{
F^{(4)} = \sum_\ga \nu^\ga \omega_\ga^{2,2} \ ,
}
where $\omega^{i,j}_\ga$ denote a basis of harmonic $(i,j)$ forms and the 
number $\nu^\ga$ are integers (or possibly half-integers, if ${p_1 \over 4}$ is 
not an integral class \witflux).
The only other non-vanishing component of $F^{(4)}$ is $F_{\mu\nu\rho m}
\sim \epsilon_{\mu\nu\rho}$ 
\bebe, where roman  (greek) indices refer to compact (non-compact) dimensions.

Expanding the fields $\Psi_M$ and $A^{(3)}_{MNP}$ in harmonics on $\x$ one obtains
terms in the three-dimensional action which are classically proportional to integrals 
over the internal manifold. In particular, terms related to the couplings 
$K_{\al\be\ga\de}$  and $Y_{\al\be\ga}$ should have an internal structure
$$
\kcl_{\al\be\ga\de} = 
\int \omega_\al^{1,1} \wedge \omega_\be^{1,1} \wedge \omega_\ga^{1,1} \wedge \omega_\de^{1,1}\ ,\qquad
\ycl_{\al\be\ga} = 
\int \omega_\al^{1,1} \wedge \omega_\be^{1,1} \wedge \omega_\ga^{2,2}\ .
$$

The relevant four field coupling in eleven 
dimensions related to $\kcl$ are the four-fermi
interactions of the gravitino $\Psi_M$  in the action of \cre\ which yield 
four-fermi terms in three dimensions from zero modes of $\Psi_M$ which are
related to harmonic $h^{1,1}$ forms similar as in the three-fold case \stroIII.

On the other hand, terms proportional to $\ycl$ arise from the three field couplings
\eqn\tfc{
\int A^{(3)} \wedge F^{(4)} \wedge F^{(4)}  \ ,\qquad
\int (\bar{\Psi}_\mu\Gamma^{\mu\nu\al\be\ga\de}\Psi_\nu + 
\ 12 \bar{\Psi}^\al\Gamma^{\ga\de}\Psi^\be)F^{(4)}_{\al\be\ga\de} \ ,
}

From \flux\ it follows that the 3pt couplings $Y_{\al\be\ga}$ are related to Chern-Simons
couplings in three dimensions
\eqn\csI{
c_{\al\be} A^\al \wedge F^\be,\qquad c_{\al\be}\equiv Y_{\al\be\ga} \nu^\ga\  ,
}
whereas the second coupling in $\tfc$ provides the $N=2$ supersymmetric completion, a mass
term for the fermionic superpartners. 

M-theory on $\x$ is dual to heterotic
string on $\z\times S^1$. It is interesting to compare the above result with the 
compactification of the heterotic string to three dimensions on Calabi--Yau three-fold
times $S^1$ performed in \nis. Starting from the dual formulation in ten dimensions
involving a six-form gauge potential instead of the usual two-form 
potential one obtains similarly gauge and gravitational Chern-Simons type of couplings 
in three dimensions\foot{The analysis of the five-dimensional duality between M-theory and 
the heterotic string in \aft\ suggests that the choice of representation is actually
not essential.}. In fact it was argued there that these couplings break supersymmetry. 
We will see momentarily that tree-level supersymmetry indeed implies the vanishing
of the coefficients $c_{\al\be}$ in the four-fold compactification. However non-zero
$c_{\al\be}$ will be generated by instanton corrections to the classical periods.

\subsubsec{$h^{2,2}$ cohomology and $F^{(4)}$ flux}
The four-form cohomology $H^{2,2}(\x)$ plays a special role in many 
respects. From the point of  topological field theory it is special since
it is part of both the vertical primary space $H_V^{k,k}(\x)$ 
as well as its horizontal primary space $H_H^{d-k,k}(\x)$.
Similarly there exist mixed boundary conditions in the construction
of supersymmetric 4-cycles \bebeII. 
Moreover recall that the elements of $H_V^{2,2}(\x)$ considered so far
are only a tiny subspace of the full $H^{2,2}(\x)$ cohomology, which has a dimension
of the order of $10^4$ in the simple four-folds we consider (see table 1).

Elements of $H^{2,2}(\x)$ are related to topological degrees 
of freedom of the vacuum, the $\fff$ flux on $\x$ characterized by the flux numbers 
$\nu^\ga$. This is in agreement
with the relation between $H^{2,2}(\x)$ and the operators of $\co 2$ in
the topological field theory. They are obtained by fusion of $\co 1$
operators and have twice the $U(1)$ charge of a marginal operator,
representing massive perturbations of the topological background rather
than massless fields. It is also in agreement with the expected spectrum
for the type IIA theory after compactification on a further circle. In particular, starting
from the dual five form field strength in ten dimensions one obtains vector gauge fields
in two dimensions which can carry only topological degrees of freedom.
Since the $\co 1$ operators are naturally related
to the fundamental space-time fermions there is also the possibility that 
the $\co 2$ operators could also correspond to condensates quadratic in
space-time fields.

The actual choice of the fluxes $\nu^\ga$ is related directly to both
internal and space-time properties of the vacuum, namely the $\fff$ flux
and the resulting Chern-Simons couplings \csI. From the internal point of view
there are several immediate questions concerning a non-zero value of a flux 
$F^{(4)}$ on a 4-cycle: i) what are the possible consistent configurations of
flux on the internal manifold given the cohomology of $\x$; ii) how does the
choice of flux influence wrappings of branes and therefore the spectrum and
physical properties of the vacuum; iii) what is its origin.

Concerning the first question consider the equation fulfilled by the
components $F_{m\bx m n \bx n}$ derived in \bebe:
\eqn\bebef{
F_{m\bx m n \bx n} g^{n \bx n} = 0 \ ,}
where
$g_{m \bx m}$ is the K\"ahler metric on $\x$. It is straightforward to show
that this equation implies the vanishing of the
integral $I(F)=\int_\x J\wedge J \wedge F$, where $J$ is the K\"ahler form.
Moreover $\fff$ has to be self-dual.
$I(F)$ is a quadratic polynomial in the $h^{1,1}$ special coordinates $t_\al$,
$ I(F) = \sum_{\al\be\ga} t_\al t_\be \nu^\ga Y^{\rm cl.}_{\al\be\ga}$.
For generic values of these moduli the $h^{1,1}(h^{1,1}+1)/2$ coefficients 
have to vanish. Since $dim(H^{2,2}_V(\x))=2h^{1,1}-2$, there is in general
no non-trivial solution for the $\nu^\ga$ within $H^{2,2}_V(\x)$ and $I(F)=0$ 
imposes a constraint on the moduli. 
In fact $I(F)$ measures the volume of a 4-cycle
if $F$ is dual to a codimension two submanifold as is the case for the 
individual terms in the expansion in \flux\ for a properly chosen basis 
$\omega^{2,2}_\ga$. Moreover $I(F)=0$ 
implies that the classical Chern-Simons couplings \csI\ vanish.
On may ask about the stability of such a vacuum, which is consistent
only for special values of the K\"ahler moduli. The following observation
indicates that vacua of this kind indeed exist. Mathematically one
expects a reduced quantum moduli space associated to the slice
of fixed $t_\al$. This kind of quantum moduli spaces can be realized
as that of toric manifolds which describe the modding of a basic toric
manifold (the physical theory without flux) by an identification of
K\"ahler moduli, that is a modding on moduli space. For an example
of this kind we refer the reader to the discussion of the 4-fold
$P_{1,1,1,1,4,4}$ in sect.6.

Allowing for general elements of $H^{2,2}(\x)$ in the expansion of $F^{(4)}$
in \flux, there will be non-trivial solutions for the $\nu^\ga$ for which all 
coefficients of the quadratic polynomial $I(F)$ vanish. The classical
coefficient $c_{\al\be}$ of the Chern-Simons couplings then still are zero for the
same reasons as before. However since such a non-trivial solution involves
relations between classical geometrical data which are generically not respected
by the quantum corrections, instantons will generate non-zero 
$c_{\al\be}$. These instanton generated Chern-Simons interactions 
in the three-dimensional space-time effective action with a potential of 
breaking supersymmetry deserve certainly further study. 
A more detailed analysis is 
necessary to determine the vacuum structure which could also involve condensates
arising from the 4-pt couplings $K_{\al\be\ga\de}$.

\subsubsec{Phase transitions and disconnected vacua}
An interesting interplay between the flux conditions $\nu^\ga$ and 
possible five-brane
wrappings inside $\x$ is implied by the Bianchi identity of the 
three-form field strength $T^{(3)}$ on the five-brane, 
$dT^{(3)}=F^{(4)}$ \town\witm.
This equation has been used in \dmw\ to argue that 
the flux of $F^{(4)}$ associated to a four-cycle $C_4$ has to be
zero in order that the five-brane can be wrapped around $C_4$. 

Said differently, 
a vacuum can be stabilized by putting appropriate $F^{(4)}$ flux on 4-cycles in
divisors which can support superpotential terms generated by wrappings of five 
branes. It seems likely that this mechanism plays a role in the situation where
a point is blown up in a four-fold compactification $\x$ which did not generate a
superpotential before the blow up \witspp\sethi. In this case one faces 
the 
contradiction that the two vacua, $\x$ and $\x^\prime$, 
which are apparently connected by a phase 
transition, differ in that a superpotential $W$ is generated only in the theory 
$\x^\prime$ with the point blown up. 
The above comments suggest the following resolution
to this problem: the moduli space $\cx M(\x)_{W=0}$ where  $W=0$ 
is connected to a version of $\x^\prime$ with moduli space $\cx M(\x^\prime)_{W=0}$,
where flux on the exceptional divisor $E$ prevents a superpotential (alternatively
one could also consider the case where some of the space-time filling 
membranes are located on $E$ and change the zero mode structure of
a five brane wrapped on $E$). On the other hand the moduli space of that
version of $\x^\prime$ with $W\neq0$ is disconnected from $\x$.
Indeed the superpotential depends now non-trivially on the volume of $E$
and is therefore no longer a flat direction.

Also the Bianchi identity for the three form field strength
$T^{(3)}$ has to be augmented by a source term for a membrane boundary
in a more complicated 
configuration of branes, $dT=\fff+Q_{1B} \delta^{(4)}$, where $\delta^{(4)}$ is
a four-form which integrates to one in the directions transverse to the 
one-dimensional intersection of the membrane with the five brane. 
Clearly it will be interesting to 
understand the possible consistent configurations and transitions amongst them.

Let us finally also mention the homology class of 3-cycles on $\x$. 
3-cycle classes are relatively rare objects in the four-fold we have encountered
as can be seen from the values $h^{1,2}=0,1$ in table 1. If a  three cycle exists 
it can contribute to the tadpole in three dimensions via the interaction 
$\int_{5B} A^{(3)} \wedge T^{(3)}$ in the presence of further membranes.
\vfill

\newsec{$N=1$ superpotentials}

In the next section we will look in some detail at the singularities 
in the moduli space of the Calabi--Yau four-folds. Physical phenomena
associated to these singularities may lead to the generation of
a superpotential in the 4d $N=1$ theories obtained from compactification
of F-theory on $\x$ or heterotic string on $\z$. It was shown in
\witspp\ that superpotential terms can be generated by wrapping
the M-theory five brane on divisors in $\x$ which fulfill a
certain set of conditions. In the following we give a 
{\it complete classification of divisors of the appropriate type from
intersection theory on the toric variety}. Here we use the fact that 
the hypersurfaces $D_i:\{x_i=0\}$ of the toric variety provide 
a complete basis of divisors on $\x$ \bat. Note that intersection
theory is anyway the minimal framework that is needed to make any sensible 
statement about the superpotential, since it determines the instanton 
action $\sim e^{-V_D}$.

The first condition derived in \witspp\ from anomaly cancellation
requires the arithmetic genus $\chi(\cx O_D)$ of 
$D$ to be one, where $\chi(\cx O_D) = h^{0,0}(D)-h^{1,0}(D)+
h^{2,0}(D)-h^{3,0}(D)$. More precisely, to have the 
correct number of fermionic zero modes to generate a superpotential,
we consider divisors with $h^{1,0}=h^{2,0}=h^{3,0}=0$. As explained in 
\witspp, $h^{3,0}(D)$ is the number of complex
structure deformations in the class $D$; therefore $D$ should
have no moduli. 

The contribution of $h^{1,0}$ and $h^{2,0}$ 
is zero for a divisor given by the section of a positive line
bundle \witspp\ by the Lefschetz theorem, which would imply
non zero Hodge numbers $h^{1,0},\ h^{2,0}$  for the Calabi--Yau
in contradiction with $SU(4)$ holonomy. On the other hand
exceptional divisors introduced by the resolution of singularities
can have non zero $h^{i,0}, \ i=1,2$; this happens e.g. in the
case of resolutions of curve singularities in three-folds with an exceptional
divisor of the form $C \times \IP^1$, where $C$ is a genus $g$ curve.
In this case 
$g$ adjoint hypermultiplets arise in the twisted world brane
theory from the $g$ holomorphic one forms on the singular curve $C$ \kmp.

A further condition on $D$ arises from the 
scaling behavior of its area when the four--dimensional limit 
of M theory on $\x$ is taken using the relation to F theory
on $\x \times \sI$ \witspp. Contributions from horizontal divisors - 
that is divisors which are sections of the elliptic fibration $\pi$,
$\pi(D)=B$ - are suppressed compared to vertical divisors - 
that is divisors which project on divisors $D^\prime$ in $B$, 
$\pi(D)=D^\prime \subset B$. Said differently, the action of
the five brane wrapped around the elliptic fiber is smaller
in the limit of vanishing fiber volume than that of a five brane 
wrapped only in the base. In the following let 
$\Dh$ denote a divisor fulfilling
the above conditions; that is it is vertical and has $h^{i,0}=0,\ i=1,2,3$.

\def\chid{\chi(\cx O_D)}
We can determine the classes of divisors $D$ which can 
contribute to the superpotential in F-theory compactification 
on $\x$ represented as before as a hypersurface in a toric variety,
from the intersection data on $\x$. 
$\chi(\cx O_D)$ can be expressed in terms of 
the intersections on $\x$ as \donI: 
\eqn\agI{
\chi(\cx O_D) \equiv \sum (-)^n h^{n,0}(D) = -{D^4+D^2c_2\over 24} = 1
}
This determines $\chi(\cx O_D)$ for any toric divisor $D=\sum a_iD_i$
in terms of  the topological intersection numbers in 
$R_2$ and $R_0$ in \toppart.
To avoid the moduli problem we require a divisor $D$ to be the single 
divisor with a certain weight under the $C^\star$ actions of $\x$.
Vertical divisors are obtained by taking linear combinations which
do not impose a constraint on the fiber.

In the examples in \witspp, the arithmetic genus of $D$ was always
less or equal than one. It was pointed out that in the case that $\chid = n$ 
with $n>1$, where naively there are too many zero modes to allow
for a non zero superpotential, strong infrared dynamics might
lead to the generation of fractional instantons with the right 
quantum numbers. In the more general four-folds we consider there are
points in the moduli space where strong infrared behavior is expected
and indeed we will also find the case with $\chid>1$.

There is a subtlety in the calculation of $\chid$ in the toric variety
which arises from the fact that the map from the divisor classes in the
toric variety to divisor classes in the Calabi--Yau hypersurface might not be
one to one. In particular this happens if there are points on a face
in the dual polyhedron $\dels$ and points in the interior of the dual
face in the original polyhedron $\del$. These correspond to perturbations,
which can not be realized as polynomial deformations.
Geometrically what happens
is that the divisor introduced by the resolution of a singularity 
in the toric variety intersects the Calabi--Yau hypersurface more than 
one time. As a consequence, a priori independent divisors associated to
resolutions of the 
Calabi--Yau hypersurface are not independent in the toric description.
In this case the arithmetic genus calculated from the toric
intersection numbers sums up the contribution of several divisors
in the Calabi--Yau which all correspond to the same class in the toric 
variety. As a simple example, consider the degree 12 hypersurface $\x$ in 
$P_{1,1,1,1,4,4}$ which was also considered in \witspp. There are
three exceptional divisors $\IP^2$ in $\x$ from blowing up the ${\bf Z}_4$ 
quotient singularities $x_i=0,\ i=1,\dots 4$. They all correspond
to the same divisor class in the standard description as a toric variety
where only 2 of 4 deformations can be represented as polynomial deformations.
As a result one obtains $\chid = 3$ from the intersections in the toric
variety and eq. \agI. From the definition of $\chid$ and $h^{i,0}=0,\ i=
1,2,3$ it is clear that the arithmetic genus three arises from the three
disconnected components of the toric divisor in $\x$, contributing each
with a one to $h^{0,0}$.

It was also explained in \witspp\ how the location of the complex 
codimension of $\Dh$ in $\x$ is related to the physical interpretation
of the effect in the heterotic theory on $\z$: 
vertical divisors $\Dh$ in $\x$ map to divisors $\Dh^\prime$ in $\z$.
If $\Dh^\prime$ is vertical with respect to the elliptic fibration 
$\pi_H:\ \z \to B_H$, it is localized in complex dimension one in 
the base and corresponds to a world sheet
instanton effect on the heterotic side. On the other hand horizontal
divisors are interpreted as space time instantons.

Note that by classifying the divisor classes which can support 
five brane compactifications we have not counted the actual 
representatives of a given class\foot{ Such a counting has been performed
in \donI\ for a special four-fold.}. 
In the case of rational curves
this counting is possible due to mirror symmetry; clearly it would be 
important to get a similar control over the representatives of a divisor 
in a given class.

In the following we determine the divisors $\Dh$ for the four-folds $X_{I-V}$
using intersections and describe their geometrical location and the
fibration structure of $\x$. They will serve as examples for various 
physical phenomena one encounters , namely

\noindent
o space-time/world-sheet instanton generated superpotentials\br
o compactifications of six-dimensional strings\br
o singularities associated with tensionless strings and generation of a quantum tension

\subsubsec{Divisors on $X_I$}
The intersection data for $X_I$ are given in \topXI. From 
\cstarXI\ we see that a  $D_i, \ i=6 \dots 9$ provide a basis
for the classes of divisors $D=\sum a_iD_i$ on $X_I$. In order
that $D$ has no complex structure moduli it has to be the
single divisor with a certain weight $\Lambda$ under the $C^\star$
symmetries \cstarXI. Moreover $D_6$ represents a restriction on the
elliptic fiber. We therefore consider divisors $D=a_7D_7+a_8D_8$. 
From the intersections we find 
$\chi(\cx O_D) = 1 =  2a_8^2$, which has no integer 
solutions\foot{By the same reasonings one can see
that there are horizontal divisors with $\chi(\cx O_D) = 1$.
Thus M-theory in three dimensions develops a superpotential; this
is a consequence of the existence of a global section of the elliptic 
fibration and is true for all of the four-folds below.}.
So naively there is no divisor of the type $\Dh$ to generate a 
superpotential; one would have to wrap half a five brane on
$D_8$ to get the correct number of fermionic zero modes. We will
come back to this question in the next section when we discuss
the singularities in the moduli space of this model. Finally note
that $D_8$ maps to a vertical divisor in $\z$ and would therefore
be associated to a {\it world sheet} instanton effect in the heterotic theory.

\subsubsec{Divisors on $X_{II}$}
The fibration structure of $X_{II}$ is
\eqn\cstarXII{
\vbox{\offinterlineskip
\halign{\hfil$#$\hfil\cr \ominus \cr  {1 \atop \ }
 \swarrow \searrow {1 \atop \ } \cr \circ \qquad \circ\cr} \vskip 15pt}
\hskip 80pt
\vbox{\offinterlineskip\tabskip=0pt
\halign{\strut\vrule#
&\hfil~$#$
&\vrule#&~
\hfil ~$#$~
&\hfil ~$#$~
&\hfil $#$~
&\hfil $#$~
&\hfil $#$~
&\hfil $#$~
&\hfil $#$~
&\hfil $#$~
&\hfil $#$~
&\vrule#\cr
\noalign{\hrule}
&  &&x_1&x_2&x_3&x_4&x_5&x_6&x_7&x_8&x_9&\cr
\noalign{\hrule}
&\lambda_1  && 0& 0& 0& 2& 3& 0& 0& 1& 0&\cr
&\lambda_2  && 0& 0& 1& 4& 6& 1& 0& 0& 0&\cr
&\lambda_3  && 1& 0& 1& 6& 9& 0& 0& 0& 1&\cr
&\lambda_4  && 0& 1& 1& 6& 9& 0& 1& 0& 0&\cr
\noalign{\hrule}}
\hrule}
}
The $C^\star$ actions \cstarXII\ make transparent the fibration structure
$$
{\bf T}^2(x_4,x_5,x_8) \to \IP^1_A(x_3,x_6) \to 
\big(\IP_{D_1}^1(x_2,x_7) \times \IP_{D_2}(x_1,x_9)\big)
$$
where the fiber structure of  $\IP^1_A \to \IP^1_{D_i}$ is that of
of the rational surfaces $\IF_1$.
In particular the dual heterotic string has as compactification manifold
the elliptically fibered Calabi--Yau with base $\IP^1 \times \IP^1$ \mv.
From \cstarXII\ we see that we have to consider divisors $D=a_6D_6$,
with $\chi(\cx O_D) = 1 = a_6^2$. So in this case there is 
a divisor, namely simply $D_6$. Since
$D_6$ maps to the 
base of the three-fold on the heterotic side, the 
physical interpretation in the heterotic theory is that this
contribution from the superpotential arises due to a {\it space-time}
instanton effect, rather than a world sheet instanton. 

\subsubsec{Divisors on $X_{III}$}
$X_{III}$ differs from $X_{II}$ only by the 
fibration structure over the base
\eqn\cstarXIII{
\vbox{\offinterlineskip\tabskip=0pt
\halign{\strut\vrule#
&\hfil~$#$
&\vrule#&~
\hfil ~$#$~
&\hfil ~$#$~
&\hfil $#$~
&\hfil $#$~
&\hfil $#$~
&\hfil $#$~
&\hfil $#$~
&\hfil $#$~
&\hfil $#$~
&\vrule#\cr
\noalign{\hrule}
&  &&x_1&x_2&x_3&x_4&x_5&x_6&x_7&x_8&x_9&\cr
\noalign{\hrule}
&\lambda_1  &&0& 0& 0& 2& 3& 0& 0& 1& 0&\cr
&\lambda_2  &&0& 0& 1& 4& 6& 1& 0& 0& 0&\cr
&\lambda_3  &&0& 1& 1& 6& 9& 0& 1& 0& 0&\cr
&\lambda_4  &&1& 2& 2& 12& 18& 0& 0& 0& 1&\cr
\noalign{\hrule}}
\hrule}
\hskip 80pt
\vbox{\offinterlineskip
\halign{$#$ \cr 
\hskip 22pt\ominus \cr 
\scriptstyle{1} \ \swarrow \vert \cr
\hskip 5pt \oslash \hskip 10pt \vert \ \scriptstyle{2} \cr
\scriptstyle{2} \ \searrow \hskip 0pt \downarrow\cr
\hskip 22pt\circ\cr}\vskip 15pt}
}
that is 
${\bf T}^2(x_4,x_5,x_8) \to \IP^1_A(x_3,x_6) \to \IP_B^1(x_2,x_7) \to \IP_D(x_1,x_9)
$
where the last two sequences $\IP^1_A \to \IP^1_B$ and $\IP^1_B \to \IP^1_D$
have the structure of rational surfaces of type $\IF_1$ and $\IF_2$, respectively.
The heterotic dual has again $\IF_2$ as the base of the elliptic fibration.
Divisors $D=a_6D_6+a_7D_7$.
have $\chi(\cx O_D) = 1 =  2a_7^2+a_6^2$. So $D_6$ is a divisor with 
the correct 
properties. On the other hand $D_7$ has similar properties as the 
divisor $D_8$ in the
first example. Similar to the previous cases $D_6$ maps to the base 
on the heterotic side while $D_7$ maps to a vertical divisor in $\z$.

\subsubsec{Divisors on $X_{IV}$}
$X_{IV}$ has been considered already in sect. 3 of \witspp; 
the fibration structure is
\eqn\cstarXIV{
\vbox{\offinterlineskip\tabskip=0pt
\halign{\strut\vrule#
&\hfil~$#$
&\vrule#&~
\hfil ~$#$~
&\hfil ~$#$~
&\hfil $#$~
&\hfil $#$~
&\hfil $#$~
&\hfil $#$~
&\hfil $#$~
&\hfil $#$~
&\hfil $#$~
&\vrule#\cr
\noalign{\hrule}
&  &&x_1&x_2&x_3&x_4&x_5&x_6&x_7&x_8&x_9&\cr
\noalign{\hrule}
&\lambda_1  &&0& 0& 0& 0& 0& 0& 2& 3& 1&\cr
&\lambda_2  &&0& 1& 0& 0& 1& 0& 4& 6& 0&\cr
&\lambda_3  &&0& 0& 1& 1& 0& 1& 6& 9& 0&\cr
&\lambda_4  &&1& 0& 0& 0& 0& 1& 4& 6& 0&\cr
\noalign{\hrule}}
\hrule}
\hskip 80pt
\vbox{\offinterlineskip
\halign{$#$ \cr 
\hskip 8pt\ominus \hskip12pt  \circ\cr 
\hskip 1.5pt \scriptstyle{1} \hskip 4pt \downarrow \cr
\hskip 1.5pt \hskip 8pt\circ \cr}\vskip 15pt}
}
reflecting the fibration structure
$
{\bf T}^2(x_7,x_8,x_9) \to (\IP^1_A(x_2,x_5) \times
(\IP_B^1(x_1,x_6) \to \IP_D(x_3,x_4)\big)
$.
There are two heterotic duals, one with base $\IF_1$, the other with base
$\IF_0$. A divisor $D=a_1D_1$ has $\chid=a_1^2$ thus $D_1$ has the deliberate properties.
$D_1$ maps to a horizontal divisor in the fibration with $\IF_0$ as the base and to 
a vertical one 
in the $\IF_1$ fibration.

\subsubsec{Divisors on $X_{V}$}
The last example has a fibration structure
\eqn\cstarXV{
\vbox{\offinterlineskip\tabskip=0pt
\halign{\strut\vrule#
&\hfil~$#$
&\vrule#&~
\hfil ~$#$~
&\hfil ~$#$~
&\hfil $#$~
&\hfil $#$~
&\hfil $#$~
&\hfil $#$~
&\hfil $#$~
&\hfil $#$~
&\hfil $#$~
&\vrule#\cr
\noalign{\hrule}
&  &&x_1&x_2&x_3&x_4&x_5&x_6&x_7&x_8&x_9&\cr
\noalign{\hrule}
&\lambda_1  &&0& 0& 0& 2& 3& 1& 0& 0& 0&\cr 
&\lambda_2  &&0& 0& 1& 4& 6& 0& 1& 0& 0&\cr 
&\lambda_3  &&0& 1& 2& 8& 12& 0& 0& 1& 0&\cr 
&\lambda_4  &&1& 1& 2& 10& 15& 0& 0& 0& 1&\cr
\noalign{\hrule}}
\hrule}
\hskip 80pt
\vbox{\offinterlineskip
\halign{$#$ \cr 
\hskip 8pt\ominus \cr
\hskip 1.5pt\scriptstyle{2} \hskip 4pt \downarrow \hskip 3pt \searrow\cr
\hskip 8pt\oslash \hskip 6.5pt  |\ \scriptstyle{2}\cr
\hskip 1.5pt\scriptstyle{1} \hskip 4pt \downarrow \hskip 3pt \swarrow\cr
\hskip 1.5pt \hskip 8pt\circ \hskip12pt  \cr
}\vskip 15pt}}
reflecting the fibration structure
$
{\bf T}^2(x_4,x_5,x_6) \to \IP^1_A(x_3,x_7) \to
\IP_B^1(x_2,x_8) \to \IP_D(x_1,x_9)
$
From \cstarXV\ we see that we have to consider divisors $D=a_7D_7+a_8D_8$;
such a divisor has $\chid=a_8^2+a_7^2-a_8a_7$; divisors with $\chid=1$
are therefore $D_7,\ D_8$ and $D_7+D_8$. $D_7$ maps to the heterotic base $B_H$
whereas $D_8$ maps to a divisor in $B_H$.

\newsec{Singularities in the moduli space}

For special values of the moduli, the Calabi--Yau manifold $\x$
develops singularities. At these points in the moduli space the 
periods and the correlation functions derived from them may 
vanish or acquire singularities. 
The singularities of the $N=2$ theories appearing at the discriminant 
locus of three-folds can often be understood in terms of the presence of massless 
BPS states at these special submanifolds in the moduli space \stroII.

The elliptic/K3/Calabi--Yau fibration structure of the four-folds we
consider is primarily a property of the four-fold $\x$; although in 
many cases also the mirror manifold $\xs$ will have a similar structure 
for appropriate choice of complex structure.
Therefore the singularities analyzed by the singularity structure
of the correlation functions are associated to K\"ahler moduli.
Since the dimension of the cycles associated to the K\"ahler moduli
is even and the type IIB theory branes have odd dimension
these singularities are generically associated to tensionless strings.
Differently than in the three-fold case also the complex structure
moduli are naturally related to even dimensional cycles. One possibility
to get massless point like states is to shrink a $p$ cycle to an
extended cycle rather then to a point; this is how e.g. gauge symmetry
enhancement can be related to the variation of K\"ahler moduli 
in the conventional type IIB compactifications \bsv,\kmI,\kmp.
On the other hand the fiber plays a special role in the F-theory compactification;
singularities in the fiber are interpreted as seven-branes located at 
complex codimension two rather than in terms of vanishing cycles. Therefore
massless point-like states associated to gauge symmetries can arise 
from singularities in the fiber reached by varying the complex structure.

It is known that non-vanishing superpotentials require 
singularities if the compact moduli space is compact \witbag. Therefore
some information about the physical origin of non-vanishing superpotentials 
from divisors $\Dh$ as determined in the previous section can be gained
by looking for the singularities associated to them. 
Five branes dual to small $SO(32)$ 
instantons with enhanced $SU(2)$ gauge symmetries at 
zero size \witsi\ have been argued to generate a superpotential 
in \ksil\ by analyzing
the singularities in the elliptic fibration of the 
dual F-theory compactification.
We will perform a similar analysis of
the singularities in the fibration structure of our four-folds
below, with the result that singularities in the fibration appear only at high
codimension. This is not too surprising in the light of the fact that the four-folds
we consider are expected to be dual to $E_8 \times E_8$ heterotic compactifications;
small instanton dynamics of $E_8$ instantons however involves tensionless 
strings \eestring\
rather than enhancement of gauge symmetries. The fact that 
the existence of a T-duality between $SO(32)$ and $E_8 \times E_8$ heterotic 
strings in three dimensions for a given compactification would
imply the relevance of tensionless string dynamics has been pointed 
out in \ksil. 

\subsec{Weierstrass model and fiber singularities}
Singularities in the elliptic fibrations have been analyzed in  detail
\mv,\berI\ using a Weierstrass model for $\x$. The Weierstrass model for $X_I$
is $y^2+x^3+xz^4f+z^6g$ where $y=x_5,\ x=x_4,\ z=x_6$ and 
\def\ss#1{{\scriptstyle{#1}}}
\eqn\wsI{\eqalign{X_I:\ 
f&=\sum_{\de,\ep} f_{(2\ep ;2\de -\ep ;8-\de ;\de ;\ep)}
{(x_1,x_9;x_2;x_3;x_7;x_8)}\ , \cr
g&=\sum_{\de,\ep}g_{(2\ep ;2\de -\ep ;12-\de ;\de ;\ep)}
{(x_1,x_9;x_2;x_3;x_7;x_8)}}}
where the sections $f$ and $g$ are polynomials in the $x_i$ with the degrees denoted by the
subscripts. The fibration becomes singular when the discriminant of the elliptic fiber
$\Delta_{\IT^2}=4 f^3 + 27 g^2 $ vanishes. From \wsI\ we see that above $x_8=0$, 
$f$ and $g$ are of the form
\eqn\wsIp{
f \sim \sum_{\de=0}^{8} x_2^{2\de}x_3^{8-\de}x_7^\de f_\de,
\qquad g \sim \sum_{\de=0}^{12} x_2^{2\de}x_3^{12-\de}x_7^\de g_\de,
}
whereas $\Delta_{\IT^2}$=0 requires $f\sim \Phi(x_i)^2,\ g \sim \Phi(x_i)^3$
for some function $\Phi(x_i)$.  Counting the number of parameters we see that a 
singular fiber appears at the first time at codimension nine. Similar arguments apply to the
other four-folds. For further reference we collect their Weierstrass forms:
\eqn\wsII{\eqalign{
X_{II}:&x_5^2+x_4^3+x_4x_8^4\sum_{\de,\ep} f_{(4+\de ;4+\de ;8-\de ;\de)}(x_1,x_9;x_2,x_7;x_3;x_6)
\cr&+x_8^6 \sum_{\de,\ep}g_{(6+\de ;6+\de ;12-\de ;\de )}(x_1,x_9;x_2,x_7;x_3;x_6)
}}
\eqn\wsIII{\eqalign{
X_{III}:&x_5^2+x_4^3+x_4x_8^4\sum_{\de,\ep} f_{(2\ep ;4+\de-\ep ;8-\de ;\de ;\ep)}(x_1,x_9;x_2;x_3;x_6;x_7)
\cr&+x_8^6 \sum_{\de,\ep}g_{(2\ep ;6+\de-\ep ;12-\de ;\de ;\ep)}(x_1,x_9;x_2;x_3;x_6;x_7)
}}
\eqn\wsIV{\eqalign{
X_{IV}:&x_8^2+x_7^3+x_7x_9^4\sum_{\de} 
f_{(8-\de ;8,12-\de,\de)}(x_1;x_2,x_5;x_3,x_4;x_6)\cr&
+x_9^6 \sum_{\de}
g_{(12-\de ;12,18-\de,\de)}(x_1;x_2,x_5;x_3,x_4;x_6)
}}
\eqn\wsV{\eqalign{
X_{V}:&x_5^2+x_4^3+x_4x_6^4\sum_{\de,\ep} 
f_{(4+\ep;2\de-\ep;8-\de;\de;\ep)}(x_1,x_9;x_2;x_3;x_7;x_8)\cr&
+x_6^6 \sum_{\de,\ep}
g_{(6+\ep;2\de-\ep;12-\de;\de;\ep)}(x_1,x_9;x_2;x_3;x_7;x_8)
}}
Analogously as in the previous case one can verify that the condition
$\Delta_{\IT^2}\sim x_i$ for the relevant $x_i$ 
requires the  restriction to a locus of high codimension 
in the complex moduli space.

\subsec{Tensionless strings and their compactifications}

In six dimensions there are several types of non-critical
strings with different number of supersymmetries and world sheet 
properties \wittalk\eestring\swsix\mv.
Clearly F-theory compactification on three-fold fibered four-folds 
will involve compactification of these non-critical strings 
to four dimensions. 
In particular non-critical strings in type IIB or F-theory 
arise if a 2-cycle $C_2$ or 4-cycle $C_4$ shrinks to zero size from wrapping 
the 3-brane and 5-brane around these cycles, respectively\foot{For a
detailed discussion of aspects of shrinking 4-cycles see e.g. \witm,\mv.}.
The world sheet properties of the string depend on the normal 
bundle of $C_i$ in the manifold. Additional structure may arise if 
the cycle wrapped by the brane is itself intersected by branes or if
lower dimensional branes live on it.

We will call the two relevant types of normal bundles
of vanishing 2-cycles and strings from wrapping three branes 
around them
type i): $\cx O(-1)\times \cx O(-1)$ and
type ii): $\cx O(-2)$.

In the present context - fibrations of elliptically fibered threefolds
with base $\IF_n$ - these types of vanishing cycles arise 
e.g. when the exceptional section $E$ of  
$\IF_n$ in the fiber $\y$
shrinks to zero size; we will restrict to the cases $n=1,2$ in the following. 

In six dimensions the type i string has $N=1$ supersymmetry.
$E$ is intersected by eight seven branes, 
leading to a current algebra of rank eight on the world sheet 
of the string \witm. This is the same tensionless string as it
appears if a M-theory fivebrane comes close to a nine brane
and becomes a small $E_8$ instanton \eestring. In fact in F-theory
compactification to six dimensions, where one has to take the limit
of small elliptic fiber, the vanishing of $E$ coincides with the 
vanishing of a whole 4-cycle $S$ in the Calabi--Yau \mv. After compactification
on a circle this theory becomes dual to M-theory on $\y$ and develops
a new exotic phase \kmv. In this five dimensional theory one can now shrink
$E$ and $S$ independently; in the first case one obtains only a 
massless hypermultiplet \witm, whereas for small volume of $S$ 
one obtains a low tension magnetic string from wrapping the five brane around $S$
and infinitely many massless electric states from wrapping membranes
around 2-cycles in $S$ which are all equally relevant in the 
limit of zero volume \mv,\witm,\kmv.
 
For $n=2$, $E$ is not intersected by any seven brane and consequently 
there is no variation of the type IIB coupling constant above it;
shrinking $E$ one obtains a tensionless string with $N=2$
supersymmetry, which becomes in the small volume limit 
the same string as in type IIB on K3 with vanishing 
2-cycle \witm. More precisely the vanishing tension requires also
the adjustment of four real scalars corresponding to a hypermultiplet
in the $N=2$ spectrum.

In compactifications of F-theory to four dimensions, new 
quantum corrections may arise and lead to the appearance 
of new phases in the lower dimensional theories. 

\subsec{Five branes and the superpotential}
We will see now that the divisors determined in sect. 6 in the four-folds can 
be divided two classes: a) divisors $\Dh_a$ with $\chi=1$ arise
from fibering the special 4-cycle $S$ of del Pezzo type
over a further $\IP^1$ in different ways.  
From the five-dimensional 
point of view the superpotential arises therefore from  
compactifications of the magnetic non-critical string to three dimensions;
b) divisors $\Dh_b$ with $\chi=2$ have the structure $K3 \times C$,
where $C$ is the base of the four-fold. $C$ can be shrunk to zero size
in the four-fold; at this point in the moduli space a $\IZ_2$ symmetry
is restored.

\subsubsec{4-cycle fibrations}
As an example for the first type 
let us identify the singularities in the K\"ahler moduli
space of $X_{III}$ associated to the contribution to the superpotential
from $D_6$. F-theory compactification on the fiber $\y_{\IF_1}$ has been
discussed in  detail in \mv,\kmv. This three-fold compactification has
a point in the moduli space where the 4-cycle $S$ in the
Calabi--Yau shrinks to zero size. $S$ is a del Pezzo surface and has a 
remarkable geometric structure; in particular there is a natural action 
of the Weyl group of $E_8$ on the representatives in $H^{1,1}$. After compactification
on a circle one can count the BPS winding states which in the dual M-theory arise
from membranes wrapping 2-cycles in $S$ \ganorI\kmv. 
In addition to
these ``electric'' winding states there is a ``magnetic'' string 
from wrapping the five brane around $S$. To see that this five brane 
is the origin of the superpotential term arising from $D_6$,
consider the Weierstrass model of $X_{III}$, eq. \wsIII, 
expanded around $x_6=0$:
\eqn\wsIIIp{\eqalign{
&y^2+x^3+xz^4\sum_{\ep=0}^{4}f_{2\ep}(x_1,x_9)x_2^{4-\ep}x_7^\ep+
z^6\sum_{\ep=0}^{6}f_{2\ep}(x_1,x_9)x_2^{6-\ep}x_7^\ep \ ,
}}
where $y=x_5,\ x=x_4,\ z=x_8$, $f_\al$ is a homogeneous polynomial of degree $\al$
and we have set $x_3$ to one. Choosing a point
on the base $\IP^1$, say $x_1=0,\ x_9=1$, this is nothing but the $E_8$
type del Pezzo surface. The complex surface $D_6$ in the four-fold is a
fibration of the vanishing 4-cycle over the base $\IP^1$. The superpotential
in four dimensions arises from the compactification of the magnetic
string in six dimensions on the base $\IP^1$ of the fibration.

\goodbreak\midinsert
\centerline{\epsfxsize 3.truein\epsfbox{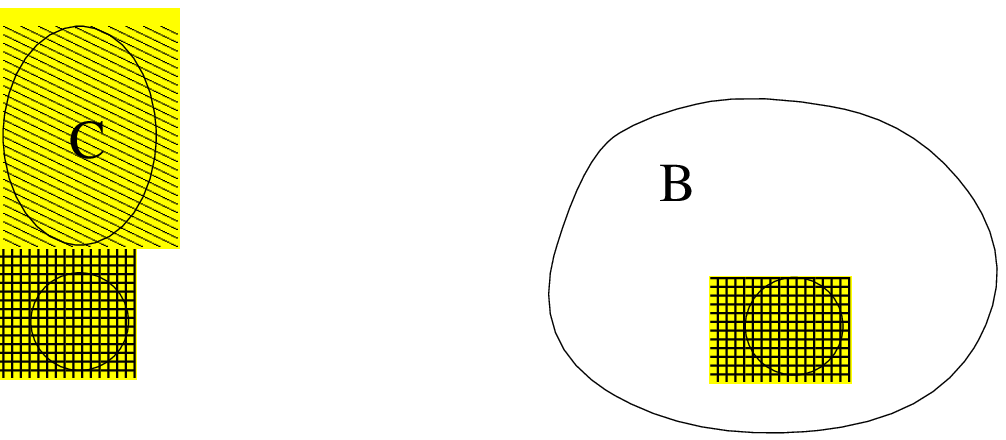}}\leftskip 2pc
\rightskip 2pc\noindent{\ninepoint  \baselineskip=8pt {\bf Fig.1}:
Generation of the superpotential by worldsheet instantons of a
solitonic string. On the F-theory side (left) wrapping the
five brane on the 4-cycle $C$ one obtains a solitonic string.
Wrapping the world-sheet further on the base $\IP^1$ of the 
6-cycle produces an instanton contribution to the superpotential.
The dual heterotic theory observes a world-sheet instanton from
wrapping the {\it fundamental} string on the same $\IP^1$ in the
base $B$ of the elliptic fibration $\pi:\z \to B$, where
$B$ common to F-theory and heterotic compactifications.}
\endinsert

Similar arguments apply to the divisors $\Dh$ with $\chi=1$ found in the 
other examples, as can be verified from the Weierstrass from given
in eqs. \wsI-\wsV; the divisors $\Dh$ are fibrations of the 4-cycle $S$ 
over a further $\IP^1$. However there are also interesting differences:
first as mentioned previously the divisors can map to horizontal 
and vertical divisors on the heterotic base with the consequence that they
can appear as world sheet or space time instanton effects in the heterotic
case. The situation that geometrically similar configurations appear
as either world sheet or space time instanton effects in the dual 
heterotic theory is particularly marked in the four-fold $X_{IV}$ with
two heterotic duals \witspp, where
the same five brane wrapping in the M-theory picture is interpreted 
differently in the two dual theories.
Additional differences arise due to various configurations of 2-cycles
contained in or intersecting $\Dh$,
which support further brane wrappings and can get zero size.

A peculiar situation is found in the four-fold $X_{V}$ where 
the divisor
$\Dh=D_7$ is of the form $S\times \IP^1$, as can be seen from eq. \wsV. 
Let $\ep^2$ be the volume of 
$S$ and $\ep^\prime$ the volume of the $\IP^1$. In the M-theory
compactification we obtain the following states: a) the five-brane
instanton with an action $\sim \ep^2\ep^\prime$; b) the magnetic
string with tension $\sim \ep^2$ from wrapping $S$; c) a new string with
tension $\sim \ep \ep^\prime$ from wrapping $\IP^1 \times C$, where $C$
is a holomorphic 
curve in $S$; three-dimensional states from wrapping d) the $\IP^1$
leading to a mass $\sim \ep^\prime$ or e) a curve $C$ in $S$ providing
a state with mass $\sim \ep$. In this case we can actually count the 
number of representatives of {\it four}-cycles contributing to $c)$ by counting
holomorphic  curves in $S$; this can be done using mirror symmetry as
in \kmv. The result is that the number of curves $C$ of degree $n$ is 
given by the coefficient of $q^n$ in the ``string partition function'', 
$\Theta_{E_8}\eta^{-12}$ where $\Theta_{E_8}$ is the $E_8$ lattice sum.
Each contributes a different string state. This is similar as in the case
of IIB on K3 with a $A_N$ singularity where each 
vanishing 2-cycle contributes a tensionless string in the six-dimensional 
theory.  

The other divisor with $\chi=1$, $D_8$, has also an interesting 
structure. Again from \wsV\ one finds that it can be either interpreted 
as a fibration of $S$ over $\IP^1$ or as a fibration of K3 over another 
$\IP^1$. The M-theory five-brane wrapped on K3 gives a string with 
the world-sheet structure of the heterotic string \hs. 
Viewing the fibration from the
one or the other perspective the five-brane instanton arises in the
compactification of the ''$S$''-string or the solitonic heterotic string.
In fact the K3 fibration has the same base $\IP^1$ as the dual heterotic
theory, where it is interpreted as a world-sheet instanton, too 
(since $D_8$ maps to a vertical divisor in $Z$); this provides a quite
explicit picture of how the instanton effects are generated in the dual theory.
It would be interesting to have a similar understanding for the ''$S$''-type
string seen in the second fibration.

\subsubsec{Divisors with $\chi=2$}
The divisors with $\chi=2$ have a different
structure. From \wsI,\wsIp\ we see that $D_8$ in $X_I$ is given by an 
equation 
$$
y^2 + x^3 +xz^4 f_8(x_3,x_7)+z^6g_{12}(x_3,x_7) \ ,
$$
where $f_8$ and $g_{12}$ are again homogeneous polynomials and $x_2$ has been set to one.
This equation defines a K3 hypersurface of degree twelve in ${\bf WP}(1,1,4,6)$.
Together with the base $\IP^1$ $D_8$ is of then of the form $K3\times C$ with
$C$ a rational curve.
The individual contributions to the arithmetic genus are $\chid=1-0+1-0=2$,
where the contribution to $h^{2,0}$ comes from the holomorphic two form of $K3$
\foot{
The same argument applies also to the models and exceptional divisors 
considered in \ksil; however in these cases the $K3$ is singular.
We did not check with eq. \agI\ since the toric construction is 
quite involved due to the large number of moduli in these models.}. 
$C$ is a curve with normal bundle of type ii which can get zero size; 
the singularity associated to this point in the moduli space where
a tensionless string appears and a $\IZ_2$ symmetry is restored will be
discussed momentarily.

\newsec{Non-critical four-dimensional strings and quantum tension}
In the previous section we have argued that different than in the cases
considered in \ksil\ the generation of a superpotential is linked to 
singularities in the moduli space due to tensionless strings rather than
strong coupling dynamics of non-abelian gauge symmetries. Apart from the
singularity which is associated to the vanishing 4-cycle producing the
$E_8$ type of tensionless string, additional singularities with light strings
arise if one can shrink 2-cycles; in fact
the divisors $\Dh$ have in common that there are points at finite 
distance in the moduli space where one can shrink 2-cycles $C_2$ contained 
in them, at least to a very small size. 
These can be either curves of type i which can be flopped or curves of
type ii which come together with a restoration of a $\IZ_2$ symmetry. 
At the point in the moduli space where the volume $V$ of $C_2$ 
vanishes, the effective action becomes
singular due to the appearance of new light modes. At least in some cases 
one can decouple (fundamental) string and gravity states and 
obtains genuine new theories which involve strings of an arbitrary
light scale and can not be described in terms of conventional field 
theory. 

Since the tension of the string is related to the volume of the 
2-cycle we have to be very careful about the definition of this 
volume $V$. It is well-known that the so called sigma model measure of $V$
\morIII, which is the definition of the volume related to the scale of
the BPS states in three-folds, is corrected by world-sheet instantons.
In general there can be of course also space-time instanton contributions.

In the cases where the instanton corrections can be calculated using
mirror symmetry, the quantum size $V$ can be inferred from arguments
related to the topology of the discriminant in the moduli space 
and the leading classical 
behavior. For other instanton corrections we will have
to use some more indirect arguments.

In any case there are two possible scenarios which will arise:
starting from a ``classical'' singularity in the moduli space
which is due to a tensionless string, the quantum corrections
might leave the string tensionless or give it a non-vanishing
tension proportional to a new dynamically generated scale 
$\Lambda$ due to to non-perturbative corrections to the quantum volume $V$ of 
$C_2$. However the scale of $\Lambda$ may be still very small compared
to all other scales in the theory and a limit can exist where
one can decouple the fundamental string and gravity effects similar as in the
``classical'' theory. In this case one obtains a theory which 
at lowest energies looks like a conventional field theory 
but at very small scale $\Lambda\ << M_{pl}$, 
generated by non-perturbative effects,
has a string with tension $\sim \Lambda^2$
bounded from below by the instanton effects which give it a non-zero
quantum tension. In fact we will see that, for certain geometries,
$\Lambda$ characterizes at the same time the scale of supersymmetry breaking.
This new combination of a field theory and a massive but
light string like spectrum
seems to be quite interesting and it is encouraging to
find it in the present context.

\subsubsec{Space-time instanton corrected tension}
World-sheet instantons can be partially analyzed using mirror symmetry,
as will be done in some detail later on. A case with 
space time instanton corrected tension can be found using the recent results
on the exact metric for hypermultiplet moduli spaces in four and
three dimensional string and field theories \seibIII\swIII\ov\sIII.

Consider type IIA compactified on a three-fold 
at a point in the K\"ahler moduli space with 
2-cycle $C_2$ of vanishing volume $V=0$,
even after taking into account world-sheet instantons as determined by 
mirror symmetry. There is a massless hypermultiplet from wrapping the 
D2 brane on $C_2$ and a singularity in the moduli space \stroII. After 
compactification to three dimension on a circle of radius $R$
the singularity is smoothed
out by D2 brane instantons on $C_2 \times S^1$; this follows 
from the exact known moduli space of $U(1)$ with $N_f=1$ in three dimensions
\seibIII\swIII. T-duality on the circle transforms type IIA to type IIB,
the D2 brane on $C_2$ in a D3 brane wrapped on $C_2\times S^1$ and 
the D2 brane wrapped on $C_2\times S^1$ into a D1 brane wrapped on $C_2$. 
The light state in three dimensions in IIB is the 3 brane
wrapped on $C_2\times S^1$. Although we can use the world sheet
instanton corrected prepotential of the type IIA theory on $\x$ 
to describe also the hypermultiplet moduli space of IIB on $\x$ 
\cec, in the latter case this is not the exact answer;
in particular the sigma model measure of $V$ is zero in 
type IIA, but not an exact quantity for the type IIB theory. This is 
just good since the
above picture implies that making $S^1$ large in the IIB theory does not 
suppress instanton effects; therefore if the tension of the string
is not zero in three dimensions then it should be non-zero already in 
four dimensions. In measuring the tension we have
neglected instanton effects which depend on the space time 
coupling \bebes. Mirror symmetry - now on $\x$
rather than the $S^1$ - transforms the situation to type IIA with a 
vanishing 3-cycle which has been solved in \ov\ with the result that there
is indeed no singularity in four dimensions, the relevant effect being 
D2 brane instantons wrapped on the vanishing 3-cycle.

For $R\to \infty$ when the instanton effects are exponentially suppressed,
the hypermultiplet of type IIA becomes very light. The state in the 
type IIB theory is the four-dimensional low tension string wrapped around
the cycle with radius $1/R \to 0$. On the other hand for $R\to 0$, the
type IIB theory becomes four-dimensional. For some value of $R$ the state
obtained from winding the string around the large circle becomes less 
relevant than the excitations of the 
low tension string itself; thus the three dimensional effective $N_f=0$
theory related to the type IIA side develops a relevant 
light string like object.

\subsec{World sheet instanton corrections}
We want now to use mirror symmetry to determine the quantum corrections
to $V$ which are calculated by the two-dimensional topological field theory.
In particular we ask about the fate of the 2-cycles of type i and ii
and the del Pezzo type 4-cycle which appear in the present context. 
For the 2-cycles the strategy is to first ensure
that the volume of a 2-cycle $C_2$ is zero taking into account instantons
which wrap $C_2$ itself and then to see how instanton effects
associated to the global embedding change the picture. Below $\x$ stands 
collectively for a Calabi--Yau manifold without specifying the dimension.

Let $t$ denote the special K\"ahler 
modulus associated to $C_2$ of $\x$; $t$ is also the sigma model measure of
the volume $V$. There exists an ordinary differential equation 
describing the relation between $t$  and the complex structure modulus
$z$ associated to it in the mirror $\xs$ of $\x$ \morIII. In the 
complex structure moduli space of $\xs$ there are singularities if
the hypersurface becomes singular itself; this happens if a solution
to the equations $f_{\Delta^\star}=df_{\Delta^\star}$ (see eq. \polyI) exists.
This locus is the discriminant locus $\Ds$ and is given by
a polynomial in the complex structure moduli $z_i$ in a given
Calabi--Yau phase. From the above map $z=z(t)$ and its inverse
one can determine the value of $t$ for a value of $z=z_0$ on the discriminant,
$\Delta(z_0)=0$.

To put the above in a concrete physical context, let $t$ be e.g. a modulus 
which plays the role of the scalar field in a vector multiplet
parameterizing the Coulomb branch of $N=2$ four-dimensional $SU(2)$
theory as in \sw\ (i.e. we 
consider type IIA on a threefold $\x$). The period related to the 2-cycle
$C_2$, 
namely $t$, and the dual 4-cycle become the periods of the field theory
in a certain limit. There is always a locus
in the compactified moduli space where one can shrink $C_2$
to zero size - and therefore $a=0$. This is due to the fact that one 
of the Calabi--Yau moduli plays the role of the quantum scale 
$\Lambda$ of the theory; one can therefore switch of any quantum 
effects by taking the limit $\Lambda \to 0$ (corresponding
usually to restriction to a boundary divisor in the Calabi--Yau moduli space). 
The precise question about the quantum volume of $C_2$ is then whether 
it stays zero for generic values of all the other $z_i$; it
can be answered by analyzing the topology of a discriminant factor 
associated to the singularity.

It is clear that the answer depends on how the cycle is embedded in
$\x$, said differently, how the modulus $t$ couples to the other 
moduli. We are lead therefore to consider a system with two moduli
and to ask whether the new modulus $y$ introduces quantum corrections
to the volume of $C_2$ and give it a everywhere non-vanishing size
(away from $\Lambda(y)=0$). In the present context the question can be addressed
by considering two moduli systems associated to the Hirzebruch 
surfaces $\IF_1$ and $\IF_2$ which provide the appropriate types of 2-cycles
and embeddings when used as a part of a Calabi--Yau three-fold or four-fold.

The one moduli systems associated to type i and type ii cycles are given
in \morIII\  and have solutions which can be written in terms of the 
elementary functions:
\eqn\onemodsol{
i):\ t= {1\over 2\pi i}
\ln z \qquad ii):\ t= {1\over 2\pi i} \ln\big({1-2z-2\sqrt{1-4z}\over 2z}
\big)}
The relevant singularities are at $z=1,\ {1\over 4}$, respectively, in both cases
$t=0$. In the second case there is a $\IZ_2$ monodromy $t\to-t$ under 
a loop of $z$ around $z=1$.
Physics wise, this monodromy can play the role of a Weyl reflection
of an enhanced gauge symmetry or the $\IZ_2$ symmetry at the 
critical point of the type ii tensionless string. 

To treat the quantum corrections we consider the system of differential
equations associated to $\IF_1$ and $\IF_2$, which have a toric description
defined by the vertices of the dual polyhedron, 
$\ns_i=(0,0),\ (-1,0),\ (0,-1),\ (0,1),\ (1,n)$
with Mori generators:
\eqn\morifn{\eqalign{
\IF_1:&\  l^{(1)} = (-2, 0, 1, 1, 0),\ l^{(2)}= (-1, 1, 0, -1, 1)\cr
\IF_2:&\  l^{(1)} = (-2, 0, 1, 1, 0),\ l^{(2)}= (0, 1, 0, -2, 1)}}
The homology of $\IF_n$ consists of the two rational curves associated
to these Mori generators; the curves associated to $l^{(2)}$ are the
base of the $\IP^1$ fibration. For $\IF_1$, $C_2^{(2)}$ is of type i, 
the other 2-cycles are of type ii. However they are differently coupled to the interior
point $\ns_0=(0,0)$ which corresponds to the hyperplane section.
The differential equations obtained from these generators coincide
with those obtained from  the Picard-Fuchs system of the Calabi--Yau $\x$ 
when restricted to the two moduli describing a $\IF_n$ surface which
is part of $\x$, $z_3=z_4=0$ in (D.11) and $z_2=z_3=0$ in \pfI.
The locus in the 
moduli space where a  cycle vanishes can be detected as previously 
by the vanishing of the discriminants, which are given by:
\eqn\discfn{\eqalign{
\Delta_{\IF_1}:\ &(1-z_2)+(36 z_2-8-27 z_2^2) z_1+16 z_1^2=\cr&
(1-4 z_1)^2+(-1+36 z_1) z_2+(-27 z_1) z_2^2\cr
\Delta_{\IF_2}:\ &
(1-4z_2) \times \big[(1-4 z_1)^2 -(64 z_1^2) z_2\big]
}}
From the above we see that the discriminant factors $z_2={\rm const.}$
associated to the singularity of the base $\IP^1$ do not change
topology under deformation with $z_1$ whereas the factors associated 
to $z_1={\rm const.}$ are quadratic factors splitting for $z_2 \neq 0$.
This split is associated to a non-zero quantum volume of the fiber $\IP^1$'s
$C_2^{(1)}$, as can be understood from the following argument.
Consider a hyperplane $H$ in the properly resolved and compactified complex 
structure moduli space $\mcs$. The discriminant $\Delta$ has in general
several factors describing codimension one loci which 
intersect $H$ in points. If $H$ is generic, the 
monodromies in $\mcs$ are generated by loops around the intersection points 
of $\Delta$ with $H$. Clearly the monodromies can not change under
variations of the position of $H$ unless $H$ becomes non-generic;
in particular this means that the vanishing cycle causing a singularity
stays the same under smooth variations of the position of $H$. On the other hand if
a point splits into two points, as it happens if $H:z_2=0$ is moved to 
$H^\prime:z_2 \neq 0$ in \discfn, the combined monodromy around the 
now two singular points will be still the same as the one around the single
point at $z_2=0$, however the individual 
monodromies are different and related to new states with different 
quantum numbers and associated  to new vanishing cycles. 

In the above case we see the cycles corresponding to the base $\IP^1$
of $\IF_n$, $C_2^{(2)}$, can still get 
zero size whereas the fibers always have a non-zero quantum volume.
The singularities associated to vanishing cycles which cause the
singularities $\Delta=0$ in the first case are therefore related to 
the vanishing of 2-cycles, whereas in the second case the
singularities are due to the {\it dual } cycles; the former singularity
due to $V_{C_2}=0$ has been wiped out by the quantum effects.

As a concrete example consider again type IIA
on a three-fold where $t$ corresponds to the scalar $a$ in the
vector multiplet of $N=2$ Yang-Mills with $SU(2)$ gauge group and no 
matter. This theory arises as a certain limit in the moduli space
of the elliptically fibered three-fold with base $\IF_2$ \kklmv,
which is also a K3 fibration with a base $\IP^1$ identical 
to the base of $\IF_2$.
If we take the large base space limit which by heterotic/type IIA
duality in four dimensions corresponds to switching off quantum effects,
we have to take therefore $z_2=0$ in \discfn\ and recover the gauge symmetry enhancement
of the six-dimensional theory on K3 from wrapping the 2 brane around
the fiber of $\IF_2$. For finite base size however the volume of 
the fiber $\IP^1$ never becomes zero again. 
The two singularities of the four-dimensional
theory - which reduce to the monopole states of the field theory in the 
appropriate limit - are given by wrapping different vanishing cycles.
In fact one can show that the vanishing periods are related to 
4-cycles instead of 2-cycles\foot{This might seem natural, since
the dual cycles of 2-cycles in the Calabi--Yau are 4-cycles; however 
from a simple dimensional counting of the
moduli spaces it is clear that one needs all kind of even dimensional
cycles and combinations of them to reproduce the singularity structure of 
the ``mirror theory''. Some consequences of the above reasonings for the 
dual heterotic theory are collected in appendix E.}. 

Although vanishing 2-cycles are replaced by vanishing 4-cycles
they still have an arbitrarily small size compared to the other scales
of the theory and in particular pass the decoupling limit in which 
one obtains the field theory - supporting the moderately massive gauge bosons of the
field theory. If we put type IIB on the same manifold, states are 
replaced by strings and  the singularities might get wiped out by 
the space-time perturbative and instanton effects as in \ov.
In fact in the four-dimensional theories with $N=2$ supersymmetries
it follows from the results in \seibIII\swIII\ that the singularity is smoothed
for the curves of type i corresponding to $U(1)$ with $N_f=1$. For
the curves of type ii the situation depends of whether they appear as
the base or as a fiber. In the first case
one gets enhanced $SU(2)$ with adjoint matter multiplets in 
a type IIA theory compactification as in \kmI\kmp; there is also a singularity 
in the three-dimensional compactification of the theory implying
in turn a true zero tension string in the four-dimensional type IIB theory
compactified on the same manifold.

\subsubsec{Hidden strings at the supersymmetry breaking scale}
On the other hand if the type ii curve appears as the fiber,
world sheet instantons split the singularity and one ends 
up again with singularities with a single hypermultiplet. 
In these cases one is left only with a low energy range 
described by a field theory modes together with very light strings,
with a tension entirely due to non-perturbative effects.
More precisely, the scale is of the order of $e^{-V_B}$,
where $V_B$ is the volume of the base $\IP^1$. An attractive
picture emerges, if this geometry appears in the context
of 3-fold fibered 4-folds, with $B$ at the same time the
base of the 3-fold fibration. As explained in sect. 2.1., any
supersymmetry breaking effect is necessarily of the order $e^{-V_B}$ - 
as is the minimal string tension.
\goodbreak\midinsert
\centerline{\epsfxsize 2.3truein\epsfbox{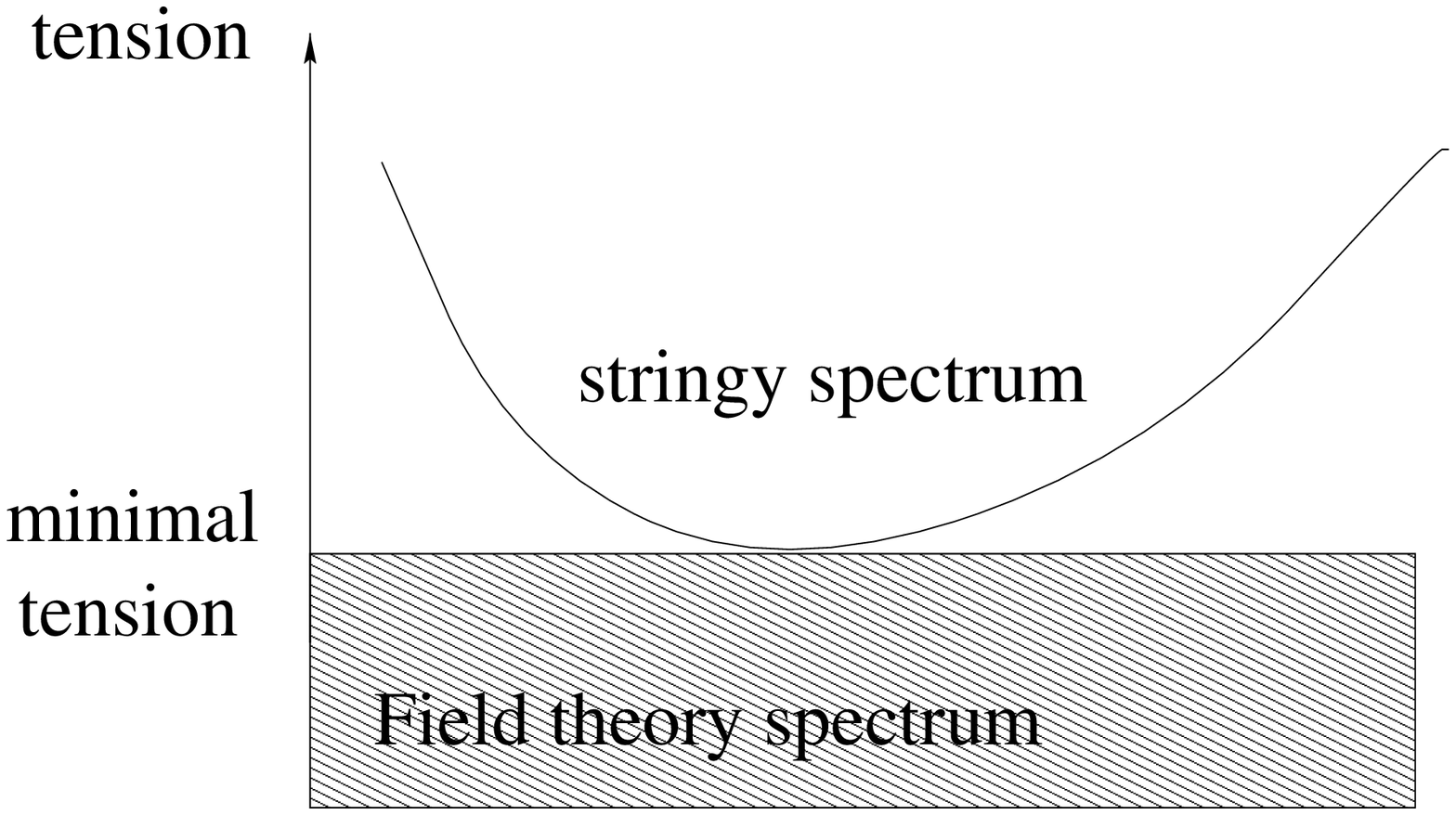}}\leftskip 2pc
\rightskip 2pc\noindent{\ninepoint\sl \baselineskip=8pt {\bf Fig.2}:
In the case with non-zero quantum tension, the spectrum at lowest
energies is that of a conventional 
field theory. Above a non-perturbatively generated 
scale $\Lambda\sim M_{SUSY}$
the spectrum is enriched by states of the solitonic string}
\endinsert

\subsubsec{Quantum tension for the $E_8$ string}
Let us now consider the generation of a quantum volume in the
case of the vanishing del Pezzo 4-cycle in four dimensions.
There are two interesting situations to consider:
a) a four-dimensional $N=1$ four-fold F-theory compactification 
with a zero size fiber; b) a four-dimensional $N=2$ three-fold
type IIA compactification with finite size fiber. As mentioned
at the beginning, the latter case is dual to the the situation considered 
in \ganorII.

In the case a) one can use the fact that if the 
elliptic fiber of the del Pezzo is identical to that of the 
Calabi--Yau, the base of the del Pezzo fibration agrees with the
base of a $F_1$ Hirzebruch surface embedded in the Calabi--Yau. 
After taking the zero size fiber limit the problem therefore
reduces to that of a type i curve considered before; we have seen
that there are no quantum correction in this case. Therefore
we expect a tensionless string as far as world sheet instantons
corrections are concerned.

In the case b) we have to determine the quantum corrections to both,
the volume of the 4-cycle $S$ and the 2-cycles embedded in it. 
A simple argument shows that neither of the two volumes can be
corrected at the singularity. First note that the volume of the
2-cycle, supporting the K\"ahler class which is varied to reach the
singularity, remains zero. 
This is already almost clear from the fact that the singular locus
in the moduli space is at a boundary where one can either flop to a 
non-geometric phase \mv\ or make a transition to the elliptically fibered
three-fold with base $\IP^2$ \witm\mv. Alternatively one can use a similar argument as 
in the previous cases based on the fact that the discriminant factor
does not change the topology under the new deformation. It is then 
straightforward to see that the monodromy under a loop around 
the singular locus requires either both volumes and periods to be zero 
or both to be non-zero. This is because the monodromy mixes 
non-trivially the vanishing period associated to the 2-cycle
with the period corresponding to the dual 4-cycle as can be e.g.
inferred from the index structure of the periods. On the other
hand the 2-cycle period has still to vanish after the monodromy 
transformation; together this implies that the dual period vanishes,
too.
Therefore
there are no world sheet quantum corrections 
from world sheet instantons to the volumes of both the 4-cycle and the 
2-cycles  in it in the four-dimensional theory at the singularity.

\subsec{Tensionless string singularities and the superpotential}
The local geometry of vanishing 2-cycles described above can lead to 
non-critical strings with low tension in the four-dimensional F-theory 
compactification\foot{A 
different representation of the $N=2$ supersymmetric string as a string living on the
3+1 dimensional intersection of two M-theory five branes has been analyzed in \hk.}.
Apart from the new physical phenomena associated to these
theories in itself, it is interesting to study their interplay with the
generation of a superpotential. In particular singularities associated to
tensionless strings can provide the poles in the superpotential required by
holomorphicity \witbag. 

In fact, the divisors considered before contain always
2-cycles or 4-cycles which can be shrunk to zero size. 
The divisors $\Dh$ with $\chi=1$ contain a curve $C$ of type i
with normal bundle $\co 1 \times \co 1$ which can be flopped in the 
Calabi--Yau. This is the 2-cycle which
one has to flop out of the del Pezzo ${\bf B}_9$, which is $\IP^2$ blown up
at nine points, to be able to shrink the remaining 4-cycle in the 
Calabi--Yau\foot{See \mv,\kmv\dkv\ for more details.}. As mentioned 
already, in the small fiber limit this flop coincides with the collapse
of the whole 4-cycle, whereas for finite size fiber (e.g in three dimensions)
$C$ can be shrunk separately. In particular $C$ is always interior to the 
base of the K3 fibration and can also be seen on the heterotic side. 

The fact that $C$ can be flopped indicates already that it can be shrunk to zero 
size also after including world sheet instantons, since it is replaced on the
other side by a different homology class and should no longer exists.
In fact it can be seen from the previously given $C^\star$ actions, that the
situation is locally described by the analysis of the $\IF_1$ case 
and the arguments of the previous section apply.

A special case where $C$ can be shrunk without shrinking the 4-cycle (at zero size
fiber), thus leading to a simple type i string appears
in the 4-fold $X_{II}$ which admits two heterotic duals on the threefold with base $\IF_1$.
In this case $\Dh=D_6$ contains two curves $C_i$ of type i which can be flopped. 
Performing a flop on  one of them, say $C_1$  one passes to a Calabi--Yau phase in which one can 
shrink the homology 2-cycle which is related to $C_2$ in the first phase,
$C^\prime_2$, without shrinking the 4-cycle. 
The heterotic dual is compactified on the elliptically 
fibered 3-fold with base $F_0=\IP_1^1 \times \IP_2^1$.
The flop maps to a point in the moduli space of $F_0$ where the volumes 
of the two $\IP^1$ are equal. According to the previous arguments, in a 
four-dimensional $N=2$ theory we would expect space time instantons to 
give this string a non-zero quantum tension. It seems very plausible that 
the same happens in the $N=1$ theory. 

Two-cycles of type ii can also be contained in the divisors $\Dh$,
in both variants discussed in sect. 8.1. In fact there is always a curve $C_2$ 
of the uncorrected type contained in the exceptional divisors with $\chid=2$;
it is at the same time the base of the Calabi--Yau three-fold fibration.
From the results in sect. 8.1. we know that world sheet instantons do 
not lead to a non-zero quantum volume (this can be also seen from the singularity
structure of the full Calabi--Yau moduli space); moreover there is
a restoration of a $\IZ_2$ symmetry as in the six-dimensional theory.
From the local geometry the string we obtain is identical to the one
obtained in the six-dimensional F-theory compactification on 
the threefold with base $\IF_2$, dual to the heterotic string on K3 with 
instanton embedding (10,14) \witm\mv. In fact there is one more
analogy: recall that the six dimensional string required the adjustment of 
a further hypermultiplet due to the fact that it is  a $N=2$ string in a $N=1$
theory. This hypermultiplet is represented by a non-polynomial deformation
which is frozen in the
toric description. We find that the same happens in the 4-fold case: there
is again the same kind of non-polynomial deformation frozen to zero at the 
$\IZ_2$ symmetric point.

In summary, the divisors $\Dh$ with $\chid = 2$ are of the form 
$K3 \times C$, where $C$ can get zero size. The tensionless string 
obtained from wrapping the three brane comes together with a $\IZ_2$ 
symmetry restored at that point and  which in turn can be spontaneously
broken by switching on the deformation frozen in the toric
description. In fact the local geometry is that of a resolved $A_1$
surface singularity which resembles a $\IZ_2$ orbifold singularity.
It is tempting to suggest that in this situation 
instanton configurations with the appropriate number of fermionic
zero modes exist.

It is an interesting question of what is the precise theory in the scaling
limit where one shrinks cycles within a divisor supporting the superpotential.
Recall that both types of divisors we have found involve a 4-cycle either
of the del Pezzo or K3 type. The 2-cycle cohomology of these 4-cycles - which 
in both cases involves a factor of the intersection form proportional to the
$E_8$ Cartan matrix - is expected to generate world brane degrees of freedom
which become light in the limit of vanishing 2-cycles. 

As an illustration of how infinitely many new contributions to the superpotential 
become relevant when the volume of a 2-cycle shrinks to zero size, consider the 
four-fold compactification analyzed in 
ref. \donI, where a precise counting of the individual representatives
in the divisor classes contributing to the superpotential has been done. 
The four-fold $X$ in this case is an elliptic fibration over the base $B$: 
$S \times \IP^1$, where, as before, $S$ denotes the elliptically fibered 
del Pezzo with $\chi =12$. Apart from the elliptic fibration $\pi: X \to B$
there is a second elliptic fibration of $X$, $\tilde{\pi}:X \to \tilde{B}$
where the fiber is the elliptic fiber of the del Pezzo and the base is 
a K3 fibration. In \donI\ the  divisors $\Dh$, vertical with respect to the 
first fibration $\pi$,  have been determined with the result that they are 
sections\foot{Note that also in this case the divisors $\Dh$
contain a K3 factor.} of the second fibration $\tilde{\pi}$.

Now consider F-theory compactified on $X$ but with the elliptic fibration
defined by $\pit$ instead of $\pi$. While the five brane wrappings on the
divisors $\Dh$ have of course still the correct numbers of zero modes to
contribute to the superpotential in the M-theory compactification on $X$,
they fail to be vertical since the intersection of the divisor with the
del Pezzo factor is a curve $C$ with volume proportional to $n t_E+t_B,
\ n\geq 1$, where $t_E$ is the size of the elliptic fiber and $t_B$ the size of the 
base $B$ of the elliptically fibered del Pezzo. Therefore in the limit of small
fiber, instantons associated to  these divisors fail to produce the necessary 
factor $\sim t_E$ to render their action finite and none of the divisors $\Dh$
contributes to the superpotential of F-theory on $X$ at a generic point
of the moduli space. However it is clear now that if, at the same time as one
takes the four-dimensional limit $R_{S^1}\sim 1/\ep \to \infty$, one adjusts
the size of $B$ to be of the order of $\ep$, all divisors $\Dh$ contribute to
the superpotential despite of the fact that they are not vertical. Said
different, infinitely many new contributions to the superpotential become
relevant in the limit, where the size of this 2-cycle becomes comparable to the
inverse of the radius of the fourth dimension.

\newsec{Discussion}
Mirror symmetry together with the powerful non-renormalization theorems 
separating the hyper and vector multiplet moduli spaces have played an
important role in the determination of exact non-perturbative quantities
of four-dimensional $N=2$ supersymmetric string theories. It has to be seen 
to which extent these concepts can be generalized and replaced in the case of
less supersymmetry.
The calculation of the correlation functions of the topological sigma model
in the first part is apart from its mathematical interest only
a first step in the determination of the $N=1$ effective action in four
dimensions. It remains to determine the precise map to the physical
quantities in the heterotic theory and also to get control over effects
which can not be encoded in the geometrical four-fold compactification of
the F-theory. This involves a better knowledge of the moduli space of
$(2,0)$ compactifications and properties of it, such as factorization
properties as in the $N=2$ case. Moreover even to the moduli dependence as 
calculated from the F-theory compactifications one should expect corrections
which can not be included in the correlation functions of the two-dimensional
topological field theory. 

At first sight it seems difficult to generalize the concepts which have been 
useful in the $N=2$ case. However things might be not as bad. Firstly the 
geometrical concepts which have been shown to be intimately linked to physical 
quantities seem to be to general and to strong not to play an important role. 
We have seen that taking a certain geometrical limit in the four-fold reproduces 
physical sensible limits where the correspondence is known. Moreover
recall that in $N=1$ orbifold compactifications - even in 
$(0,2)$ cases - the perturbative moduli dependence of the holomorphic quantities 
arises from so-called $N=2$ subsectors and is determined by the
one-loop result. And although orbifolds are special constructions,  they 
have turned out to have remarkably generic properties in the past. Our
results on the large base limit imply that the full world-sheet instanton 
corrections to the gauge 
couplings are encoded properly in the four-fold dual to such $N=1$ orbifolds.
It will be very interesting to see 
to which extent these properties generalize to heterotic compactifications 
on smooth Calabi--Yau manifolds.

An omnipresent feature of the four-fold compactifications dual to $E_8 \times
E_8$ heterotic strings is the presence of points in the moduli space where
non-critical strings obtain classically zero tension. We have seen that 
divisors generating the superpotential involve precisely the vanishing cycles 
which support these strings. 
An interesting issue is the non-perturbative generation of a minimal tension for extended 
BPS states which are classically tensionless at certain points in the moduli 
space. As discussed, such an effect can result effectively in field theories 
with a (non-critical) string spectrum starting 
at a new low energy scale. The generation of a small, exponentially suppressed
scale related to the quantum size of a geometric cycle might also be relevant
for providing a lower cut-off on the size of those six-cycles which generate
superpotential terms from wrapped five-branes and the scale of supersymmetry
breaking.

\vskip 1.5cm

\noindent
{\bf Acknowledgments}\br
I wish to thank P. Berglund, R. Minasian and S. Theisen for 
valuables conversations.
I thank also K. Becker, M. Becker, S. Ferrara, S. Katz, 
W.Lerche, D.R. Morrison, H. P. Nilles and 
E. Verlinde for helpful discussions and comments.
\vfill
\appendix{A}{\bf Holomorphic structure}
Let $\gc i _\al$ be a basis of the middle cohomology with 
intersection form $\eta^{(i)}_{\al\be}$ as in \topbas.
The holomorphic (4,0) can be decomposed as 
$$
\Om = \gc 0  Z^0 + \gc 1 _\al Z^\al + \gc 2 _\al H^\al + 
\gc 3 _\al \cx G^\al + \gc 4 \cx G^0
$$
$Z^M,\ M=1,\dots \hoo$ are coordinates on the moduli space
while $H^\al$ and $\cx G^\al$ are sections depending on the $Z^M$.
From the usual orthogonality relations 
$$
0 = \mx {\Om}{\Om} = \mx {\Om}{\p_M \Om} = \mx {\Om}{\p_M\p_N \Om} = 
\mx {\Om}{\p_M\p_N\p_Q \Om}
$$
one obtains
$$\eqalign{
2 Z \cdot \cx G + H \cdot H   &= 0\ ,\cr
(\eta^{(1)}\cx G)_M+Z \cdot \p_M \cx G + H\cdot \p_M H  &=0 \ ,\cr
H   \cdot \p _M \p _N H + Z \cdot \p_M \p_N \cx G &= 0\ , \cr
H   \p_M\p_N\p_P H+ Z\cdot \p_M\p_N\p_P\cx G &= 0\ ,\cr
H   \p_M\p_N\p_P\p_Q H+ Z\cdot \p_M\p_N\p_P\p_Q\cx G &= K_{MNPQ}\ ,\cr
}$$
where a dot denotes contraction with the appropriate metric $\eta^{(i)}
_{\al\be}$ and $K_{MNPQ} = \mx {\Om}{\p_M \p_N \p_P \p_Q \Om}$.
The first equation reflects the algebraic dependence of the 
components of the period vector which is always the case for even (complex) 
dimensions. The remaining equations can be considered as differential 
equations determining the sections $\cx G_M$ in terms of $H$.
Taking derivatives of these relations we find
$$\eqalign{
\p_R(\eta^{(1)}\cx G)_M+\p_M (\eta^{(1)}\cx G)_R+ \p_M H \cdot \p_R H &=0\ ,\cr
\p_M\p_N(\eta^{(1)}\cx G)_R+\p_RH  \cdot \p_M\p_N H &=0\ ,\cr
K_{RMNP}+\p_M\p_N\p_P(\eta^{(1)}\cx G)_R+\p_RH  \cdot  \p_M\p_N\p_PH &=0\ ,\cr
\p_S\p_M\p_N(\eta^{(1)}\cx G)_R+\p_S\p_RH  \cdot  \p_M\p_NH+\p_RH  \cdot  
\p_S\p_M\p_N H&=0\ ,\cr
}$$
and therefore
\eqn\facII{
K_{RMNP}=\p_R\p_MH  \cdot  \p_N\p_PH \ .
}
The last equation reflects the factorization of four point functions in terms 
of the fundamental three point couplings, see eqs. \facI,\yukI,\sympmet.

\appendix{B}{Non-holomorphic equations}
Conventions are as in \stromI.
$\cx D$ is the covariant derivative acting on $\cx H \oplus \cx L$:
$$
\cx D_m = \d_m +q\ \cx K_m 
$$
where $\d$ is the metric compatible covariant derivative on $\cx H$ 
and $q$ is the charge w.r.t. the $U(1)$ line bundle with the
K\"ahler class as its first Chern class.
Define $G_{m_1,m_2,\dots}=\cx D _{m_1} \cx D _{m_2} \dots \Om$ 
and as above $\mx{A}{B} = \int  A \wedge B$.
We have
$$
\eqalign{
\mx {\Om}{ \bx \Om} &= e^{-\cx K}\ ,\cr
\mx {G_m}{\bx \Om} &= 0\ ,\cr
\mx {G_{mn}}{\bx \Om} &= \mx {G_{mn}}{G_{\bx m}} = 0\ ,\cr
\mx {G_{mnp}}{\bx \Om} &= \mx {G_{mnp}}{G_{\bx m}} = 0,\ 
\mx {G_{\bx m \bx n }}{G_{mnp}} = e^{-\cx K} \cx D_m R_{n\bx m p\bx n} \cr
}$$
and
$$\eqalign{
\mx {G_{m}}{G_{\bx m}} &= -e^{-\cx K} g_{m \bx m}\equiv -e^{-\cx K}
\p_m \p_{\bx m} \cx K\ ,\cr
\mx {G_{mn}}{G_{\bx m \bx n}} &= e^{-\cx K}
(R_{m\bx m n \bx n}-g_{m\bx m}g_{n \bx n} - g_{m\bx n}g_{n \bx m})\ .
}$$

\appendix{C}{Solutions to the Picard--Fuchs equations}
The solutions to the Picard-Fuchs system for Calabi--Yau 3-folds
periods with several moduli have been considered in detail in 
\hkty, to which we refer for details. Picard-Fuchs 
equations related to the case of $d>3$ have been considered 
in \lsw.

In an expansion around the large complex structure point
$z_i=0,\ \forall i$, there is a unique power series
solution $w_0$ given by
$$\eqalign{
w_0 &= W(x;\nrho)|{\nrho=0}\ ,\quad
w(x;\nrho) = \sum c(\nv,\nrho) \prod_{i=1}^{h^{1,3}} z_i^{n_i}\ ,\cr
c(\nv,\nrho) &= {\Gamma(1-\sum_{i=1}^{h^{1,3}}
l^{(i)}_0 (n_i+r_i)) \over 
\prod_{j=1}^{(d+1)+h^{1,3}}\Gamma(1+\sum_{i=1}^{h^{1,3}}
l^{(i)}_j (n_i+r_i))}} \ ,
$$
where a subscript $j,\ j = 0 \dots h^{1,3}+d+1$ 
at the Mori generator $l^{(i)}_j$ denotes the $j$-th entry.
Single, double, triple and quartic logarithmic solutions are 
obtained from the system of derivatives of $w(x;\nrho)$
at $\nrho=0$:
$$\
\partial_{r_i} w(x;\nrho)|{\nrho=0},\ 
\partial_{r_i}\partial_{r_j} w(x;\nrho)|{\nrho=0},\ \dots
$$
such that the leading logarithmic pieces are annihilated
by the principal parts of the Picard-Fuchs operators,
as in \normI. The mirror map is defined by the relation between
the special coordinates $t_\al$ by $t_i={\omega_i\over \omega_0}$, where 
$\omega_0$ is the unique power series solution and $\omega_i$ the
single logarithmic solutions. It was observed
in the second paper in \hkty\ that the solutions 
obtained in this way for 3-folds, 
contain the information about the topological 
intersections defined as in \toppart\ for 3-folds,
that is $\chi$ and $\int J_i\wedge c_2$. More precisely 
the constant piece of the double logarithmic solutions
corresponding to $\cx F_i=\partial_{t_i} \cx F$ 
coincides with $-24 \int J_i\wedge c_2$
and the constant piece of the triple logarithmic
solution obtained in this way is $-i\zeta (3) \chi/(2\pi)^3$.
An analogous phenomenon can be observed in the 4-fold case. The
leading piece of the quartic logarithmic solution $\Pi^{(4)}$ is
obtained from $R_0$ in \iform\ 
by multiplying each coefficient with the number
of possible different permutations of its indices
and replacing $K_\al$ by ${1\over 2 \pi i}\ln(z_\al)$. 
In a basis for $\Pi^{(3)}_\al$
where the leading cubic pieces represent derivatives of $\Pi^{(4)}$
w.r.t. $t_\al \sim \ln(z_\al)$, the linear and quadratic pieces
of $\Pi^{(3)}$ are proportional to $\zeta(3) R_3$ and $\pi^2 R_2$ in
\toppart, again with the coefficients multiplied by the number of 
possible different permutations of the index structure.

\appendix{D}{Toric data for the  other 4-folds}
\subsec{The 4-fold $X_{II}$}
\subsubsec{Basic data and properties}
The dual polyhedron for the 4-fold $X_{II}$ is the convex hull of the
negative unit vertices $\ns_i, \ i=1\dots 5$ and
$$
\ns_6 = (0,0,1,4,6),\ 
\ns_7= (0,1,1,6,9),\ 
\ns_8=(0,0,0,2,3),\ 
\ns_9= (1,0,1,6,9)
$$
There are three Calabi--Yau phases, related by flops.
First there is one phase in which corresponds to the K3 phase
in the fiber, with Mori generators
\eqn\moriXIIi{\eqalign{
l^{(1)}=(0, 1, 0, 0, 0, 0, -1, 0, -1, 1)
,\ &l^{(2)}=(0, 0, 0, 1, 0, 0, 1, 0, -2, 0)
,\cr l^{(3)}= (-6, 0, 0, 0, 2, 3, 0, 0, 1, 0)
,\ &l^{(4)}= (0, 0, 1, 0, 0, 0, -1, 1, -1, 0)
\ .}}
By flops of the curves associated to the first or fourth Mori generator
one reaches two other phases, where one can shrink 4-cycles in the
$\IF_1$ fiber. These two phases are actually isomorphic and related
by the exchange of the $\IP^1$'s associated to $l^{(1)}$ and $l^{(4)}$.
The Mori generators for one of the two are 
\eqn\moriXIIii{\eqalign{
l^{(1)}=(0, 0, 1, 1, 0, 0, 0, 1, -3, 0)
,\ &l^{(2)}=(-6, 0, 1, 0, 2, 3, -1, 1, 0, 0)
,\cr l^{(3)}= (0, 0, -1, 0, 0, 0, 1, -1, 1, 0)
,\ &l^{(4)}= (0, 1, -1, 0, 0, 0, 0, -1, 0, 1)
\ .}}
The topological intersection numbers are:
\eqn\topXIIi{\eqalign{
 R _0 =& (3 K_3^2 K_2+K_3 K_2^2+K_4 K_3 K_2+8 K_3^3+2 K_4 K_3^2) K_1+K_4 
K_3 K_2^2+\cr& 6 K_3^2 K_2^2+3 K_4 K_3^2 K_2+8 K_4 K_3^3+2 K_3 K_2^3+
52 K_3^4+18 K_3^3 K_2\cr
 R _2 =&(24 K_4+92 K_3+36 K_2) K_1+92 K_3 K_4+36 K_2 K_4+72 K_2^2+\cr&206 
K_2 K_3+596 K_3^2\cr
 R _3 =& -480 K_1-480 K_4-1080 K_2-3136 K_3
}}
and
\eqn\topXIIii{\eqalign{
 R _0 =& (K_2 K_1^2+9 K_3^3+8 K_2^3+3 K_3 K_2 K_1+9 K_3^2 K_2
+3 K_3^2 K_1
+9 K_3 K_2^2\cr&+K_3 K_1^2+3 K_2^2 K_1) K_4+18 K_2^3 K_1+18 K_3 K_2^
2 K_1+6 K_3 K_2 K_1^2\cr&+18 K_3^2 K_2 K_1+2 K_2 K_1^3+6 K_3^2 K_1^2
+54 K_3^2 K_2^2+18 K_3^3 K_1+\cr&54 K_3^3 K_2+2 K_3 K_1^3+6 K_2^2 
K_1^2+52 K_2^4+54 K_3 K_2^3+54 K_3^4\cr
 R _2 =&K_1^2+52 K_2^4+54 K_3 K_2^3+54 K_3^4
(102 K_3+92 K_2+36 K_1) K_4+596 K_2^2\cr&+206 K_1 K_3+618 K_2 K_3+72
 K_1^2+206 K_1 K_2+618 K_3^2\cr
 R _3 =& -1080 K_1-3258 K_3-480 K_4-3136 K_2 \ ,
}}
respectively.

\subsubsec{Cohomology classes of genus zero curves on $X_{II}$}
For the physical interpretation of the K\"ahler moduli 
let us determine the cohomology classes of rational curves associated
to the K\"ahler moduli for the first phase. 
The Stanley Reisner ideal is given by 
$$
SR:\  \{x_1 x_9, x_2 x_7, x_3 x_6, x_4 x_5 x_8\}
$$
while the typical Batyrev-Cox polynomial reads
\eqn\BCXII{\eqalign{
&x_9^{18 }x_8^{6 }x_6^{12 }x_2^{18}+x_9^{18 }x_8^{6 }x_6^{12 }x_7^{18}+
x_8^{6 }x_6^{12 }x_2^{18 }x_1^{18}+x_8^{
6 }x_6^{12 }x_1^{18 }x_7^{18}+x_9^{6 }x_3^{12 }x_8^{6 }x_2^{6
}\cr&+x_9^{6 }x_3^{12 }x_8^{6 }x_7^{6}+x_3^{12 }x_8^{6
 }x_2^{6 }x_1^{6}+x_3^{12 }x_8^{6 }x_1^{6 }x_7^{6}+x_4^{3}+x_5^{2}
}}
respecting the  $C^\star$ actions \cstarXII.
It follows that 
$t_1$ is the area of the base $\IP^1_{D_2}$, 
$t_2$ is area of the fiber $\IP^1_{A}$, 
$t_3$ is the area of a rational curve in the elliptic fiber and
$t_4$ is area of the base $\IP^1_{D_1}$.

\subsubsec{Counting of rational curves on $X_{II}$}
From the Mori generators \moriXIIi\ one obtains the Picard-Fuchs system
$$\eqalign{
\cx L_1    &=   z_1 (-\theta_2+\theta_1+\theta_4) 
(-\theta_3+\theta_4+\theta_1+2 \theta_2)-\theta_1^2
   \cr
\cx L_2    &=   z_2 (\theta_1+2 \theta_2+1-\theta_3+\theta_4) 
(-\theta_3+\theta_4+\theta_1+2 \theta_2
)+\theta_2 (-\theta_2+\theta_1+\theta_4)   \cr
\cx L_3    &=   12 z_3 (5+6 \theta_3) (1+6 \theta_3)+\theta_3 
(-\theta_3+\theta_4+\theta_1+2 
\theta_2)   \cr
\cx L_4    &=   z_4 (-\theta_2+\theta_1+\theta_4) 
(-\theta_3+\theta_4+\theta_1+2 \theta_2)-\theta_4^2
}
$$
We chose the following basis for the operators $\co 2_\al$:
\eqn\ringXIIi{
K_3^2+K_4 K_3,\  K_1 K_4,\  K_2^2+K_2 K_4,\  K_2^2+K_1 K_2,\  
K_1 K_3+K_3^2,\  K_2 K_3+2 K_3^2}
The results for the 3-pt functions are collected in the tables F.II.i.
A different basis for the operators $\co 2 _\al$ can be chosen
such as to reproduce the Gromov--Witten invariants of the 3-fold fiber;
they are determined by the matrix $N_\al^{\ \mu}$ in \fourptlim\ which 
in the present case is given by 
$$
N_\al^{\ \mu} = \pmatrix{0&0&0&0&0&0\cr0&0&-{1\over 3}&{2 \over 3}& 0&0\cr
-{1 \over 7}&0&0&0&{6 \over 7}& -{2 \over 7}\cr
0&1&0&0&0&0}
$$

\subsec{The 4-fold $X_{III}$}
\subsubsec{Basic data and properties}

The dual polyhedron for the 4-fold $X_{III}$ is the convex hull of the
negative unit vertices $\ns_i, \ i=1\dots 5$ and
$$
\ns_6 = (0,0,1,4,6),\ \ns_7= (0,1,1,6,9),\ \ns_8=(0,0,0,2,3),\ \ns_9= (1,2,2,12,18)
$$
There are two Calabi--Yau phases corresponding to the two phases of the 3-fold
fiber discussed in detail in \kmv. The first phase admits a K3 fibration and has
the following Mori generators $l^{(\al)}$:
\eqn\moriIII{\eqalign{
l^{(1)}=(0, 0, 1, 0, 0, 0, -1, 1, -1, 0)&,
\ l^{(2)}=(0, 0, 0, 1, 0, 0, 1, 0, -2, 0),\cr
\ l^{(3)}=(-6, 0, 0, 0, 2, 3, 0, 0, 1, 0)&,
\ l^{(4)}=(0, 1, 0, 0, 0, 0, 0, -2, 0, 1)}}
The second phase has its Mori cone spanned by the generators
\eqn\moriIIIii{\eqalign{
l^{(1)}=(-6, 0, 1, 0, 2, 3, -1, 1, 0, 0)&,
\ l^{(2)}=(0, 0, 1, 1, 0, 0, 0, 1, -3, 0),\cr
\ l^{(3)}=(0, 0, -1, 0, 0, 0, 1, -1, 1, 0)&,
\ l^{(4)}=(0, 1, 0, 0, 0, 0, 0, -2, 0, 1)}}
We will restrict our discussion of the first phase in the following.
For the topological invariants \toppart\
of the first phase with  Mori generators \moriIII\ we find
\eqn\topXIII{\eqalign{
 R_0 =& (8 K_3^3+K_3 K_2 K_1+K_3 K_2^2+2 K_3^2 K_1+3 K_3^2 K_2) K_4+6 K_3^2 K_2 K_1+2 K_3 K_2 K_1^2
\cr&+2 K_3 K_2^2 K_1+16 K_3^3 K_1+2 K_3 K_2^3+4 K_3^2 K_1^2+6 K_3^2 K_2^2+18 K_3^3 K_2+52 K_3^4
\cr
 R_2 =& (92 K_3+36 K_2+24 K_1) K_4+72 K_1 K_2+72 K_2^2+184 K_1 K_3+48 K_1^2\cr
&+596 K_3^2+206 K_2 K_3\cr
R_3 =& -960 K_1-3136 K_3-1080 K_2-480 K_4 \cr
}}
Again the coefficients of $K_4$ in the above expression are precisely the 
intersection invariants of the Calabi--Yau fiber \kmv.

\subsubsec{Genus zero curves on $X_{III}$}
The Stanley Reisner ideal 
is given by 
$$
SR:\  \{x_1x_9, x_2x_7, x_3x_6, x_4x_5x_8\}
$$
and the  Batyrev-Cox polynomial reads
\eqn\BCIII{\eqalign{
&x_9^{36} x_8^6 x_7^{18} x_6^{12}+x_8^6 x_7^{18} x_6^{12} x_1^{36}+x_9^{12} x_8^6 x_7^6 x_3^{12}
+\cr&x_8^6 x_7^6 x_1^{12} x_3^{12}+x_8^6 x_6^{12} x_2^{18}+x_8^6 x_3^{12} x_2^6+x_4^3+x_5^2
}}
respecting the $C^\star$ actions \cstarXIII.
From the above data we can determine the areas associated to
the K\"ahler moduli $t_\al$:
$t_1$ is the area of the fiber $\IP^1_B$, 
$t_2$ is the area of the fiber $\IP_A^1$,
$t_3$ is the area of a curve in the elliptic fiber
and $t_4$ is area of the base of the Calabi--Yau fibration.

\subsubsec{Counting of rational curves on $X_{III}$}
From the Mori generators \moriIII\ one obtains the Picard-Fuchs system
\eqn\pfIII{\eqalign{
\cx L _1    &=   z_1 (-\theta_2+\theta_1) (-\theta_3+\theta_1+2 \theta_2)-\theta_1 (\theta_1-2 
\theta_4)\ ,   \cr
\cx L _2    &=   z_2 (\theta_1+2 \theta_2+1-\theta_3) (-\theta_3+\theta_1+2 \theta_2)+\theta_2 (-
\theta_2+\theta_1) \ ,   \cr
\cx L _3    &=   12 z_3 (5+6 \theta_3) (1+6 \theta_3)+\theta_3 (-\theta_3+\theta_1+2 \theta_2)
\ ,    \cr
\cx L _4    &=   z_4 (\theta_1-2 \theta_4) (\theta_1-2 \theta_4-1)-\theta_4^2\ .   \cr
}}
The results for the 3-pt functions are collected in the tables F.III.i in a 
basis with $\co 2$ operators corresponding to
\eqn\ringIII{K_1 K_2+K_2^2,\  -2 K_1 K_3+K_2 K_3,\  K_1 K_3+K_3^2,\  2 K_1^2+K_1 
K_4,\  K_2 K_4,\  K_3 K_4 \ .}
The matrix $N_\al ^{\ \mu}$ in \fourptlim\ is given by
$$
N_\al^{\ \mu} = \pmatrix{0&0&0&{1\over 5}&0&0\cr
0&0&0&0&1&0\cr
0&0&0&0&0&1\cr
0&0&0&0&0&0}
$$
\subsec{The 4-folds $X_{IV},\ X_V$}
The dual polyhedron for the 4-fold $X_{IV}$ is the convex hull of the
negative unit vertices $\ns_i, \ i=1\dots 5$ and 
$$\eqalign{
X_{IV}&:(0,0,1,4,6),\ (0,1, 0,4,6),\ (1, 0, 1, 6, 9), (0, 0, 0, 2, 3)\cr
X_{V}&:(0, 0, 0, 2, 3), (0, 0, 1, 4, 6), (0, 1, 2, 8, 12),(1, 1, 2, 10, 15)
}$$
In both cases there is a K3 fibered Calabi--Yau phase; by the same methods
as before the cohomology classes of curves dual to the Mori generators
can be seen to be related  in this phase to the volumes of the $\IP^1$
factors and a curve in the elliptic fiber.

\appendix{E}{Heterotic U-duality symmetries}
The discussion in sect. 8.1. applies also to the $N=2$ 
compactifications of type II on threefolds and their heterotic
duals and implies some general properties of the non-perturbative
duality group of the heterotic theory. 
The duality between type II compactified on a Calabi--Yau
3-fold and heterotic string on K3 $\times \bf{T}^2$ maps
the monodromy group of the Calabi--Yau compactification
to the non-perturbative U-duality group of vector moduli
space of the heterotic theory. Perturbatively the heterotic
T-duality group is generated by monodromies in the vector moduli
space around singular points, where there are classically 
enhanced gauge symmetries or massless hypermultiplets, together
with the monodromies around infinity reflecting the axionic
shift symmetries of the imaginary parts of the scalar fields.

From the analysis in sect 8.1 we can infer 
some general properties of these non-perturbative U-duality groups.
Firstly, for generators of the U-duality group which have a perturbative 
origin we can go backwards and determine the type of the quantum monodromy -
uncorrected or deformed due to a non-zero volume of the classical vanishing
cycle - from continuously connecting to the field theory limit switching
off string and gravity effects. 
Therefore elements of the perturbative T-duality symmetries which are generated
perturbatively by Weyl reflections of enhanced gauge symmetries 
with non asymptotic free spectrum descend to exact non-perturbative
U-dualities; on the other hand perturbative T-duality symmetries 
associated to enhanced gauge symmetries with asymptotic free spectrum 
are broken in the non-perturbative theory as in the case  of the 
mirror symmetry $T\leftrightarrow U$ of the torus in standard compactifications 
observed in
the examples in \klm\foot{This breaking
is not merely the known quantum shift of the one-loop correction 
\antI\ but involves a true topology change of the discriminant
locus as is clear from the relation to the field theory}.

On the other hand 
we can also make precise the circumstances under which the 
heterotic theory has non-perturbative duality symmetries 
exchanging the dilaton with geometric moduli\foot{These symmetries
have been explained geometrically in the F-theory context in \mv\
as the exchange of base $\IP^1$'s and the volumes associated to it
However they appear similarly in 3-folds which do not admit an elliptic
fibration; in these cases a similar geometrical origin and the relation to 
the six dimensional heterotic/heterotic duality of \dmw\ is not clear.
See also \std\ for related works.} which appears in the same examples.
The dual type IIA Calabi--Yau compactification is given by a K3 
fibration with the dilaton being the size of the base $\IP^1$ 
of the fibration. If this base $\IP^1$ arises as part of the resolution 
of a $A_N$ curve singularity, which in fact is a canonical 
construction representing a large class of K3 fibrations, it is of the 
uncorrected type ii and can be shrunk to zero size with the restoration
of a $\IZ_2$ symmetry representing the exchange element. More precisely
the full group of Weyl reflections of $A_N$ 
generate a whole subgroup of exact non-perturbative duality symmetries of
the heterotic theory exchanging the dilaton with $N-1$ geometric moduli.
\vfill
\appendix{F}{Invariants for the examples}
\subsec{Invariants for $X_I$}
\def\ss#1{{\scriptscriptstyle #1}}

{\vbox{\ninepoint{
$$
\vbox{\offinterlineskip\tabskip=0pt\halign{\strut\vrule#
&\hfil~$\ss{#}$~\hfil &\hfil~$\ss{#}$~\hfil &\vrule#
&\hfil ~$\ss{#}$~ \hfil&\hfil ~$\ss{#}$~ \hfil&\hfil ~$\ss{#}$~ \hfil&\vrule#
&\hfil ~$\ss{#}$~ \hfil&\hfil ~$\ss{#}$~ \hfil&\hfil ~$\ss{#}$~ \hfil&\vrule#
&\hfil ~$\ss{#}$~ \hfil&\hfil ~$\ss{#}$~ \hfil&\hfil ~$\ss{#}$~ \hfil&\hfil ~$\ss{#}$~ \hfil
&\vrule#\cr
\noalign{\hrule}
&   &n_1    &&  0& & &&     1 & & &&  2 & & &&\cr
&   &n_2 &&     0&1&2&& 0  &1 & 2&& 0 & 1& 2&&\cr
\noalign{\hrule}
&0&0&&      &480&960&&    -1&0&0&&    0&0&0&&      \cr
&0&1&&     0&480&565776&&    0&480&676656&&    0&0&0&&      \cr
&0&2&&     0&0&960&&    0&1440&-452160&&    0&0&960&&      \cr
&1&0&&     0&0&0&&    -1&0&0&&    -4&0&0&&      \cr
&1&1&&     0&0&0&&    0&480&676656&&    0&1440&2362608&&      \cr
&1&2&&     0&0&0&&    0&1440&-452160&&    0&0&8640&&      \cr
&2&0&&     0&0&0&&    0&0&0&&    0&0&0&&      \cr
&2&1&&     0&0&0&&    0&0&0&&    0&0&0&&      \cr
&2&2&&     0&0&0&&    0&0&0&&    0&0& &&      \cr
\noalign{\hrule}&n_3&n_4\cr}
}$${\bf Table I.2}: Numbers of invariants ${1 \over 2} N_1(\nv)$ for $X_I$  }
\vskip7pt}}

{\vbox{\ninepoint{
$$
\vbox{\offinterlineskip\tabskip=0pt\halign{\strut\vrule#
&\hfil~$\ss{#}$~\hfil &\hfil~$\ss{#}$~\hfil &\vrule#
&\hfil ~$\ss{#}$~ \hfil&\hfil ~$\ss{#}$~ \hfil&\hfil ~$\ss{#}$~ \hfil&\vrule#
&\hfil ~$\ss{#}$~ \hfil&\hfil ~$\ss{#}$~ \hfil&\hfil ~$\ss{#}$~ \hfil&\vrule#
&\hfil ~$\ss{#}$~ \hfil&\hfil ~$\ss{#}$~ \hfil&\hfil ~$\ss{#}$~ \hfil&\hfil ~$\ss{#}$~ \hfil
&\vrule#\cr\noalign{\hrule}
&   &n_1    &&  0& & &&     1 & & &&  2 & & &&\cr
&   &n_2 &&     0&1&2&& 0  &1 & 2&& 0 & 1& 2&&\cr
\noalign{\hrule}
&0&0&&      &0&0&&    0&0&0&&    0&0&0&&      \cr
&0&1&&     0&0&0&&    -2&480&282888&&    0&0&0&&      \cr
&0&2&&     0&0&0&&    -4&1440&-226080&&    0&0&960&&      \cr
&1&0&&     0&0&0&&    0&0&0&&    0&0&0&&      \cr
&1&1&&     0&0&0&&    -2&480&282888&&    -4&960&565776&&      \cr
&1&2&&     0&0&0&&    -4&1440&-226080&&    4&-1920&895680&&      \cr
&2&0&&     0&0&0&&    0&0&0&&    0&0&0&&      \cr
&2&1&&     0&0&0&&    0&0&0&&    0&0&0&&      \cr
&2&2&&     0&0&0&&    0&0&0&&    0&0& &&      \cr
\noalign{\hrule}&n_3&n_4\cr}
}$${\bf Table I.2}: Numbers of invariants ${1 \over 5} N_2(\nv)$ for $X_I$  }
\vskip7pt}}
{\vbox{\ninepoint{
$$
\vbox{\offinterlineskip\tabskip=0pt\halign{\strut\vrule#
&\hfil~$\ss{#}$~\hfil &\hfil~$\ss{#}$~\hfil &\vrule#
&\hfil ~$\ss{#}$~ \hfil&\hfil ~$\ss{#}$~ \hfil&\hfil ~$\ss{#}$~ \hfil&\vrule#
&\hfil ~$\ss{#}$~ \hfil&\hfil ~$\ss{#}$~ \hfil&\hfil ~$\ss{#}$~ \hfil&\vrule#
&\hfil ~$\ss{#}$~ \hfil&\hfil ~$\ss{#}$~ \hfil&\hfil ~$\ss{#}$~ \hfil&\hfil ~$\ss{#}$~ \hfil
&\vrule#\cr\noalign{\hrule}
&   &n_1    &&  0& & &&     1 & & &&  2 & & &&\cr
&   &n_2 &&     0&1&2&& 0  &1 & 2&& 0 & 1& 2&&\cr
\noalign{\hrule}
&0&0&&      &480&960&&    0&0&0&&    0&0&0&&      \cr
&0&1&&     0&480&565776&&    0&480&565776&&    0&0&0&&      \cr
&0&2&&     0&0&960&&    0&1440&-452160&&    0&0&960&&      \cr
&1&0&&     0&0&0&&    -2&0&0&&    -4&0&0&&      \cr
&1&1&&     0&0&0&&    0&480&787536&&    0&1440&2362608&&      \cr
&1&2&&     0&0&0&&    0&1440&-452160&&    0&0&8640&&      \cr
&2&0&&     0&0&0&&    0&0&0&&    0&0&0&&      \cr
&2&1&&     0&0&0&&    0&0&0&&    0&0&0&&      \cr
&2&2&&     0&0&0&&    0&0&0&&    0&0& &&      \cr
\noalign{\hrule}&n_3&n_4\cr}
}$${\bf Table I.3}:Numbers of invariants $N_3(\nv)$ for $X_I$}
\vskip7pt}}

{\vbox{\ninepoint{
$$
\vbox{\offinterlineskip\tabskip=0pt\halign{\strut\vrule#
&\hfil~$\ss{#}$~\hfil &\hfil~$\ss{#}$~\hfil &\vrule#
&\hfil ~$\ss{#}$~ \hfil&\hfil ~$\ss{#}$~ \hfil&\hfil ~$\ss{#}$~ \hfil&\vrule#
&\hfil ~$\ss{#}$~ \hfil&\hfil ~$\ss{#}$~ \hfil&\hfil ~$\ss{#}$~ \hfil&\vrule#
&\hfil ~$\ss{#}$~ \hfil&\hfil ~$\ss{#}$~ \hfil&\hfil ~$\ss{#}$~ \hfil&\hfil ~$\ss{#}$~ \hfil
&\vrule#\cr\noalign{\hrule}
&   &n_1    &&  0& & &&     1 & & &&  2 & & &&\cr
&   &n_2 &&     0&1&2&& 0  &1 & 2&& 0 & 1& 2&&\cr
\noalign{\hrule}
&0&0&&        &0&0&&    0&0&0&&    0&0&0&&      \cr
&0&1&&     -2&480&282888&&    -2&480&282888&&    0&0&0&&      \cr
&0&2&&     0&0&960&&    -11&3840&-676080&&    0&0&960&&      \cr
&1&0&&     0&0&0&&    0&0&0&&    0&0&0&&      \cr
&1&1&&     0&0&0&&    -2&480&282888&&    -6&1440&848664&&      \cr
&1&2&&     0&0&0&&    -11&3840&-676080&&    -4&0&895680&&      \cr
&2&0&&     0&0&0&&    0&0&0&&    0&0&0&&      \cr
&2&1&&     0&0&0&&    0&0&0&&    0&0&0&&      \cr
&2&2&&     0&0&0&&    0&0&0&&    0&0& &&      \cr
\noalign{\hrule}&n_3&n_4\cr}
}$${\bf Table I.4}: Numbers of invariants ${1\over10}N_4(\nv)$ for $X_I$}
\vskip7pt}}

{\vbox{\ninepoint{
$$
\vbox{\offinterlineskip\tabskip=0pt\halign{\strut\vrule#
&\hfil~$\ss{#}$~\hfil &\hfil~$\ss{#}$~\hfil &\vrule#
&\hfil ~$\ss{#}$~ \hfil&\hfil ~$\ss{#}$~ \hfil&\hfil ~$\ss{#}$~ \hfil&\vrule#
&\hfil ~$\ss{#}$~ \hfil&\hfil ~$\ss{#}$~ \hfil&\hfil ~$\ss{#}$~ \hfil&\vrule#
&\hfil ~$\ss{#}$~ \hfil&\hfil ~$\ss{#}$~ \hfil&\hfil ~$\ss{#}$~ \hfil&\hfil ~$\ss{#}$~ \hfil
&\vrule#\cr\noalign{\hrule}
&   &n_1    &&  0& & &&     1 & & &&  2 & & &&\cr
&   &n_2 &&     0&1&2&& 0  &1 & 2&& 0 & 1& 2&&\cr
\noalign{\hrule}
&0&0&&      &480&960&&    0&0&0&&    0&0&0&&      \cr
&0&1&&     0&480&621216&&    0&480&621216&&    0&0&0&&      \cr
&0&2&&     0&0&960&&    0&1440&-452160&&    0&0&960&&      \cr
&1&0&&     0&0&0&&    0&0&0&&    0&0&0&&      \cr
&1&1&&     0&0&0&&    0&480&621216&&    0&1440&1863648&&      \cr
&1&2&&     0&0&0&&    0&1440&-452160&&    0&0&8640&&      \cr
&2&0&&     0&0&0&&    0&0&0&&    0&0&0&&      \cr
&2&1&&     0&0&0&&    0&0&0&&    0&0&0&&      \cr
&2&2&&     0&0&0&&    0&0&0&&    0&0& &&      \cr
\noalign{\hrule}&n_3&n_4\cr}
}$${\bf Table I.5}: Numbers of invariants ${1\over20}N_5(\nv)$ for $X_I$}
\vskip7pt}}

{\vbox{\ninepoint{
$$
\vbox{\offinterlineskip\tabskip=0pt\halign{\strut\vrule#
&\hfil~$\ss{#}$~\hfil &\hfil~$\ss{#}$~\hfil &\vrule#
&\hfil ~$\ss{#}$~ \hfil&\hfil ~$\ss{#}$~ \hfil&\hfil ~$\ss{#}$~ \hfil&\vrule#
&\hfil ~$\ss{#}$~ \hfil&\hfil ~$\ss{#}$~ \hfil&\hfil ~$\ss{#}$~ \hfil&\vrule#
&\hfil ~$\ss{#}$~ \hfil&\hfil ~$\ss{#}$~ \hfil&\hfil ~$\ss{#}$~ \hfil&\hfil ~$\ss{#}$~ \hfil
&\vrule#\cr\noalign{\hrule}
&   &n_1    &&  0& & &&     1 & & &&  2 & & &&\cr
&   &n_2 &&     0&1&2&& 0  &1 & 2&& 0 & 1& 2&&\cr
\noalign{\hrule}
&0&0&&      &0&0&&    0&0&0&&    0&0&0&&      \cr
&0&1&&     -2&480&282888&&    -2&480&282888&&    0&0&0&&      \cr
&0&2&&     0&0&960&&    -8&2880&-452160&&    0&0&960&&      \cr
&1&0&&     0&0&0&&    0&0&0&&    0&0&0&&      \cr
&1&1&&     0&0&0&&    -2&480&282888&&    -6&1440&848664&&      \cr
&1&2&&     0&0&0&&    -14&4800&-900000&&    -4&0&895680&&      \cr
&2&0&&     0&0&0&&    0&0&0&&    0&0&0&&      \cr
&2&1&&     0&0&0&&    0&0&0&&    0&0&0&&      \cr
&2&2&&     0&0&0&&    0&0&0&&    0&0& &&      \cr
\noalign{\hrule}&n_3&n_4\cr}
}$${\bf Table I.6}: Numbers of invariants $N_6(\nv)$ for $X_I$}
\vskip7pt}}

\subsec{Invariants for $X_{II}$}

{\vbox{\ninepoint{
$$
\vbox{\offinterlineskip\tabskip=0pt\halign{\strut\vrule#
&\hfil~$\ss{#}$~\hfil &\hfil~$\ss{#}$~\hfil &\vrule#
&\hfil ~$\ss{#}$~ \hfil&\hfil ~$\ss{#}$~ \hfil&\hfil ~$\ss{#}$~ \hfil&\vrule#
&\hfil ~$\ss{#}$~ \hfil&\hfil ~$\ss{#}$~ \hfil&\hfil ~$\ss{#}$~ \hfil&\vrule#
&\hfil ~$\ss{#}$~ \hfil&\hfil ~$\ss{#}$~ \hfil&\hfil ~$\ss{#}$~ \hfil&\hfil ~$\ss{#}$~ \hfil
&\vrule#\cr\noalign{\hrule}
&   &n_1    &&  0& & &&     1 & & &&  2 & & &&\cr
&   &n_2 &&     0&1&2&& 0  &1 & 2&& 0 & 1& 2&&\cr
\noalign{\hrule}
&0&0&&      &0&0&&    0&0&0&&    0&0&0&&      \cr
& &1&&     0&0&0&&    0&0&0&&    0&0&0&&      \cr
& &2&&     0&0&0&&    0&0&0&&    0&0&0&&      \cr
&1&0&&     3600&3600&0&&    876&-7200&-14400&&    0&0&18000&&      \cr
& &1&&     996&-7200&-14400&&    0&0&144000&&    0&0&0&&      \cr
& &2&&     0&0&18000&&    0&0&0&&    0&0&0&&      \cr
&2&0&&     7200&5535504&7200&&    34380&1729980&5086800&&    -69132&-68760&-6934320&& \
     \cr
& &1&&     37980&1733580&5086800&&    -384372&-144720&-60056640&&    0&0&1565100&&    
  \cr
& &2&&     -82752&-75960&-6948720&&    0&0&1691100&&    0&0& &&      \cr
\noalign{\hrule}&n_3&n_4\cr}
}$${\bf Table II.1}: Numbers of invariants $N_1(\nv)$ for $X_{II}$ }
\vskip7pt}}
{\vbox{\ninepoint{
$$
\vbox{\offinterlineskip\tabskip=0pt\halign{\strut\vrule#
&\hfil~$\ss{#}$~\hfil &\hfil~$\ss{#}$~\hfil &\vrule#
&\hfil ~$\ss{#}$~ \hfil&\hfil ~$\ss{#}$~ \hfil&\hfil ~$\ss{#}$~ \hfil&\vrule#
&\hfil ~$\ss{#}$~ \hfil&\hfil ~$\ss{#}$~ \hfil&\hfil ~$\ss{#}$~ \hfil&\vrule#
&\hfil ~$\ss{#}$~ \hfil&\hfil ~$\ss{#}$~ \hfil&\hfil ~$\ss{#}$~ \hfil&\hfil ~$\ss{#}$~ \hfil
&\vrule#\cr\noalign{\hrule}
&   &n_1    &&  0& & &&     1 & & &&  2 & & &&\cr
&   &n_2 &&     0&1&2&& 0  &1 & 2&& 0 & 1& 2&&\cr
\noalign{\hrule}
&0&0&&      &0&0&&    1&3&5&&    0&0&-12&&      \cr
& &1&&     1&3&5&&    1&-4&-101&&    1&-4&28&&      \cr
& &2&&     0&0&-12&&    1&-4&28&&    2&-12&72&&      \cr
&1&0&&     0&0&0&&    252&-960&-1920&&    0&0&4800&&      \cr
& &1&&     252&-960&-1920&&    -120&1320&46200&&    -120&1320&
-14040&&      \cr
& &2&&     0&0&4800&&    -120&1320&-14040&&    -360&4440&-37440&&    
  \cr
&2&0&&     0&0&0&&    5130&118170&339120&&    -18504&-20520&-947280&\
&      \cr
& &1&&     5130&118170&339120&&    -107604&-404640&-11674080&&    20520
&-376920&4579470&&      \cr
& &2&&     -18504&-20520&-947280&&    20520&-376920&4579470&&    55200
&-1261440& &&      \cr
\noalign{\hrule}&n_3&n_4\cr}
}$${\bf Table II.2}: Numbers of invariants $N_2(\nv)$ for $X_{II}$  }
\vskip7pt}}
{\vbox{\ninepoint{
$$
\vbox{\offinterlineskip\tabskip=0pt\halign{\strut\vrule#
&\hfil~$\ss{#}$~\hfil &\hfil~$\ss{#}$~\hfil &\vrule#
&\hfil ~$\ss{#}$~ \hfil&\hfil ~$\ss{#}$~ \hfil&\hfil ~$\ss{#}$~ \hfil&\vrule#
&\hfil ~$\ss{#}$~ \hfil&\hfil ~$\ss{#}$~ \hfil&\hfil ~$\ss{#}$~ \hfil&\vrule#
&\hfil ~$\ss{#}$~ \hfil&\hfil ~$\ss{#}$~ \hfil&\hfil ~$\ss{#}$~ \hfil&\hfil ~$\ss{#}$~ \hfil
&\vrule#\cr\noalign{\hrule}
&   &n_1    &&  0& & &&     1 & & &&  2 & & &&\cr
&   &n_2 &&     0&1&2&& 0  &1 & 2&& 0 & 1& 2&&\cr
\noalign{\hrule}
&0&0&&      &-6&0&&    0&13&42&&    0&0&-58&&      \cr
& &1&&     0&17&54&&    0&0&-387&&    0&0&0&&      \cr
& &2&&     0&0&-80&&    0&0&0&&    0&0&0&&      \cr
&1&0&&     0&1440&0&&    0&-3960&-15480&&    0&0&22320&&      \cr
& &1&&     0&-5040&-19440&&    0&0&178920&&    0&0&0&&      \cr
& &2&&     0&0&30240&&    0&0&0&&    0&0&0&&      \cr
&2&0&&     0&848664&2880&&    0&636390&2765160&&    0&-30780&-4343580
&&      \cr
& &1&&     0&918270&3495600&&    0&-144720&-46296360&&    0&0&1290060
&&      \cr
& &2&&     0&-30780&-5845320&&    0&0&1517940&&    0&0& &&      \cr
\noalign{\hrule}
\noalign{\hrule}&n_3&n_4\cr}
}$${\bf Table II.3}: Numbers of invariants $N_3(\nv)$ for $X_{II}$ }
\vskip7pt}}

{\vbox{\ninepoint{
$$
\vbox{\offinterlineskip\tabskip=0pt\halign{\strut\vrule#
&\hfil~$\ss{#}$~\hfil &\hfil~$\ss{#}$~\hfil &\vrule#
&\hfil ~$\ss{#}$~ \hfil&\hfil ~$\ss{#}$~ \hfil&\hfil ~$\ss{#}$~ \hfil&\vrule#
&\hfil ~$\ss{#}$~ \hfil&\hfil ~$\ss{#}$~ \hfil&\hfil ~$\ss{#}$~ \hfil&\vrule#
&\hfil ~$\ss{#}$~ \hfil&\hfil ~$\ss{#}$~ \hfil&\hfil ~$\ss{#}$~ \hfil&\hfil ~$\ss{#}$~ \hfil
&\vrule#\cr\noalign{\hrule}
&   &n_1    &&  0& & &&     1 & & &&  2 & & &&\cr
&   &n_2 &&     0&1&2&& 0  &1 & 2&& 0 & 1& 2&&\cr
\noalign{\hrule}
&0&0&&      &-6&0&&    0&17&54&&    0&0&-80&&      \cr
& &1&&     0&13&42&&    0&0&-387&&    0&0&0&&      \cr
& &2&&     0&0&-58&&    0&0&0&&    0&0&0&&      \cr
&1&0&&     0&1440&0&&    0&-5040&-19440&&    0&0&30240&&      \cr
& &1&&     0&-3960&-15480&&    0&0&178920&&    0&0&0&&      \cr
& &2&&     0&0&22320&&    0&0&0&&    0&0&0&&      \cr
&2&0&&     0&848664&2880&&    0&918270&3495600&&    0&-30780&-5845320
&&      \cr
& &1&&     0&636390&2765160&&    0&-144720&-46296360&&    0&0&1517940
&&      \cr
& &2&&     0&-30780&-4343580&&    0&0&1290060&&    0&0& &&      \cr
\noalign{\hrule}&n_3&n_4\cr}
}$${\bf Table II.4}: Numbers of invariants $N_4(\nv)$ for $X_{II}$}
\vskip7pt}}

{\vbox{\ninepoint{
$$
\vbox{\offinterlineskip\tabskip=0pt\halign{\strut\vrule#
&\hfil~$\ss{#}$~\hfil &\hfil~$\ss{#}$~\hfil &\vrule#
&\hfil ~$\ss{#}$~ \hfil&\hfil ~$\ss{#}$~ \hfil&\hfil ~$\ss{#}$~ \hfil&\vrule#
&\hfil ~$\ss{#}$~ \hfil&\hfil ~$\ss{#}$~ \hfil&\hfil ~$\ss{#}$~ \hfil&\vrule#
&\hfil ~$\ss{#}$~ \hfil&\hfil ~$\ss{#}$~ \hfil&\hfil ~$\ss{#}$~ \hfil&\hfil ~$\ss{#}$~ \hfil
&\vrule#\cr\noalign{\hrule}
&   &n_1    &&  0& & &&     1 & & &&  2 & & &&\cr
&   &n_2 &&     0&1&2&& 0  &1 & 2&& 0 & 1& 2&&\cr
\noalign{\hrule}
&0&0&&      &0&0&&    0&0&0&&    0&0&0&&      \cr
& &1&&     0&0&0&&    0&0&0&&    0&0&0&&      \cr
& &2&&     0&0&0&&    0&0&0&&    0&0&0&&      \cr
&1&0&&     3600&3600&0&&    996&-7200&-14400&&    0&0&18000&&       \cr
& &1&&     876&-7200&-14400&&    0&0&144000&&    0&0&0&&      \cr
& &2&&     0&0&18000&&    0&0&0&&    0&0&0&&      \cr
&2&0&&     7200&5535504&7200&&    37980&1733580&5086800&&    -82752&\
-75960&-6948720&&      \cr
& &1&&     34380&1729980&5086800&&    -384372&-144720&-60056640&&
0&0&1691100&&      \cr
& &2&&     -69132&-68760&-6934320&&    0&0&1565100&&    0&0& &&    
  \cr
\noalign{\hrule}&n_3&n_4\cr}
}$${\bf Table II.5}: Numbers of invariants $N_5(\nv)$ for $X_{II}$}
\vskip7pt}}

{\vbox{\ninepoint{
$$
\vbox{\offinterlineskip\tabskip=0pt\halign{\strut\vrule#
&\hfil~$\ss{#}$~\hfil &\hfil~$\ss{#}$~\hfil &\vrule#
&\hfil ~$\ss{#}$~ \hfil&\hfil ~$\ss{#}$~ \hfil&\hfil ~$\ss{#}$~ \hfil&\vrule#
&\hfil ~$\ss{#}$~ \hfil&\hfil ~$\ss{#}$~ \hfil&\hfil ~$\ss{#}$~ \hfil&\vrule#
&\hfil ~$\ss{#}$~ \hfil&\hfil ~$\ss{#}$~ \hfil&\hfil ~$\ss{#}$~ \hfil&\hfil ~$\ss{#}$~ \hfil
&\vrule#\cr\noalign{\hrule}
&   &n_1    &&  0& & &&     1 & & &&  2 & & &&\cr
&   &n_2 &&     0&1&2&& 0  &1 & 2&& 0 & 1& 2&&\cr
\noalign{\hrule}
&0&0&&      &0&0&&    0&0&0&&    0&0&0&&      \cr
& &1&&     0&0&0&&    0&0&0&&    0&0&0&&      \cr
& &2&&     0&0&0&&    0&0&0&&    0&0&0&&      \cr
&1&0&&     7320&7320&0&&    1248&-14640&-29280&&    0&0&36600&&    
  \cr
& &1&&     1248&-14640&-29280&&    0&0&292800&&    0&0&0&&      \cr
& &2&&     0&0&36600&&    0&0&0&&    0&0&0&&      \cr
&2&0&&     14640&11858544&14640&&    48240&3495960&10343160&&    
-101256&-96480&-14013120&&      \cr
& &1&&     48240&3495960&10343160&&    -512496&-192960&-121912560&& 0
&0&2170800&&      \cr
& &2&&     -101256&-96480&-14013120&&    0&0&2170800&&    0&0& &&    
  \cr
\noalign{\hrule}&n_3&n_4\cr}
}$${\bf Table II.6}: Numbers of invariants $N_6(\nv)$ for $X_{II}$ }
\vskip7pt}}

\subsec{Invariants for $X_{III}$}

{\vbox{\ninepoint{
$$
\vbox{\offinterlineskip\tabskip=0pt\halign{\strut\vrule#
&\hfil~$\ss{#}$~\hfil &\hfil~$\ss{#}$~\hfil &\vrule#
&\hfil ~$\ss{#}$~ \hfil&\hfil ~$\ss{#}$~ \hfil&\hfil ~$\ss{#}$~ \hfil&\vrule#
&\hfil ~$\ss{#}$~ \hfil&\hfil ~$\ss{#}$~ \hfil&\hfil ~$\ss{#}$~ \hfil&\vrule#
&\hfil ~$\ss{#}$~ \hfil&\hfil ~$\ss{#}$~ \hfil&\hfil ~$\ss{#}$~ \hfil&\hfil ~$\ss{#}$~ \hfil
&\vrule#\cr\noalign{\hrule}
&   &n_1    &&  0& & &&     1 & & &&  2 & & &&\cr
&   &n_2 &&     0&1&2&& 0  &1 & 2&& 0 & 1& 2&&\cr
\noalign{\hrule}
&0&0&&      &-8&0&&    0&20&64&&    0&0&-92&&      \cr
&0&1&&     0&0&0&&    0&20&64&&    0&0&-516&&      \cr
&0&2&&     0&0&0&&    0&0&0&&    0&0&-92&&      \cr
&1&0&&     0&1920&0&&    0&-6000&-23280&&    0&0&35040&&      \cr
&1&1&&     0&0&0&&    0&-6000&-23280&&    0&0&238560&&      \cr
&1&2&&     0&0&0&&    0&0&0&&    0&0&35040&&      \cr
&2&0&&     0&1131552&3840&&    0&1036440&4173840&&    0&-41040&-6792600&&    
  \cr
&2&1&&     0&0&0&&    0&1036440&4173840&&    0&-192960&-61728480&&      \cr
&2&2&&     0&0&0&&    0&0&0&&    0&-41040& &&      \cr
\noalign{\hrule}&n_3&n_4\cr}
}$${\bf Table III.1}: Numbers of invariants $N_1(\nv)$ for $X_{III}$ }
\vskip7pt}}
{\vbox{\ninepoint{
$$
\vbox{\offinterlineskip\tabskip=0pt\halign{\strut\vrule#
&\hfil~$\ss{#}$~\hfil &\hfil~$\ss{#}$~\hfil &\vrule#
&\hfil ~$\ss{#}$~ \hfil&\hfil ~$\ss{#}$~ \hfil&\hfil ~$\ss{#}$~ \hfil&\vrule#
&\hfil ~$\ss{#}$~ \hfil&\hfil ~$\ss{#}$~ \hfil&\hfil ~$\ss{#}$~ \hfil&\vrule#
&\hfil ~$\ss{#}$~ \hfil&\hfil ~$\ss{#}$~ \hfil&\hfil ~$\ss{#}$~ \hfil&\hfil ~$\ss{#}$~ \hfil
&\vrule#\cr\noalign{\hrule}
&   &n_1    &&  0& & &&     1 & & &&  2 & & &&\cr
&   &n_2 &&     0&1&2&& 0  &1 & 2&& 0 & 1& 2&&\cr
\noalign{\hrule}
&0&0&&      &0&0&&    0&0&0&&    0&0&0&&      \cr
&0&1&&     0&0&0&&    0&0&0&&    0&0&0&&      \cr
&0&2&&     0&0&0&&    0&0&0&&    0&0&0&&      \cr
&1&0&&     -840&-840&0&&    -1248&1680&3360&&    0&0&-4200&&      \cr
&1&1&&     0&0&0&&    -1248&1680&3360&&    0&0&-33600&&      \cr
&1&2&&     0&0&0&&    0&0&0&&    0&0&-4200&&      \cr
&2&0&&     -1680&-344016&-1680&&    -48240&-443880&-1186920&&    101256&96480&\
1778880&&      \cr
&2&1&&     0&0&0&&    -48240&-443880&-1186920&&    512496&192960&14331600&&    
  \cr
&2&2&&     0&0&0&&    0&0&0&&    101256&96480& &&      \cr
\noalign{\hrule}&n_3&n_4\cr}
}$${\bf Table III.2}: Numbers of invariants $N_2(\nv)/5$ for $X_{III}$ }
\vskip7pt}}
{\vbox{\ninepoint{
$$
\vbox{\offinterlineskip\tabskip=0pt\halign{\strut\vrule#
&\hfil~$\ss{#}$~\hfil &\hfil~$\ss{#}$~\hfil &\vrule#
&\hfil ~$\ss{#}$~ \hfil&\hfil ~$\ss{#}$~ \hfil&\hfil ~$\ss{#}$~ \hfil&\vrule#
&\hfil ~$\ss{#}$~ \hfil&\hfil ~$\ss{#}$~ \hfil&\hfil ~$\ss{#}$~ \hfil&\vrule#
&\hfil ~$\ss{#}$~ \hfil&\hfil ~$\ss{#}$~ \hfil&\hfil ~$\ss{#}$~ \hfil&\hfil ~$\ss{#}$~ \hfil
&\vrule#\cr\noalign{\hrule}
&   &n_1    &&  0& & &&     1 & & &&  2 & & &&\cr
&   &n_2 &&     0&1&2&& 0  &1 & 2&& 0 & 1& 2&&\cr
\noalign{\hrule}
&0&0&&      &0&0&&    0&0&0&&    0&0&0&&      \cr
&0&1&&     0&0&0&&    0&0&0&&    0&0&0&&      \cr
&0&2&&     0&0&0&&    0&0&0&&    0&0&0&&      \cr
&1&0&&     4080&4080&0&&    1248&-8160&-16320&&    0&0&20400&&      \cr
&1&1&&     0&0&0&&    1248&-8160&-16320&&    0&0&163200&&      \cr
&1&2&&     0&0&0&&    0&0&0&&    0&0&20400&&      \cr
&2&0&&     8160&6101280&8160&&    48240&1969920&5765040&&    -101256&-96480&-7\
896000&&      \cr
&2&1&&     0&0&0&&    48240&1969920&5765040&&    -512496&-192960&-68122080&&  \
    \cr
&2&2&&     0&0&0&&    0&0&0&&    -101256&-96480& &&      \cr
\noalign{\hrule}&n_3&n_4\cr}
}$${\bf Table III.3}: Numbers of invariants $N_3(\nv)$ for $X_{III}$ }
\vskip7pt}}

{\vbox{\ninepoint{
$$
\vbox{\offinterlineskip\tabskip=0pt\halign{\strut\vrule#
&\hfil~$\ss{#}$~\hfil &\hfil~$\ss{#}$~\hfil &\vrule#
&\hfil ~$\ss{#}$~ \hfil&\hfil ~$\ss{#}$~ \hfil&\hfil ~$\ss{#}$~ \hfil&\vrule#
&\hfil ~$\ss{#}$~ \hfil&\hfil ~$\ss{#}$~ \hfil&\hfil ~$\ss{#}$~ \hfil&\vrule#
&\hfil ~$\ss{#}$~ \hfil&\hfil ~$\ss{#}$~ \hfil&\hfil ~$\ss{#}$~ \hfil&\hfil ~$\ss{#}$~ \hfil
&\vrule#\cr\noalign{\hrule}
&   &n_1    &&  0& & &&     1 & & &&  2 & & &&\cr
&   &n_2 &&     0&1&2&& 0  &1 & 2&& 0 & 1& 2&&\cr
\noalign{\hrule}
&0&0&&      &0&0&&    1&3&5&&    0&0&-12&&      \cr
&0&1&&     0&0&0&&    1&3&5&&    1&-4&-101&&      \cr
&0&2&&     0&0&0&&    0&0&0&&    0&0&-12&&      \cr
&1&0&&     0&0&0&&    252&-960&-1920&&    0&0&4800&&      \cr
&1&1&&     0&0&0&&    252&-960&-1920&&    -120&1320&46200&&      \cr
&1&2&&     0&0&0&&    0&0&0&&    0&0&4800&&      \cr
&2&0&&     0&0&0&&    5130&118170&339120&&    -18504&-20520&-947280&&      \cr
&2&1&&     0&0&0&&    5130&118170&339120&&    -107604&-404640&-11674080&&    
  \cr
&2&2&&     0&0&0&&    0&0&0&&    -18504&-20520& &&      \cr
\noalign{\hrule}&n_3&n_4\cr}
}$${\bf Table III.4}: Numbers of invariants ${1\over 5} N_4(\nv)$ for $X_{III}$}
\vskip7pt}}

{\vbox{\ninepoint{
$$
\vbox{\offinterlineskip\tabskip=0pt\halign{\strut\vrule#
&\hfil~$\ss{#}$~\hfil &\hfil~$\ss{#}$~\hfil &\vrule#
&\hfil ~$\ss{#}$~ \hfil&\hfil ~$\ss{#}$~ \hfil&\hfil ~$\ss{#}$~ \hfil&\vrule#
&\hfil ~$\ss{#}$~ \hfil&\hfil ~$\ss{#}$~ \hfil&\hfil ~$\ss{#}$~ \hfil&\vrule#
&\hfil ~$\ss{#}$~ \hfil&\hfil ~$\ss{#}$~ \hfil&\hfil ~$\ss{#}$~ \hfil&\hfil ~$\ss{#}$~ \hfil
&\vrule#\cr\noalign{\hrule}
&   &n_1    &&  0& & &&     1 & & &&  2 & & &&\cr
&   &n_2 &&     0&1&2&& 0  &1 & 2&& 0 & 1& 2&&\cr
\noalign{\hrule}
&0&0&&      &-2&0&&    0&3&10&&    0&0&-12&&      \cr
&0&1&&     0&0&0&&    0&7&22&&    0&0&-129&&      \cr
&0&2&&     0&0&0&&    0&0&0&&    0&0&-34&&      \cr
&1&0&&     0&480&0&&    0&-960&-3840&&    0&0&4800&&      \cr
&1&1&&     0&0&0&&    0&-2040&-7800&&    0&0&59640&&      \cr
&1&2&&     0&0&0&&    0&0&0&&    0&0&12720&&      \cr
&2&0&&     0&282888&960&&    0&118170&678240&&    0&-10260&-947280&&      \cr
&2&1&&     0&0&0&&    0&400050&1408680&&    0&-48240&-15432120&&      \cr
&2&2&&     0&0&0&&    0&0&0&&    0&-10260& &&      \cr
\noalign{\hrule}&n_3&n_4\cr}
}$${\bf Table III.5}: Numbers of invariants $N_5(\nv)$ for $X_{III}$}
\vskip7pt}}

{\vbox{\ninepoint{
$$
\vbox{\offinterlineskip\tabskip=0pt\halign{\strut\vrule#
&\hfil~$\ss{#}$~\hfil &\hfil~$\ss{#}$~\hfil &\vrule#
&\hfil ~$\ss{#}$~ \hfil&\hfil ~$\ss{#}$~ \hfil&\hfil ~$\ss{#}$~ \hfil&\vrule#
&\hfil ~$\ss{#}$~ \hfil&\hfil ~$\ss{#}$~ \hfil&\hfil ~$\ss{#}$~ \hfil&\vrule#
&\hfil ~$\ss{#}$~ \hfil&\hfil ~$\ss{#}$~ \hfil&\hfil ~$\ss{#}$~ \hfil&\hfil ~$\ss{#}$~ \hfil
&\vrule#\cr\noalign{\hrule}
&   &n_1    &&  0& & &&     1 & & &&  2 & & &&\cr
&   &n_2 &&     0&1&2&& 0  &1 & 2&& 0 & 1& 2&&\cr
\noalign{\hrule}
&0&0&&      &0&0&&    0&0&0&&    0&0&0&&      \cr
&0&1&&     0&0&0&&    0&0&0&&    0&0&0&&      \cr
&0&2&&     0&0&0&&    0&0&0&&    0&0&0&&      \cr
&1&0&&     480&480&0&&    252&-960&-1920&&    0&0&2400&&      \cr
&1&1&&     0&0&0&&    372&-960&-1920&&    0&0&19200&&      \cr
&1&2&&     0&0&0&&    0&0&0&&    0&0&2400&&      \cr
&2&0&&     960&565776&960&&    10260&236340&678240&&    -18504&-20520&-947280&
&      \cr
&2&1&&     0&0&0&&    13860&239940&678240&&    -128124&-48240&-8065440&&       
 \cr
&2&2&&     0&0&0&&    0&0&0&&    -32124&-27720& &&      \cr
\noalign{\hrule}&n_3&n_4\cr}
}$${\bf Table III.6}: Numbers of invariants $N_6(\nv)$ for $X_{III}$ }
\vskip7pt}}

\listrefs
\end